\documentclass[11pt,preprint]{aastex}
\usepackage{emulateapj5,apjfonts,psfig}

\renewcommand{\tabcolsep}{3pt}

\newcommand{\NH}{{$N_{\rm H}$}}
\newcommand{\LX}{{$L_{\rm X}$}}

\newcommand{\LHa}{$L_{{\rm H} \alpha}$}

\newcommand{\eps}{ergs s$^{-1}$}
\newcommand{\pcm}{cm$^{-2}$}
\newcommand{\ps}{s$^{-1}$}

\newcommand{\chandra}{{\it Chandra}}
\newcommand{\asca}{{\it ASCA}}
\newcommand{\Ginga}{{\it Ginga}}
\newcommand{\Einstein}{{\it Einstein}}
\newcommand{\rosat}{{\it ROSAT}}
\newcommand{\exosat}{{\it EXOSAT}}
\newcommand{\sax}{{\it BeppoSAX}}

\newcommand{\gtsima}{$\; \buildrel > \over \sim \;$}
\newcommand{\simgt}{\lower.5ex\hbox{\gtsima}}
\newcommand{\ltsima}{$\; \buildrel < \over \sim \;$}
\newcommand{\simlt}{\lower.5ex\hbox{\ltsima}}

\slugcomment{To appear in {\it The Astrophysical Journal Supplement series.}}

\lefthead{Terashima et al.}
\righthead{X-ray Properties of LINERs and Low-luminosity Seyferts}

\begin{document}

\title{X-ray Properties of LINERs and Low-luminosity Seyfert Galaxies
Observed with {\it ASCA}. I. Observations and Results}

\author{
Yuichi Terashima\altaffilmark{1,2}, 
Naoko Iyomoto\altaffilmark{1},
Luis~C.~Ho\altaffilmark{3}, and 
Andrew~F.~Ptak\altaffilmark{4,5}
}

\altaffiltext{1}{Institute of Space and Astronautical Science, 3-1-1 Yoshinodai,
Sagamihara, Kanagawa 229-8510, Japan.}

\altaffiltext{2}{Astronomy Department, University of Maryland, College Park, 
MD 20742.}

\altaffiltext{3}{The Observatories of the Carnegie Institution of Washington, 
813 Santa Barbara St., Pasadena, CA 91101-1292..}

\altaffiltext{4}{Department of Physics, Carnegie Mellon University, 5000 
Forbes Ave., Pittsburgh, PA 15213.}

\altaffiltext{5}{Department of Physics and Astronomy, Johns Hopkins 
University, 3400 North Charles St., Baltimore, MD 21218-2686.}

\begin{abstract}

This paper presents a comprehensive study of the X-ray properties of 
low-ionization nuclear emission-line regions (LINERs) and low-luminosity 
Seyfert galaxies based on observations obtained with the {\asca} satellite. 
We analyzed data of 53 observations of 21 LINERs and 17 low-luminosity 
Seyferts.  X-ray emission has been detected in all but one object. The X-ray 
luminosities in the 2--10 keV band range from $4\times10^{39}$ {\eps} to 
$5\times10^{41}$ {\eps}, which are 1--3 orders of magnitude smaller than in 
classical Seyfert galaxies. The X-ray spectra of most objects are well 
described by a canonical model which consists of (1) a soft component from a 
thermal plasma with $kT\,<\,1$ keV and (2) a hard component represented by a 
power law with a photon index of $\Gamma\approx1.8$ or thermal bremsstrahlung 
emission with $kT\,\approx$ 10 keV. Several objects do not require the soft 
thermal component, and their continua are well fitted by a single power-law
model. Some objects show heavy absorption with column densities in excess
of $10^{23}$ {\pcm}. We detect in several objects Fe~K line emission with 
equivalent widths ranging from 50 eV to 2 keV.

Variability on timescales less than a day is uncommon in our sample. By 
comparing multiple observations made with {\asca} or with published 2--10 keV 
observations from other satellites, we show that at least eight objects are 
variable on timescales of a week to several years.  We find that the 
morphologies of many objects, both in the soft and hard bands, are consistent 
with being pointlike relative to the telescope point-spread function; a few
are clearly extended in either or both energy bands.

The second paper of this series will discuss the physical interpretation of 
the X-ray emission and its implications for low-luminosity active galactic 
nuclei.

\end{abstract}


\section{Introduction}

It is widely accepted that many bright galaxies possess low-level nuclear
activity. The Palomar optical spectroscopic survey of nearby galaxies 
by Ho, Filippenko, \& Sargent (1995, 1997a, 1997b) showed that more than 
40\% of the 486 galaxies with $B_T\le12.5$ mag and $\delta>0^{\circ}$ have
optical spectra classified as Seyfert nuclei, low-ionization nuclear
emission-line regions (LINERs; Heckman 1980), or transition objects (spectra 
intermediate between LINERs and \ion{H}{2} nuclei)\footnote{In this 
paper, we use the classifications based on optical emission-line ratios given 
by Ho et al. (1997a).}. It is not clear, however, what fraction of these
objects are genuine active galactic nuclei (AGNs), which we take to be objects 
powered by nonstellar processes.  Extensive observations have been made in
various wavelengths to clarify the nature of these classes of objects (for 
reviews see Ho 1999, 2002). X-ray observations play a key role in this regard.
This paper presents the first systematic analysis of a large sample of LINERs 
and low-luminosity Seyfert galaxies observed with the {\it ASCA} satellite.

The gross spectroscopic characteristics of LINERs can be explained by a
variety of ionization mechanisms, including traditional photoionization by 
a low-luminosity AGN (LLAGN), photoionization by a population of unusually 
hot stars, and mechanical heating by shock waves. Recent reviews on this topic 
have been given by Filippenko (1996) and Barth (2002).  If LINERs are 
low-luminosity versions of AGNs, which are powered by accretion onto massive 
black holes, we expect them to emit nonthermal, hard X-rays. On the other 
hand, X-rays from thermal plasmas are expected from shock-heated gas. Hot 
stars also have soft thermal spectra in the X-ray band. Thus, X-ray 
observations can help discriminate between the various models for LINERs.  We 
note, however, that hard X-rays (hereinafter defined as $E\,>\,2$ keV) are not 
a signature unique to AGNs: potential contributors to the hard-energy band 
include X-ray binaries, supernovae, and hot ($kT\approx10$ keV) gas produced 
by starburst activity.  An unambiguous search for AGNs using X-ray data thus 
requires careful consideration of the spectral, structural, and variability 
properties of the sources.

Many Seyfert galaxies have been identified in the Palomar survey. The
median H$\alpha$ luminosity of these sources is {\LHa}
$\approx\,2\times 10^{39}$ {\eps}, which is more than 2--3 orders of
magnitude lower than in classical Seyfert galaxies (e.g., Dahari \&
De~Robertis 1988; Whittle 1992). X-ray observations of such
low-luminosity Seyferts (hereinafter LLSeyferts) are necessary to
determine whether they are genuine AGNs and whether they are similar
to luminous AGNs.

In addition to studying the origin of activity in LINERs and LLSeyferts, 
the X-ray properties of these sources themselves hold great interest. By 
defining an LLAGN sample using strong AGN candidates based on detailed study 
of individual objects, we can investigate the nature of LLAGNs as a class, 
compare them with more luminous AGNs,  and potentially constrain models for 
accretion flows.  For example, the spectral properties (continuum shape, 
absorption column, Fe~K-line strength, presence or absence of ionized 
absorbers) and variability characteristics can be used to probe the structure 
of the central engine, in a region of parameter space that is likely to be 
quite unique because of the extreme conditions involved.  In the regime of low 
mass accretion rate, it has been suggested that advection-dominated accretion 
flows may be present (Narayan \& Yi 1994; see reviews by Kato, Fukue, \& 
Mineshige 1998 and Quataert 2001).  Comparisons between observations and 
theoretical predictions are critical for testing these and other ideas.

X-ray observations of LINERs and LLSeyferts, both in the soft and hard 
energy bands, have been published by a number of authors.  However, most, if 
not all, previous studies have been limited either by sample size 
or restricted energy bandwidth. In the soft X-ray band, data based on 
observations with {\Einstein}\ and {\rosat}\ have been presented by Halpern \& 
Steiner (1983), Koratkar et al. (1995), Komossa, B\"ohringer, \& Huchra 
(1999), Roberts \& Warwick (2000), Lira, Lawrence, \& Johnson (2000), and 
Halderson et al. (2001).  The primary disadvantages of observations in the soft 
X-ray band are contamination by emission from hot gas, often observed 
in galaxies, and decreased sensitivity to absorbed AGNs.  The number of 
hard X-ray observations of LINERs and LLSeyferts prior to {\asca} is very 
limited. Because of sensitivity limitations, only a few of the brightest 
objects have been observed (e.g., M81 and NGC 3998).  The significantly 
improved sensitivity and imaging capability in the broad energy band 
(0.5--10 keV) of {\asca} provide an unprecedent opportunity to study 
systematically objects with low-level activity.

This paper presents a comprehensive analysis of {\asca}
observations of LINERs and LLSeyferts.  In a subsequent paper
(Terashima et al. 2001, hereinafter Paper II), we discuss the origin of
the X-ray emission using the X-ray properties derived here, along with
comparisons with data from other wavelengths, and the broader implications 
for LLAGNs. Previous {\asca} results based on smaller samples can be found in
Serlemitsos, Ptak, \& Yaqoob (1996), Terashima (1998, 1999a, 1999b),
Awaki (1999), and Ptak et al. (1999). The references for
{\asca} results on individual objects are given in Terashima, Ho, \&
Ptak (2000a) and section 9 in this paper. This paper is organized as
follows. We define the sample in section 2. Section 3 briefly presents the 
observations and data reduction. The results of a spectral analysis, a 
summary of the best-fit spectral parameters, and a tabulation of the derived 
fluxes and luminosities are presented in sections 4, 5, and 6, respectively.  
Sections 7 and 8 are devoted to timing and imaging analyses, respectively. 
Notes on individual objects are given in section 9. Section 10 summarizes the 
main results.

\vspace{1cm}

\section{The Sample}

The galaxies in this study were chosen from the Palomar survey of nearby 
galaxies.  We selected objects classified by Ho et al. (1997a) as LINERs and 
Seyfert galaxies which were either in the {\asca} data archives as of 1999 
December or belonged to our own proprietary observing programs.  We decided 
to analyze Seyferts with {\LHa} $<\,10^{41}$ {\eps}, roughly the luminosity of 
NGC 4051, the lowest luminosity ``classical'' Seyfert galaxy.  Luminous 
Seyferts which have been extensively studied were excluded.  The list of 
galaxies we analyzed is shown in Table 1, where we list the Hubble types, 
distances, heliocentric velocities, and optical spectroscopic classifications. 
We use the distances adopted by Ho et al. (1997a), which are taken from Tully 
(1988) and are based on a Hubble constant of $H_0$ = 75 km s$^{-1}$ Mpc$^{-1}$.
The velocities are taken from the NASA/IPAC Extragalactic Database (NED).

We omitted a few objects with ambiguous spectral classification.  The LINER~2 
NGC 4486 (M87) was also excluded because of the complexity of its X-ray 
emission (Matsumoto et al. 1996; Reynolds et al. 1996; Allen, Di~Matteo, \& 
Fabian 2000). The X-ray spectrum of M87 consists of multiple components, 
namely emission from an AGN and hot gas from the host galaxy and the Virgo 
cluster. Properly measuring the spectral parameters of the AGN component 
requires detailed image analysis and a sophisticated modeling of the thermal 
emission, including the effects of multiple temperatures, metal abundances, 
and temperature and metalicity gradients.  This is beyond the scope of this 
paper.

The two transition objects NGC 4569 and NGC 4192, which were analyzed
by Terashima et al. (2000b), were included. Several objects in the
southern hemisphere (NGC 1097, 1365, 1386, and 4941), which have good
{\asca} data, were also used. The optical classifications for these
objects were taken from Phillips et al. (1984), Storchi-Bergmann,
Baldwin, \& Wilson (1993), V\'eron-Cetty \& V\'eron (1986), and
Storchi-Bergmann \& Pastoriza (1989).

The final sample consists of nine LINER~1s, 12 LINER~2s, eight Seyfert~1s, and 
nine Seyfert~2s. Some of these objects were observed at least twice with 
{\asca}. Whenever possible we analyzed the multiple observations to achieve 
better photon statistics and to search for variability.  In total there 
are 53 observations, as summarized in Table~2. Section 9 gives comments on
the observations which were not used.

\section{Observations and Data Reduction}

A log of the {\asca} observations is shown in Table 2. It gives the start date 
of the observations, observation modes, count rates obtained for the SIS and 
GIS detectors, and net exposure times after data screening. The GIS detectors 
(Ohashi et al. 1996; Makishima et al. 1996) were operated in the PH mode with 
the nominal bit assignment for all the observations. The spread discriminator 
was not turned on in the observations of NGC 3079, 4258, and 5194. The 
clocking and telemetry modes of the SIS detectors (Burke et al. 1994; 
Yamashita et al. 1999) are also summarized in Table 2. In the SIS observations, 
in which both the faint and bright modes were used, the two data sets were 
combined after the faint-mode data were converted to the format of the 
bright-mode data; the only exception was the 1995 observation of NGC 4579 
(see Terashima et al. 1998a).

The data were screened using the following set of criteria: (1) elevation 
angle above the Earth's limb greater than 5$^{\circ}$, (2) cut-off rigidity
greater than 6 GeV c$^{-1}$, (3) avoidance of South Atlantic Anomaly, and
(4) elevation angle above the day Earth's limb greater than 25$^{\circ}$ 
(SIS only).  X-ray spectra and light curves were extracted from a circular 
region with a typical radius of 3$^\prime$--4$^\prime$ for SIS and 6$^\prime$ 
for GIS. A smaller extraction radius was used in some cases to avoid a 
bright nearby source.  Background spectra were accumulated from a source-free 
region in the same field.

\newpage

\section{Spectral Analysis}

We fitted the SIS and GIS spectra simultaneously using the XSPEC 
spectral-fitting package (version 10). The spectra of the two SIS
detectors were combined, as were those from the two GIS detectors.
The quoted errors are at the 90\% confidence level for one interesting
parameter ($\Delta\chi^2$ = 2.7). The Galactic absorption column
densities are derived from \ion{H}{1} measurements by Murphy et
al. (1996), when available, or else from those by Dickey \& Lockman
(1990).  The results of the spectral fits are presented in
Figures~1{\it a}--1{\it l}.  In each figure, the left column displays
the SIS spectra, the right column the GIS spectra. Within each plot,
the upper panel shows the data and the best-fit model, and the lower
panel shows the residuals of the fit. For multi-component models, we
plot each component with a different line type in the upper panel.

\subsection{Simple Models}

We fitted the spectra with a simple power-law model modified by
photoelectric absorption along the line of sight.  We added a Gaussian
component to those objects which showed an indication of a linelike
emission feature around 6 keV.  The results of the fits are shown in
Table 3. Errors are not shown for the fits with poor reduced $\chi^2$
because the error estimations using the $\Delta \chi^2 = 2.7$ criterion
are not valid. Acceptable fits were obtained 
for only several objects: NGC 3031, 3147, 3507, 3998, 4203, 4579 (1998 
observation), 4639, and 5033. The other objects show 
significant residuals in the soft-energy band and/or a more complicated 
continuum shape. In many objects emission lines are seen in the region 
0.6--2 keV, indicative of the presence of a sub-keV thermal plasma.  
In the next subsection we evaluate a model consisting of a hard component and 
a soft thermal plasma component.

%- partial covering model

Several objects show both a heavily absorbed hard component and a less absorbed
soft component. In order to fit such a spectral shape, we tried a partially 
covered power-law model. This model is equivalent to one consisting of a 
combination of an absorbed power law plus and a scattered power law, as is 
often observed in Seyfert 2 galaxies (e.g., Turner et al. 1997; Awaki et al. 
2000) if one assumes that the power-law slopes of these two components are 
same. Although the power-law slopes of the soft and hard components could be 
different, we assumed that they have the same slope because of limited photon 
statistics. We considered this model for NGC 1052, 2273, 2639, 2655, 4565, and 
4941; the results are shown in Table 4. Some of these objects (NGC 1052, 2639, 
2655, and 4565) seem to have a softer component accompanied by emission lines 
from a thermal plasma. In the next subsection, we examine a multi-component 
model which consists of a partially covered power law and a soft thermal plasma.

We also tried a thermal bremsstrahlung model instead of a power-law
model.  Again, we added a Gaussian line when a linelike feature was
seen around 6 keV. We did not attempt the thermal bremsstrahlung model
for those objects with a heavily absorbed continuum shape, since the
temperature cannot be well determined and a heavily absorbed continuum
is likely to indicate the presence of an obscured AGN. The result of
these fits are given in Table 5. Again errors are not shown for the
fits with poor reduced $\chi^2$. This thermal model resulted in
significantly worse fits for those objects which were well fitted with
a power-law (plus Gaussian) model, except for the relatively faint
objects NGC 3507 and NGC 4639, for which $\chi^2$ values similar to
those of the power-law fits were obtained.  Note that these two
galaxies are fainter than the rest of objects which are well fitted
with a simple power-law model.

\subsection{Two-component Models}

Only a few objects were represented successfully by a simple power-law model;
the majority indicated the presence of emission from a soft thermal plasma in 
addition to the hard component. Therefore, we investigated multiple-component 
models.  We first adopted a model consisting of an absorbed power law plus 
a Raymond-Smith (RS; Raymond \& Smith 1977) thermal plasma modified by 
Galactic absorption.  The abundance was allowed to vary only if a meaningful
constraint could be obtained. In many cases the abundance was not
well constrained and was fixed at 0.1 solar or 0.5 solar, values 
typically observed for thermal plasmas in normal and starburst
galaxies. A Gaussian line was also added if a linelike feature was seen
around 6 keV. When the continuum shape suggests the presence of both a
heavily absorbed and a lightly absorbed component, we used a partially
covered power-law model instead of a simple power law; this applied to 
NGC 1052, 2655, and 4565. Both models were tried for NGC 2639.  The
results of the fits are summarized in Table 6.  The fits were successful 
for all the objects except NGC 4258 (1993 observation) and NGC 4636, which 
show very complicated spectra. 

Thus, a model consisting of an RS plasma plus a hard power-law component 
can be applied to nearly all objects, with a few objects not requiring the 
soft component. This ``canonical'' model successfully represents the spectra 
of the LINERs and LLSeyferts discussed here. Ptak et al. (1999) found that 
this model was also applicable to starburst galaxies.

In objects whose continuum was not heavily absorbed, we also evaluated the 
possibility that in the two-component model the power law could be substituted 
by a thermal bremsstrahlung model.  In all the cases, we obtained 
$\chi^2$ values (Table~7) similar to those for the RS plasma plus 
power-law model.

Some objects show apparently weak Fe~L emission compared to other
emission lines such as the K lines from O, Ne, Mg, and Si.  For these
objects, we fitted the abundances of Fe and the $\alpha$-processed
elements separately. We fixed the abundances of the $\alpha$-processed 
elements at 0.5 solar because the photon statistics are too limited to 
allow these abundances to be constrained individually. The abundance of Fe
was allowed to vary.  We assumed a power-law shape for the hard component in
these fits. Only objects with a relatively good signal-to-noise ratio for the
soft component were fitted with this model. The results are tabulated in
Table 8.  The $\chi^2$ values show improvement for NGC 1097, 3079, 4636, and
5194, compared to the case of a RS plasma with solar abundance ratios.

\subsection{Complex Models}

The spectra of only two objects, NGC 4258 (1993 observation) and NGC 4636, 
cannot be fitted with the models described above.  As discussed in section 9, 
we applied more complex models to these objects. 

 \clearpage

\begin{figure*}[t]
\figurenum{1{\it a}}
\psfig{file=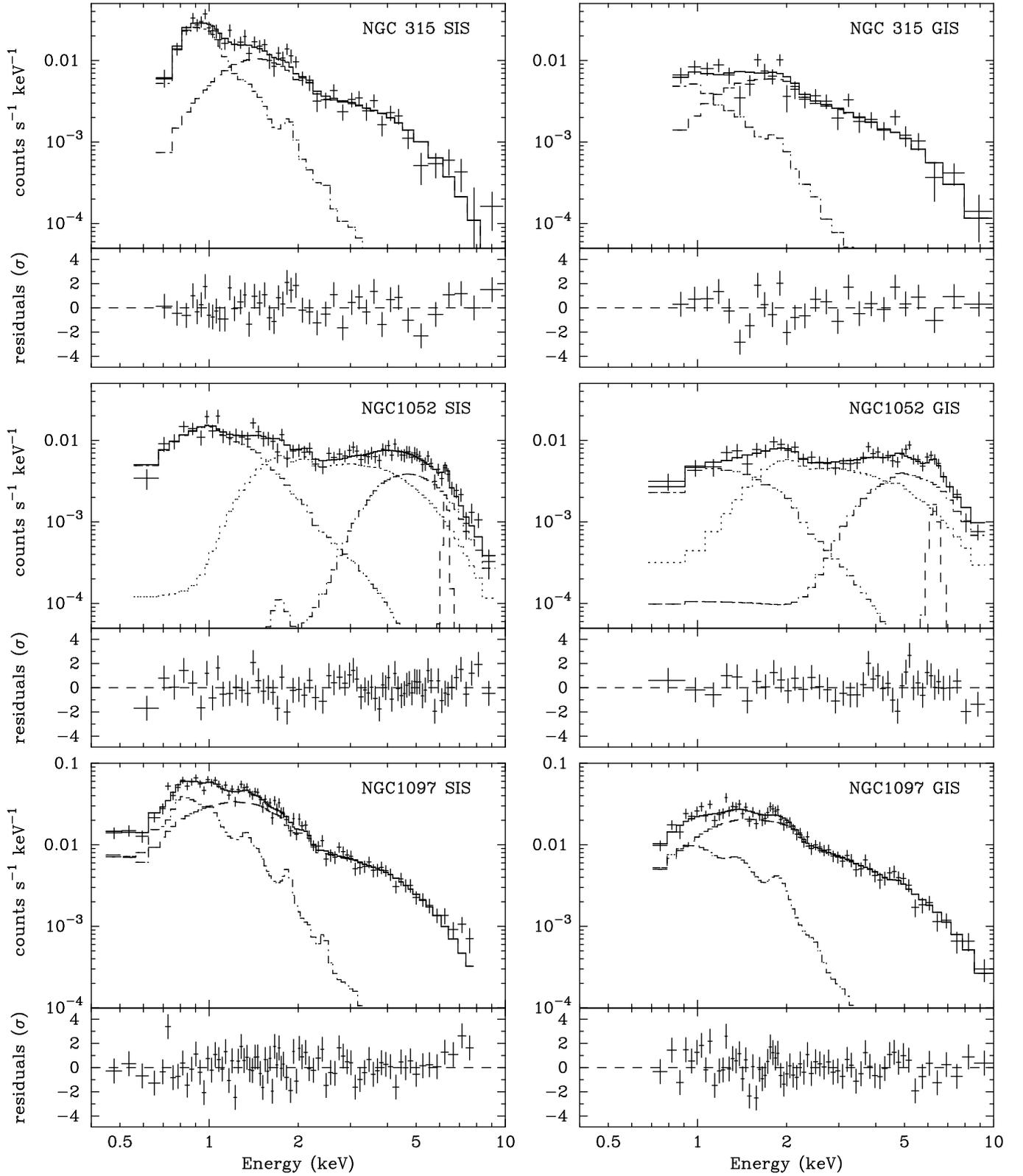,width=18.5cm,angle=0}
\figcaption[]{
{\asca} SIS ({\it left}) and GIS ({\it right}) spectra.  In each plot, the
upper panel shows the data and the model, and the lower panel shows the
residuals of the fit. ({\it a}) NGC 315 ({\it top}), NGC 1052 ({\it middle}),
NGC 1097 ({\it bottom}).
}
\end{figure*}
 
\clearpage
\begin{figure*}[t]
\figurenum{1{\it b}}
\psfig{file=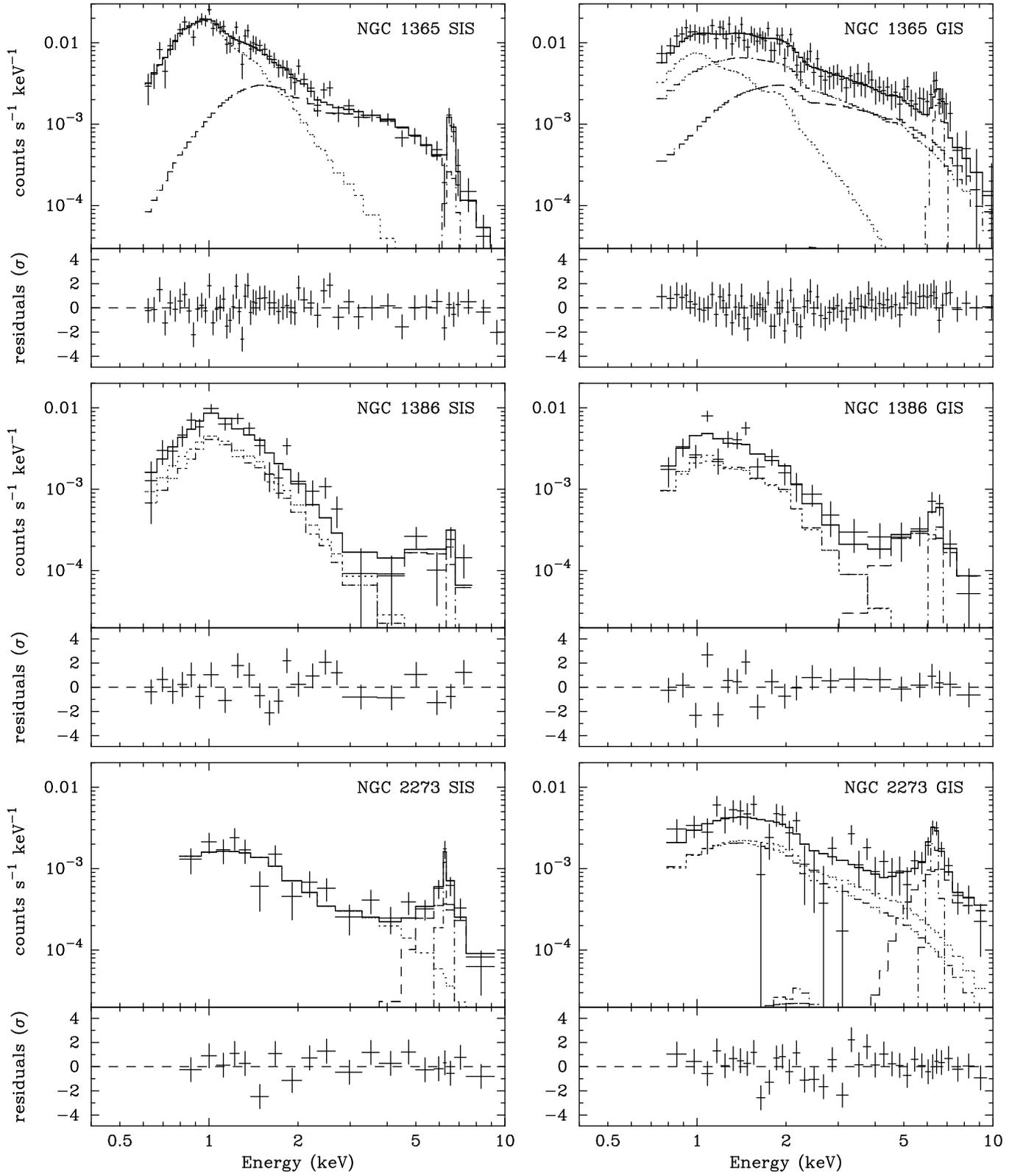,width=18.5cm,angle=0}
\figcaption[]{
Same as in Figure 1{\it a}.  ({\it b}) NGC 1365 ({\it top}), NGC 1386
({\it middle}), NGC 2273 ({\it bottom}).
}
\end{figure*}

\clearpage 
\begin{figure*}[t]
\figurenum{1{\it c}}
\psfig{file=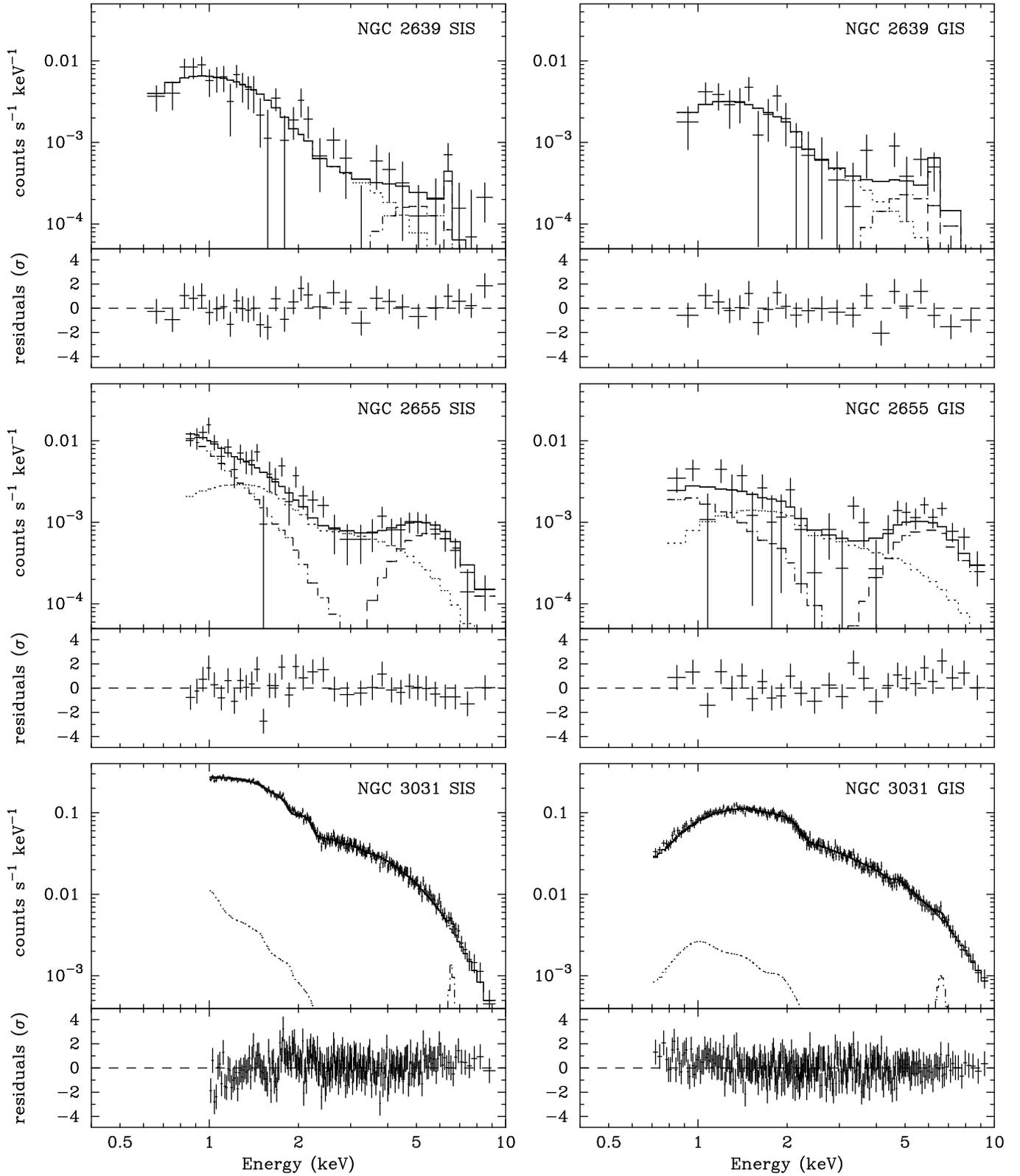,width=18.5cm,angle=0} 
\figcaption[]{
Same as in Figure 1{\it a}.  ({\it c}) NGC 2639 ({\it top}), NGC 2655
({\it middle}), NGC 3031 ({\it bottom}).
} 
\end{figure*}

\clearpage 
\begin{figure*}[t]
\figurenum{1{\it d}}
\psfig{file=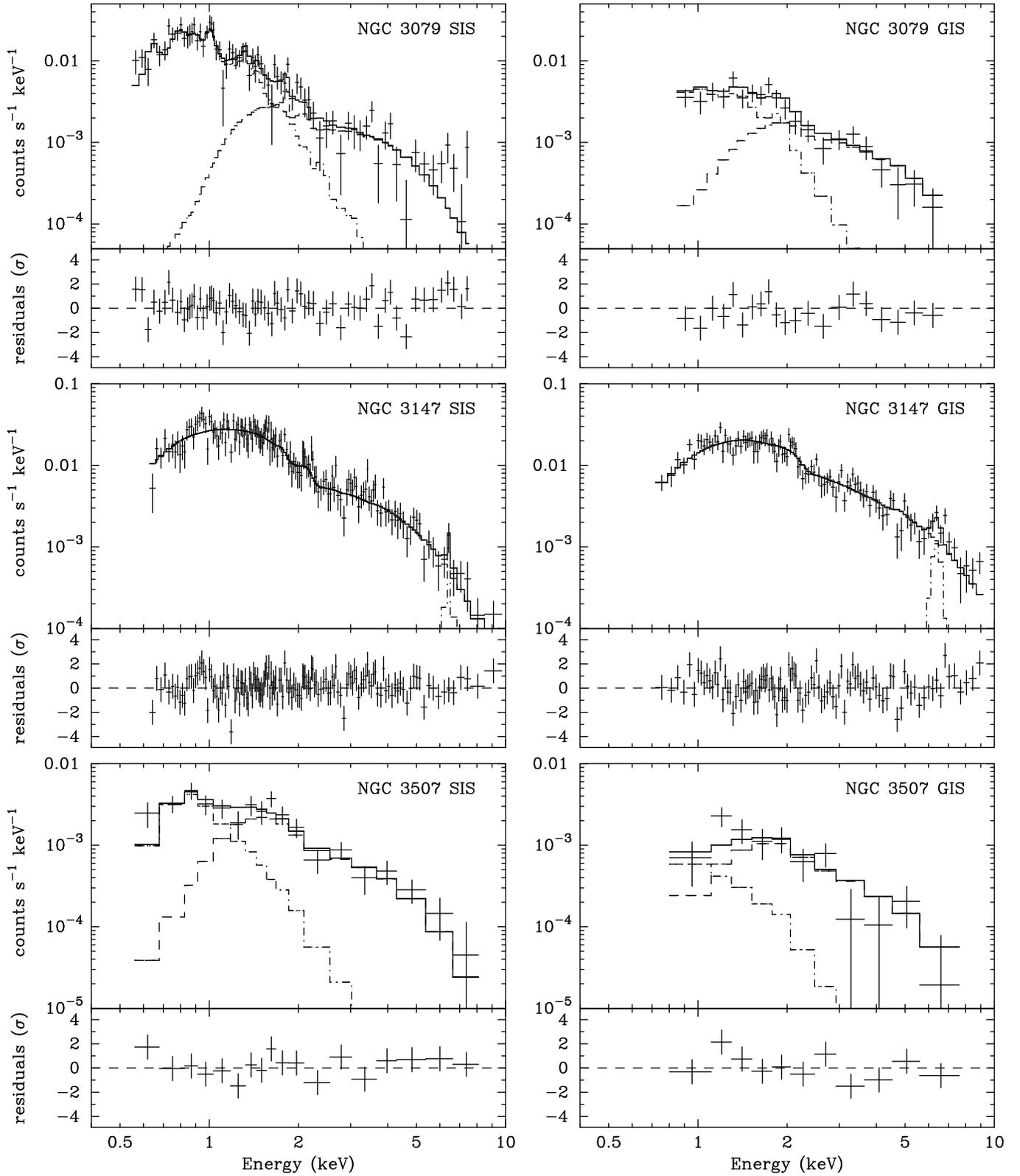,width=18.5cm,angle=0} 
\figcaption[]{
Same as in Figure 1{\it a}.  ({\it d}) NGC 3079 ({\it top}), NGC 3147
({\it middle}), NGC 3507 ({\it bottom}).
} 
\end{figure*}

\clearpage 
\begin{figure*}[t]
\figurenum{1{\it e}}
\psfig{file=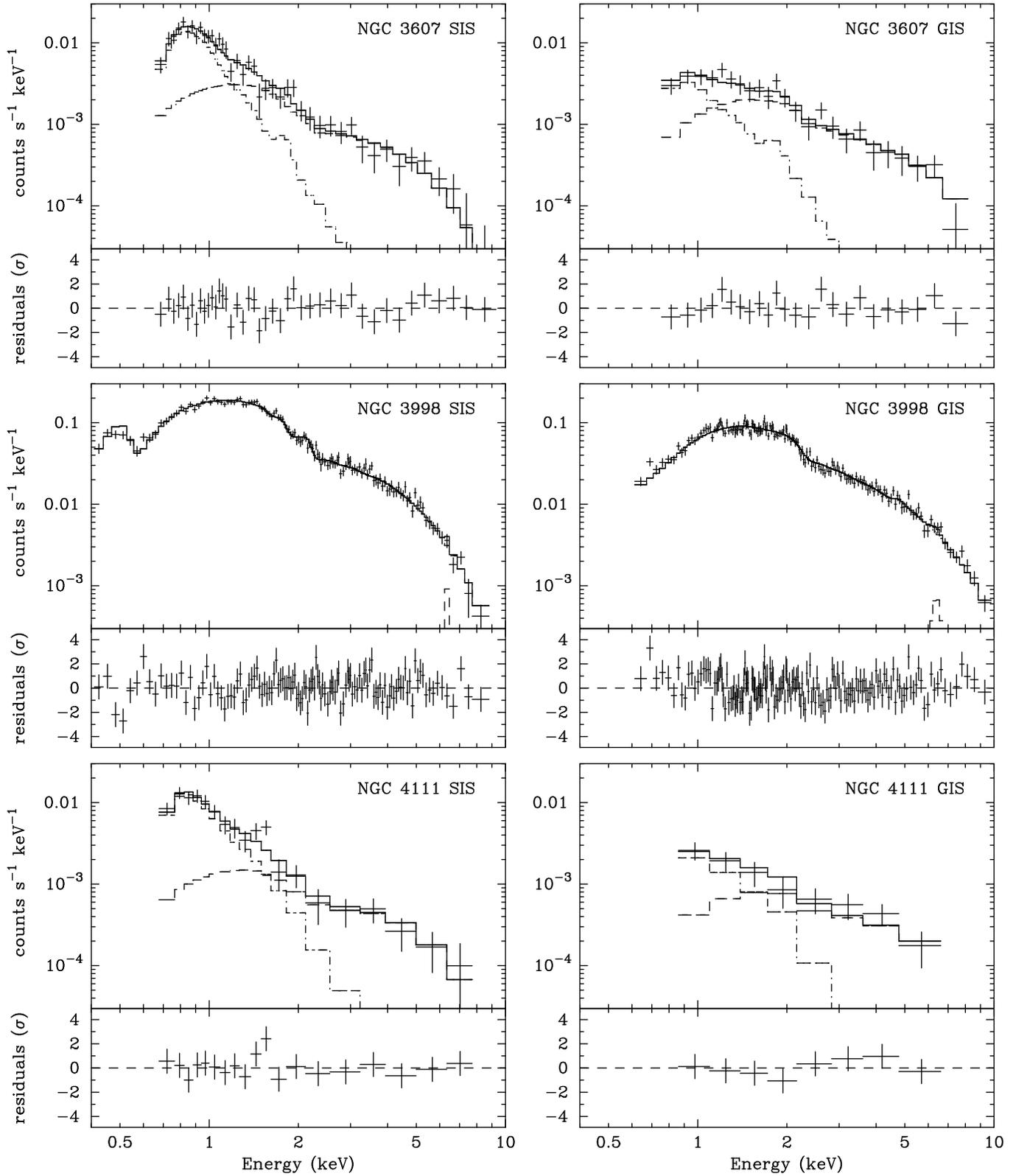,width=18.5cm,angle=0} 
\figcaption[]{
Same as in Figure 1{\it a}.  ({\it e}) NGC 3607 ({\it top}), NGC 3998
({\it middle}), NGC 4111 ({\it bottom}).
} 
\end{figure*}

\clearpage 
\begin{figure*}[t]
\figurenum{1{\it f}}
\psfig{file=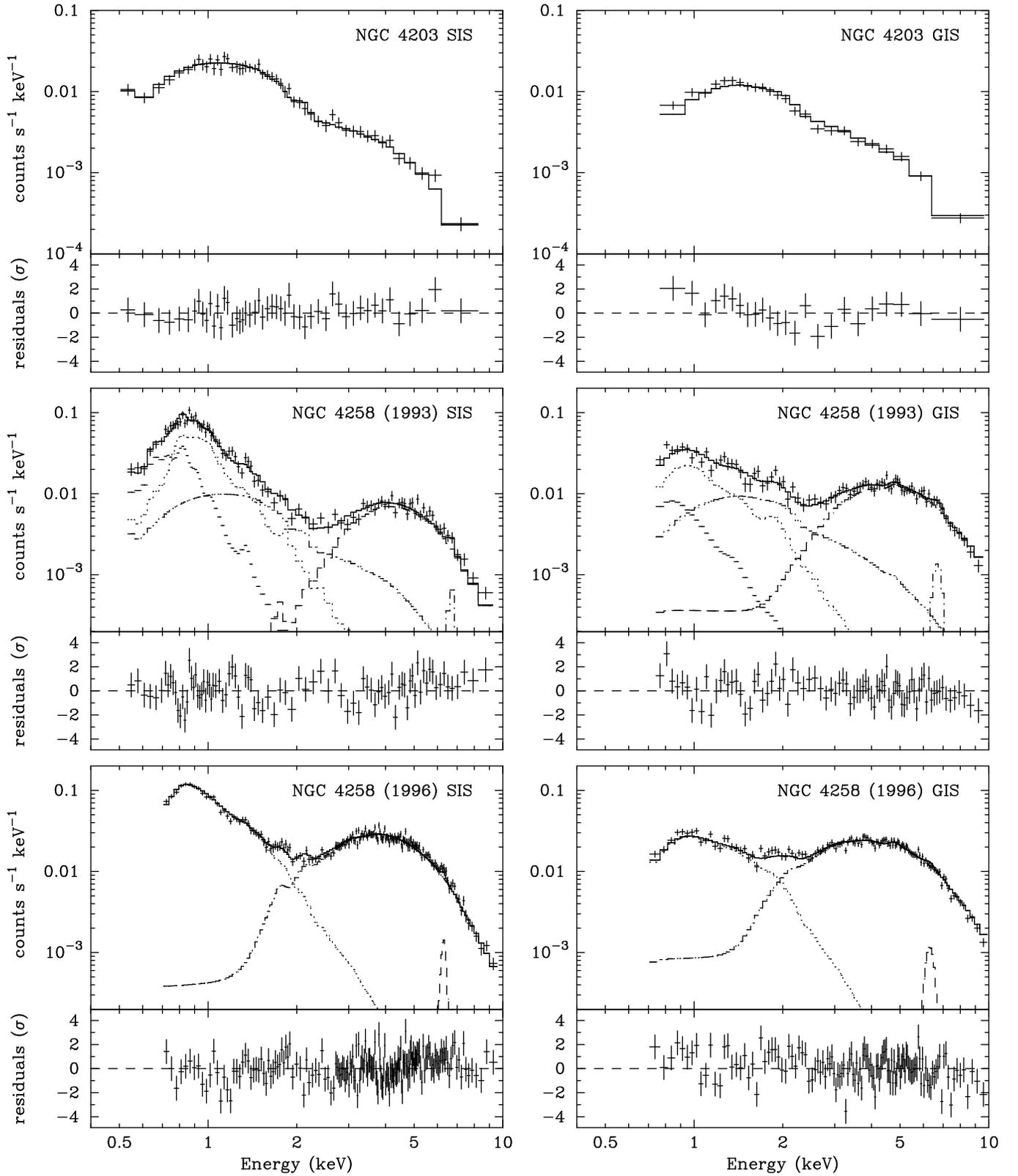,width=18.5cm,angle=0} 
\figcaption[]{
Same as in Figure 1{\it a}.  ({\it f}) NGC 4203 ({\it top}), NGC 4258 (1993)
({\it middle}), NGC 4258 (1996) ({\it bottom}).
} 
\end{figure*}

\clearpage 
\begin{figure*}[t]
\figurenum{1{\it g}}
\psfig{file=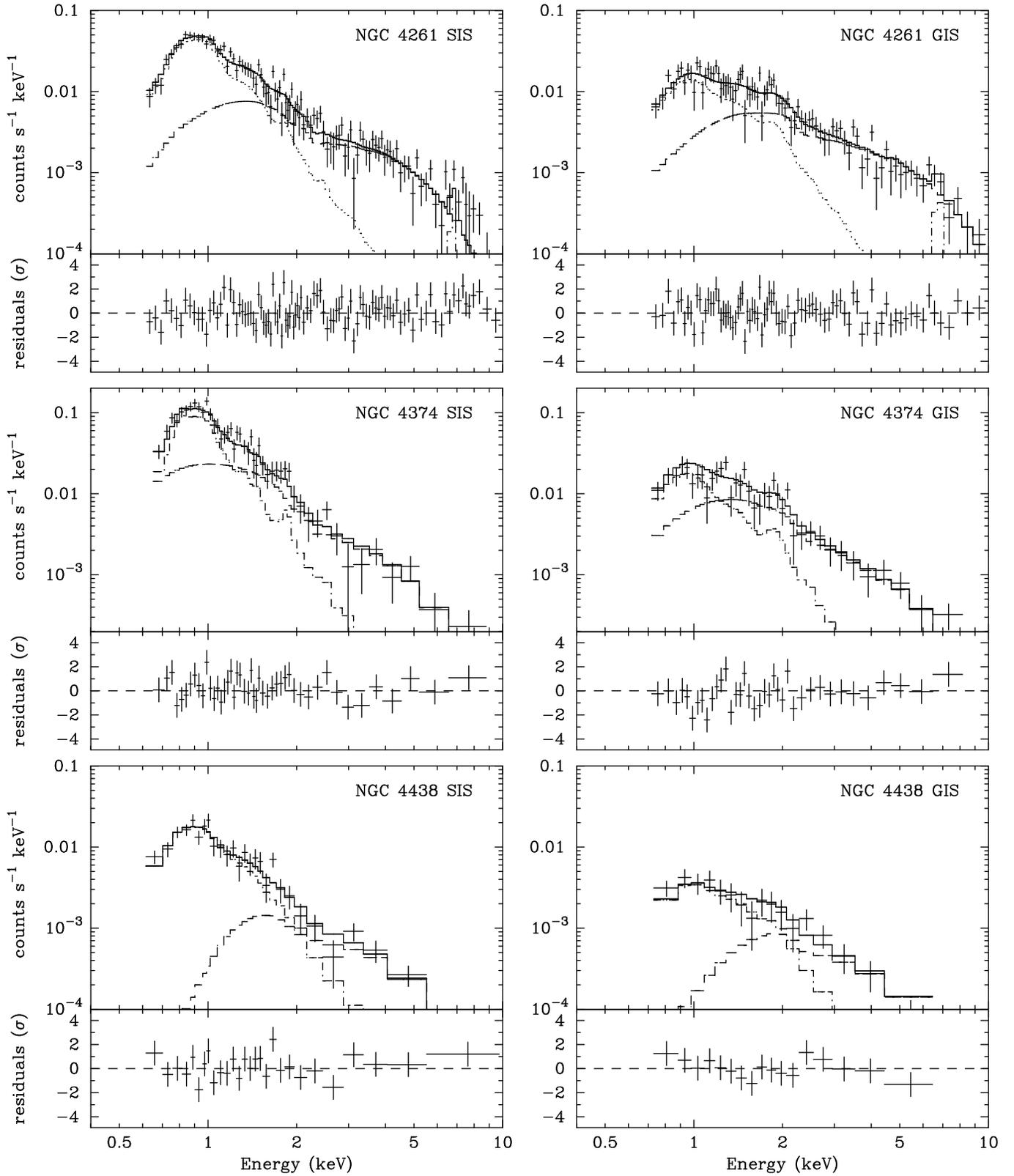,width=18.5cm,angle=0} 
\figcaption[]{
Same as in Figure 1{\it a}.  ({\it g}) NGC 4261 ({\it top}), NGC 4374
({\it middle}), NGC 4438 ({\it bottom}).
} 
\end{figure*}

\clearpage 
\begin{figure*}[t]
\figurenum{1{\it h}}
\psfig{file=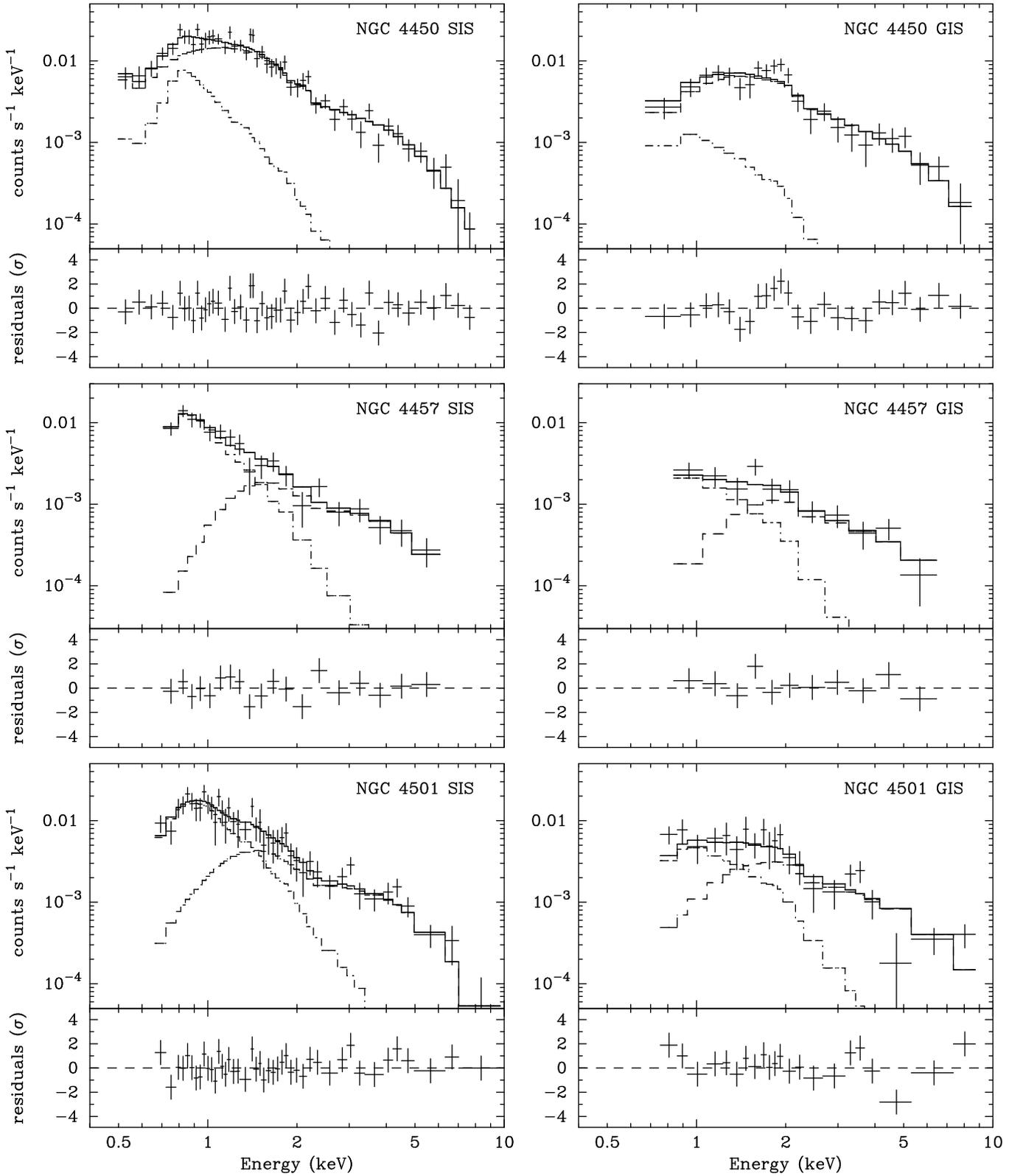,width=18.5cm,angle=0} 
\figcaption[]{
Same as in Figure 1{\it a}.  ({\it h}) NGC 4450 ({\it top}), NGC 4457
({\it middle}), NGC 4501 ({\it bottom}).
} 
\end{figure*}

\clearpage 
\begin{figure*}[t]
\figurenum{1{\it i}}
\psfig{file=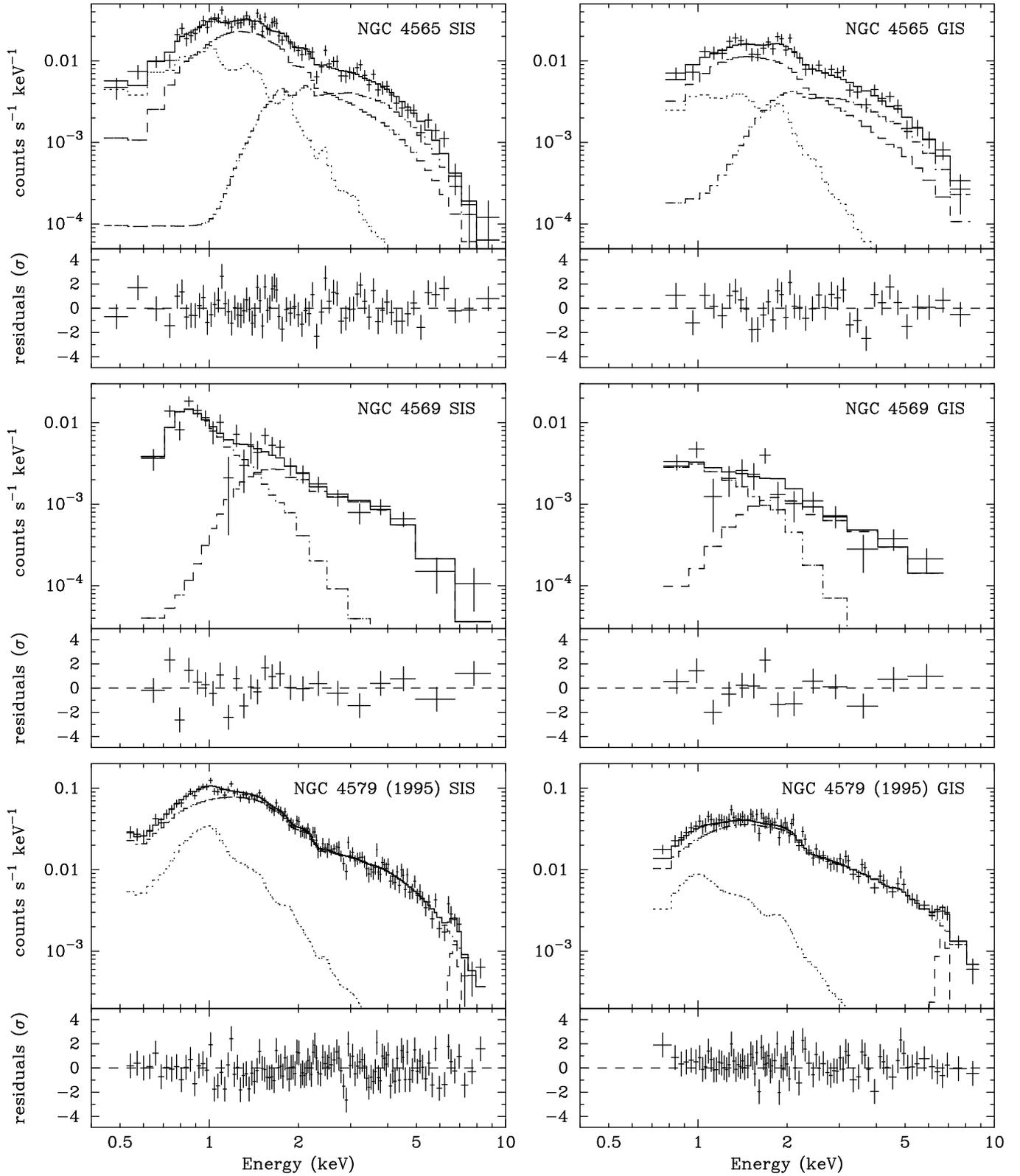,width=18.5cm,angle=0} 
\figcaption[]{
Same as in Figure 1{\it a}.  ({\it i}) NGC 4565 ({\it top}), NGC 4569
({\it middle}), NGC 4579 (1995) ({\it bottom}).
} 
\end{figure*}

\clearpage 
\begin{figure*}[t]
\figurenum{1{\it j}}
\psfig{file=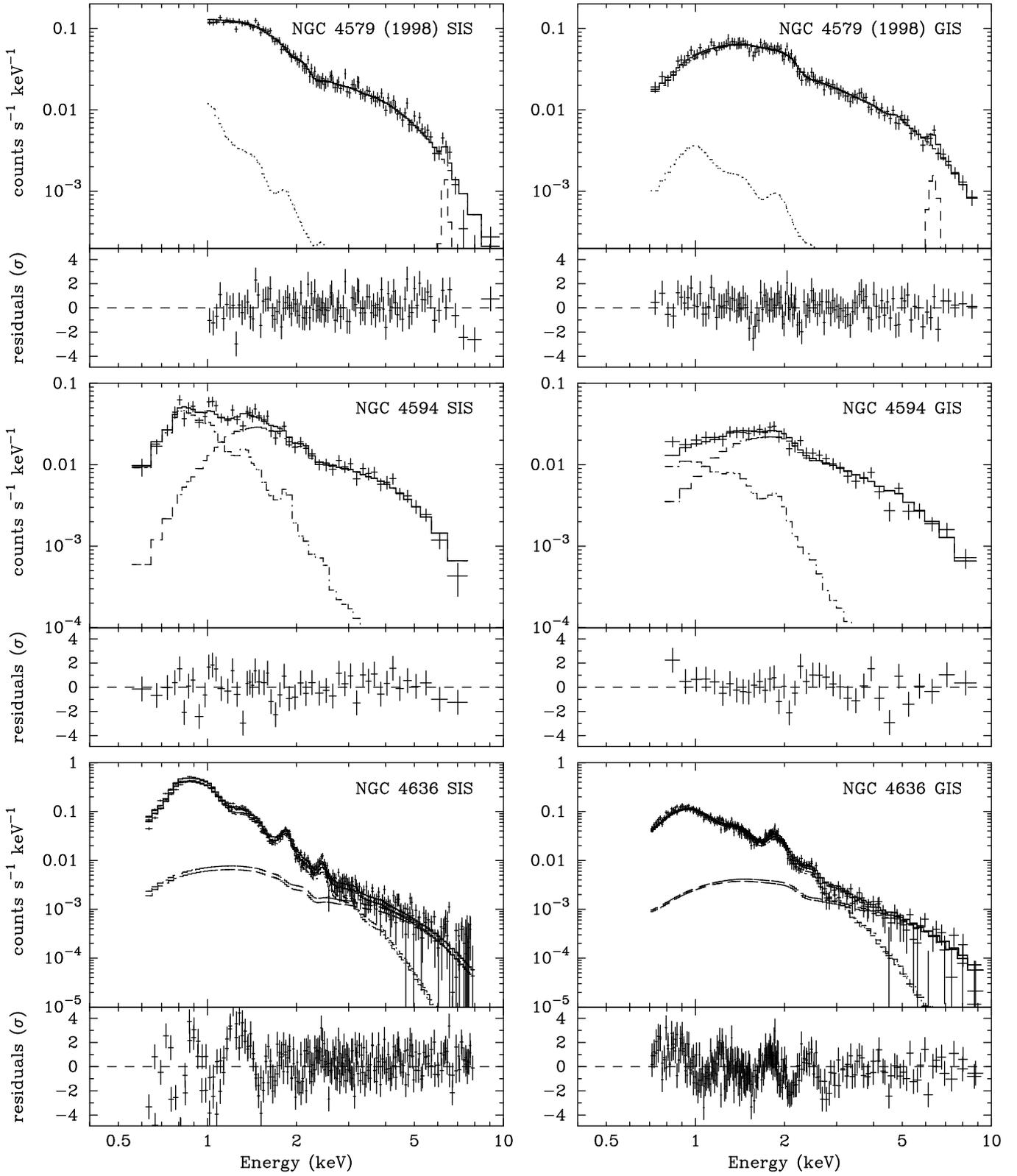,width=18.5cm,angle=0} 
\figcaption[]{
Same as in Figure 1{\it a}.  ({\it j}) NGC 4579 (1998) ({\it top}), NGC 4594
({\it middle}), NGC 4636 ({\it bottom}).
} 
\end{figure*}

\clearpage 
\begin{figure*}[t]
\figurenum{1{\it k}}
\psfig{file=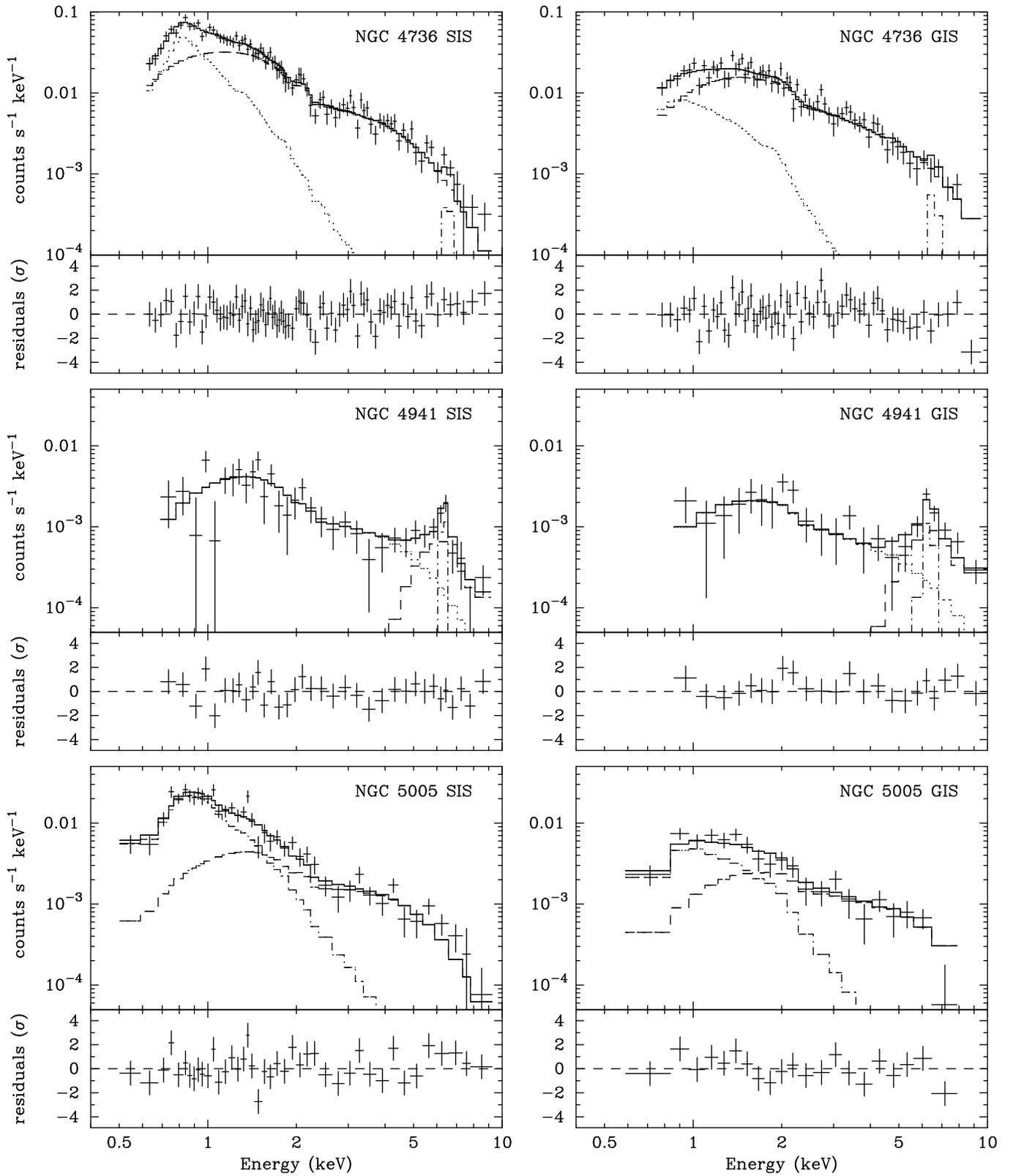,width=18.5cm,angle=0} 
\figcaption[]{
Same as in Figure 1{\it a}.  ({\it k}) NGC 4736 ({\it top}), NGC 4941
({\it middle}), NGC 5005 ({\it bottom}).
} 
\end{figure*}

\clearpage 
\begin{figure*}[t]
\figurenum{1{\it l}}
\psfig{file=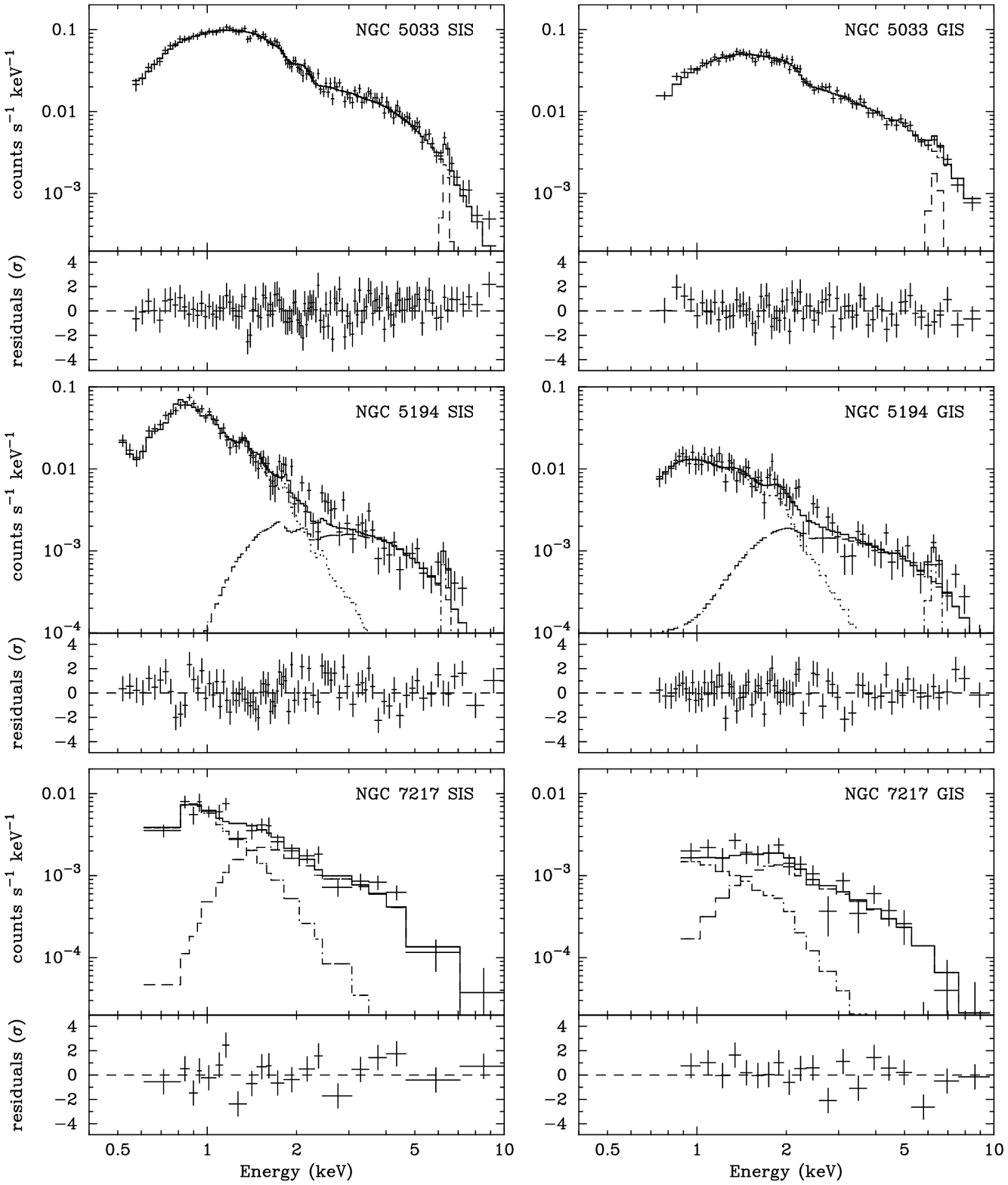,width=18.5cm,angle=0} 
\figcaption[]{
Same as in Figure 1{\it a}.  ({\it l}) NGC 5033 ({\it top}), NGC 5194
({\it middle}), NGC 7217 ({\it bottom}).
}
\end{figure*}
\clearpage

\noindent

\subsection{Fe K Lines}

Tables 9 and 10 summarize the parameters of the Fe K emission line.
Adding a Gaussian component to the best-fit continuum model improved
the $\chi^2$ values ($\Delta \chi^2 > 3$) for the 16 objects shown in
Table 9.  The additional parameters in the fit are the center energy,
line width, and line intensity. In cases where the photon statistics
are limited, we assumed the line to be narrow. For NGC 2639, the line
center energy was fixed at 6.4 keV in the source rest frame. Table 10
gives upper limits for the rest of the objects in which the Fe K
emission line is not clearly seen.

\section{Summary of the Spectral Results}

We summarize the best-fit models and parameters for all the spectral classes.
Tables 11 and 12 present results for the soft and hard components, 
respectively. These tables show that the spectra of most of the
objects analyzed here are well represented by the canonical model and
that a small fraction of LINER 1s, Seyfert 1s and Seyfert 2s do not
require the soft RS component. Thus, this canonical spectrum appears to be 
widely applicable to LINERs and LLSeyferts.

Figure~2  shows histograms of the spectral parameters (photon index, 
absorption column, and temperature for the RS component) for each spectral 
class.  Since the errors of the spectral parameters for many objects are 
large, we present {\it weighted} histograms, where the same area of a 
rectangle is assigned to each data set and the width of the rectangle is the 
error range in the spectral fits.  In generating the histograms of photon 
indices and absorption column densities, we excluded objects with very low 
signal-to-noise ratios or cases where the photon index was not well 
constrained (NGC 404, 1386, 4192, and 7743).  The histogram of 
the temperature of the RS component is created using objects which clearly 
show the presence of the soft thermal component.  The objects for which the 
temperature was not well constrained were excluded.

The distribution of photon indices for the four spectral classes are similar 
to each other, although LINER~2s and Seyfert~2s have somewhat broader
distribution than LINER~1s and Seyfert~1s. This difference is possibly
due to the larger errors in the type~2 objects.

The distributions of the absorption column densities appear different from
each other. Very large column densities --- in excess of $10^{23}$ {\pcm} ---
are observed in several Seyfert~2s, one LINER~1 (NGC 1052), and one Seyfert~1 
(NGC 4258, 1993 observation).  No LINER~2s in the present sample are obscured 
by $N_{\rm H}\,>\,10^{23}$ {\pcm}.  We note that in objects with very high 
columns ($N_{\rm H}\,>\,10^{24}$ {\pcm}), only scattered emission would be 
seen, and the {\it apparent} (observed) column densities would be much lower 
than the true column densities.  Such is the situation in NGC 1365 and 
NGC 5194.

The temperature of the RS component is confined to the range 0.5--0.8
keV. This distribution is similar to that presented by Ptak et al. (1999) 
for a smaller, more heterogeneous sample of LINERs, Seyferts, and starburst 
galaxies.

Apart from the absorption column densities, which exhibit greater diversity, 
the spectral parameters of the LINERs and Seyferts in our sample are rather 
homogeneous.  The canonical model we have adopted, however, can also be 
applied to starburst galaxies and normal galaxies (e.g., Ptak et al. 1999). 
The apparent similarity of the spectra, therefore, does not in itself imply 
that the different classes of objects share the same X-ray production 
mechanism, nor does it necessarily mean that they are physically related.  
Recall that, with very few exceptions, the hard component of the spectral fit 
could just as well be represented by a power law or a thermal bremsstrahlung 
model.  Therefore, we need to analyze carefully all the available information 
--- morphology, variability, and multiwavelength information --- before we 
can draw firm conclusions regarding the nature of these sources.  This is 
the subject of Paper II.

\section{Fluxes and Luminosities}

The total X-ray fluxes and luminosities, and the respective quantities for the 
the hard and soft components separately, are listed in Tables 13 and 14.
The best-fit model was assumed to calculate them. As for the hard
component, we assumed a power law rather than a thermal bremsstrahlung model
for all the cases, since the power-law model gave better or similar results
compared to a thermal bremsstrahlung model. The total (soft plus hard 
components) fluxes and luminosities in the 0.5--2 keV and 2--10 keV bands are 
shown in columns (2) and (3).  The observed and absorption-corrected fluxes 
and luminosities for the power-law component are given in columns (4) and (5) 
for the 2--10 keV band, and in columns (6) and (7) for the 0.5--4 keV band.
Since the estimated fluxes and luminosities of the soft component depend on 
the assumed abundance, we show two cases (abundance = 0.1 solar and 0.5 solar) 
if the abundance was assumed in the fits. The fluxes and luminosities of the 
hard component depend only weakly on the assumed abundance of the soft 
component.

\section{Variability}

X-ray variability is one of the most important characteristics of AGNs. 
We searched for X-ray variability by inspecting light curves constructed by 
binning the data to a bin size of 5760~s, one orbital period of {\asca}. 
When observations were performed more than twice, we searched for long-term
variability between multiple observations, usually separated by intervals 
of a week to a few years. We also compared the {\asca} luminosities obtained 
in this paper with published results from observations by other missions. An 
X-ray variability study of the brighter objects in our sample, concentrating 
on timescales within a day, has been presented by Ptak et al. (1998) and Awaki 
et al. (2001).

\begin{figure*}[t]
\figurenum{2}
%%BoundingBox: 130 20 530 750
%\plotone{fig2.ps}
\psfig{file=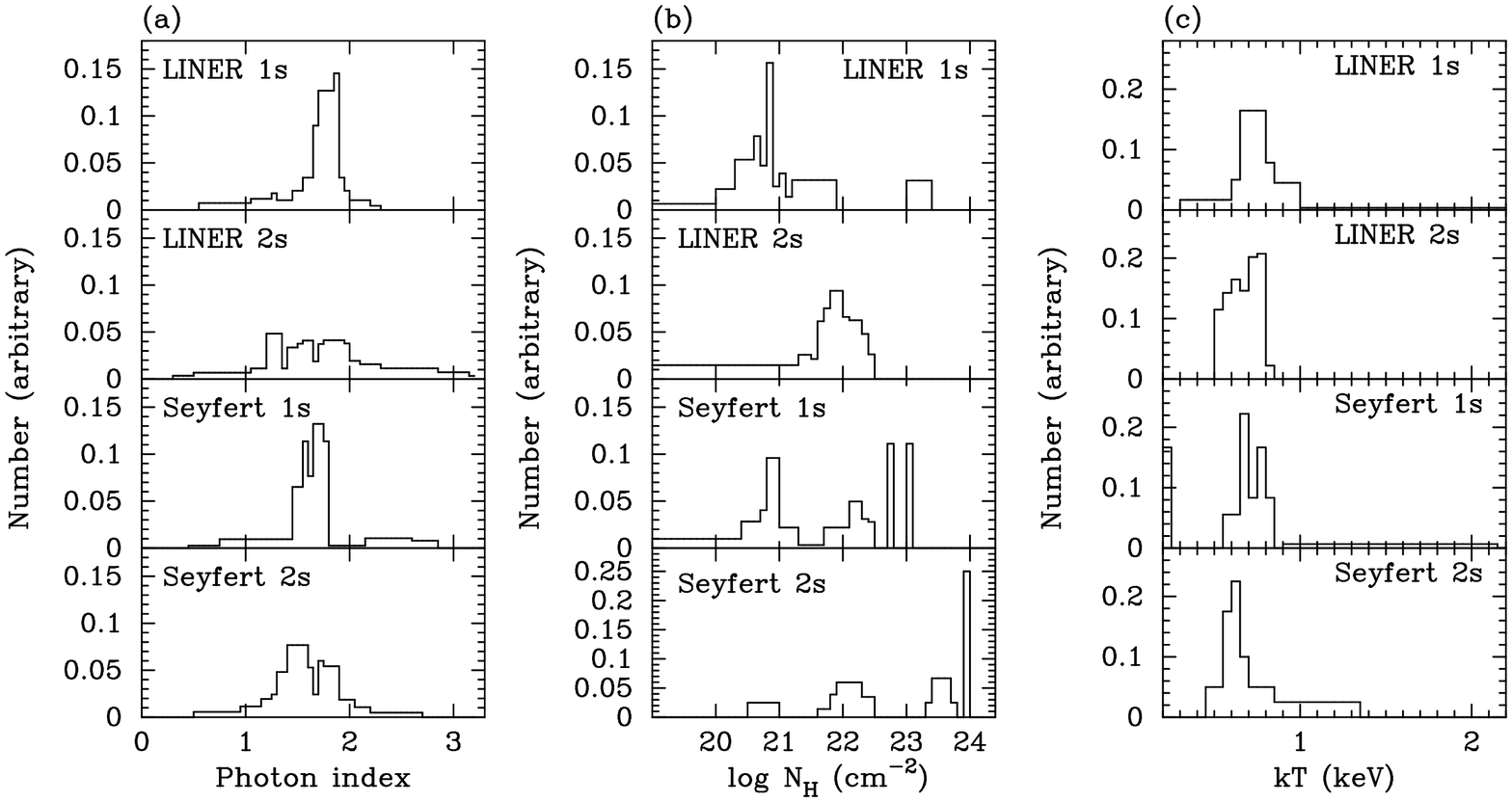,width=18.5cm,angle=270}
\figcaption[]{
Weighted histograms of the best-fit parameters for the whole sample.
({\it a}) Photon index.
({\it b}) Absorption column density.
({\it c}) Temperature of RS plasma.
}
\end{figure*}

\subsection{Short-term Variability}

X-ray variability within each {\it ASCA}\ observation, which has a typical
duration of one day, has been detected in NGC 3031 (Ishisaki et al. 1996;
Iyomoto \& Makishima 2001) and NGC 5033 (Terashima, Kunieda, \& Misaki 1999). 
Other objects showed no significant variability in a single-day observation. A 
typical upper limit on the root-mean-square amplitude is $\sim$10\% for 
relatively bright objects ($\sim$0.1 counts s$^{-1}$ detector$^{-1}$) and 
$\sim50\%$ for faint objects ($\sim$0.03 counts s$^{-1}$ detector$^{-1}$ or 
less). Short-term variability (less than one day) has been reported in a few 
objects from {\it BeppoSAX} observations (NGC 1365, Risaliti, Maiolino, \& 
Bassani 2000; NGC 3031, Pellegrini et al. 2000a; NGC 4258, Fiore et al. 2001).

\subsection{Long-term Variability}

We searched for long-term variability by comparing X-ray fluxes in multiple 
{\asca} observations. Eleven objects (NGC 404, 1365, 3031, 4111, 4192, 4258, 
4438, 4569, 4579, 4639, and 4941) were observed with {\asca} more than 
twice. NGC 3031 (see Ishisaki et al. 1996; Iyomoto \& Makishima 2001),
NGC 4258 (Reynolds, Nowak, \& Maloney 2000), and NGC 4579 (Terashima et al. 
2000c) show significant variability on timescales of a week to three years. No 
significant variability was seen in other objects. See Table~2 for count 
rates in individual observations.

We compared the fluxes obtained here with {\asca} and those measured
with other satellites. Although there are flux measurements in the
soft-energy band, it is difficult to compare them with the fluxes in
the harder band because of the presence of thermal emission in the
soft X-ray band and because of possible calibration uncertainties.
Therefore, we used the energy band of 2--10 keV for the comparison.
To minimize contamination from extranuclear sources, we used mainly
data from imaging detectors, the exceptions being NGC 3998, 5033, and
5194.  The 2--10 keV fluxes of NGC 3998 and NGC 5194 were obtained
with {\Ginga}. NGC 5033 was observed with {\it EXOSAT}. NGC 1052, 1365, 1386, 
2273, 3079, 3998, 4258, 4565, 4941, 5005, and 5194 were observed with 
{\it BeppoSAX}. Table 15 provides a summary of these fluxes.

Seven out of these 12 galaxies (NGC 1365, 3031, 3998, 4258, 4941,
5005, 5194) clearly show long-term variability on a timescale of a
month to several years. The variability in other objects is not clear,
given the uncertainties in calibration and spectral modeling.

\section{Image Analysis}

Next, we examine whether the X-ray images are consistent with the point-spread 
function (PSF) of {\asca}. Only SIS images were used since the
combination of SIS and the X-ray telescope (XRT; Serlemitsos et
al. 1995) has better spatial resolution than GIS+XRT. In order to
compare the observed SIS images with the PSF, we constructed radial
profiles of the surface brightness when there is no nearby bright
source. Then we fitted the radial profile with a model PSF plus a
constant background using a $\chi^2$-minimization technique. The free
parameters in this fit are the normalization of the PSF and the
background level. We also tried to apply a model two-dimensional
Gaussian convolved with the PSF plus a constant background to
constrain the source extent (technique described in Ptak 1998). The free
parameters in this fit are the normalization of the Gaussian, the
width of the Gaussian ($\sigma$), and the background level. Examples of
fits can be found in 

\begin{figure*}[t]
\figurenum{3}
%%BoundingBox: 50 50 550 700
%\plotone{fig3.ps}
\psfig{file=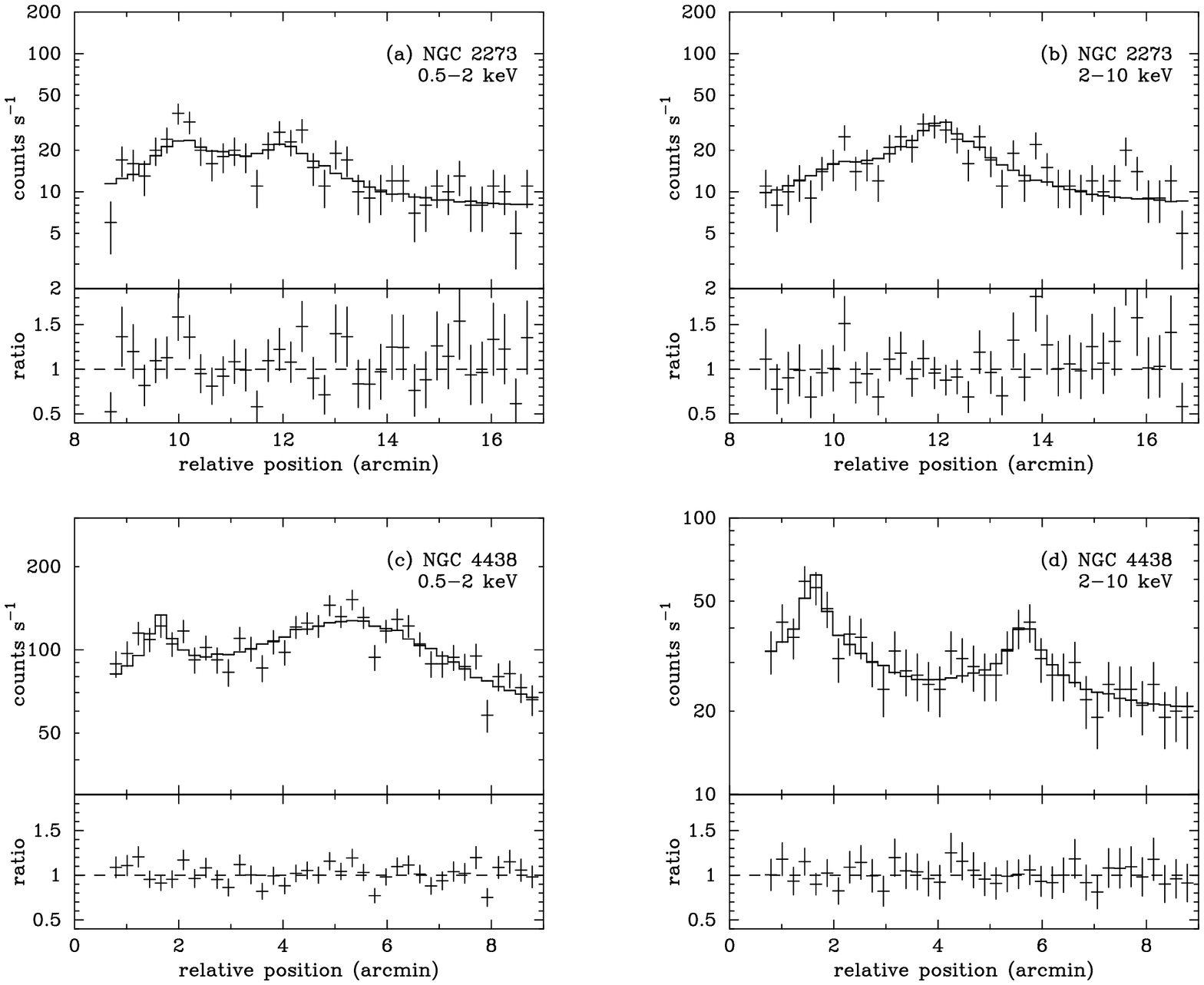,width=18.5cm,angle=270}
\figcaption[]{
Examples of fits to the one-dimensional projection of the surface brightness.
The model consists of two PSFs plus a constant background.
({\it a}) NGC 2273 in 0.5--2 keV.
({\it b}) NGC 2273 in 2--10 keV.
({\it c}) NGC 4438 in 0.5--2 keV.
({\it d}) NGC 4438 in 2--10 keV.
}
\end{figure*}

\noindent
Ptak (1998) and Terashima et al. (2000b).

In some cases there is a nearby source of nonnegligible brightness
compared to the main target. We constructed a one-dimensional
projection along the two sources. The one-dimensional profile of the
surface brightness was fitted with a model consisting of two PSFs plus a
constant background. The peak positions for the two sources were fixed,
and the free parameters are the two normalizations of the PSF and the
background level. If the main target was clearly extended, we utilize instead 
a two-dimensional Gaussian convolved with the PSF.  The resulting 
two-dimensional model is then projected into one dimension using the same 
procedure as applied to the actual data.  Figure~3 gives examples of the 
fits to the one-dimensional projections for NGC 2273 and NGC 4438 in 
the two energy bands 0.5--2 keV and 2--10 keV.  The source 2$^\prime$ south of 
NGC 2273 (left peak in Fig. 3{\it a}\ and 3{\it b}) coincides with the 
radio source 064542.0+605210 within positional accuracy, but the source near 
NGC 4438 located at 
(12$^h$27$^m$57.4$^s$, 13$^{\circ}$2$^\prime$18$^{\prime\prime}$) (J2000) 
(left peak in Fig. 3{\it c}\ and 3{\it d}) has no counterpart in NED. 
The soft-band image of NGC 4438 is extended compared to the PSF, while the 
hard-band image is consistent with a point source.  NGC 2273 is pointlike in 
both energy bands.  The nucleus of NGC 2273 is brighter in the hard band than 
the adjacent radio source, whose spectrum can be well fitted by a power law 
with a photon index of 1.9. This fact indicates that the spectrum of NGC 2273 
is very hard and in agreement with the results of the spectral fits.

The objects which clearly show extended emission in the hard-energy band 
are NGC 3607, 4111, 4374, 4569, 4636, 5194, and probably 7217.  Some of the
objects whose images are consistent with the PSF are faint. We comment
on the constraints on the source extent from Gaussian fits in the next
section for such objects. Some information from {\rosat} HRI images
and recent {\chandra} observations are given in section 9.

\section{Notes on Individual Objects}

This section presents some notes on individual objects, based mainly on 
data from the hard X-ray band.   The optical spectroscopic classification
is given in parentheses after the object name. \\

\noindent {\it NGC 315 (L1.9)}. --- An analysis of the {\it ASCA}\ data is 
presented in Matsumoto et al. (2001). A {\rosat} HRI image shows a compact core 
and probably an extended component. The contribution of the point source
to the HRI flux depends on modeling details (Canosa et al. 1999; Worrall \& 
Birkinshaw 2000). Extended emission is seen in the {\rosat} PSPC image 
(Worrall \& Birkinshaw 2000). The {\asca} images in the soft and hard bands
are consistent with being pointlike.\\

\noindent {\it NGC 404 (L2)}. --- An analysis of the {\it ASCA}\ data is 
presented in Terashima et al. (2000c). No clear X-ray emission was detected 
in our {\it ASCA}\ observations. Komossa et al. (1999) and Roberts \& Warwick 
(2000) reported a {\rosat} HRI detection with an X-ray luminosity of
$5\times10^{37}$ {\eps} in the 0.1--2.4 keV band. Lira et al. (2000) also 
reported a similar result. This luminosity is slightly higher than our upper 
limit which is obtained assuming a power-law spectrum with $\Gamma=2$.  The 
{\rosat} detection and luminosity suggest that this source has a softer X-ray 
spectrum than that we assumed, since {\rosat} is more sensitive in the softer 
energy band than {\asca}.  Alternatively the source could have faded since the 
the {\rosat}\ observation.   No spectral information was obtained from the 
{\rosat} observation.

We analyzed a recent 24~ks {\chandra} observation and found a
compact nucleus with a very soft spectrum which is well represented by a
thermal plasma model with $kT\,\approx$ 0.8 keV. A hint of a hard component
is also seen. If the hard component is modeled by a power law with a
photon index of 2.0, the observed flux in the 2--10 keV band becomes
$\sim3\times10^{-14}$ {\eps} {\pcm}, which corresponds to a luminosity
of $2\times10^{37}$ {\eps}.\\

\noindent {\it NGC 1052 (L1.9)}. --- {\asca} results are presented in 
Guainazzi \& Antonelli (1999) and Weaver et al. (1999). Detection of hard 
X-rays up to $\sim200$ keV with {\sax} is reported in Guainazzi et al. (2000).

A heavily absorbed continuum, which indicates the presence of an obscured AGN, 
is clearly seen in the {\asca} spectrum. Systematic undulations in the 
residuals were seen in the fits using a partially covered power-law
model or a RS plus absorbed power-law model. The best-fit model we obtained 
is a RS plus partially absorbed power-law model. The resulting photon index of 
$1.67^{+0.57}_{-0.40}$ is similar to those found in luminous AGNs and in other 
objects presented in this paper. Since the photon index depends
on the assumed spectral model (see Tables in this paper and Weaver et
al. 1999), the very flat spectral slope obtained by Guainazzi \& Antonelli
(1999) and Guainazzi et al. (2000) ($\Gamma\,\approx\,1.4$) should be taken 
with caution. The thermal plasma component in the best-fit model plausibly 
can be attributed to a hot gaseous halo associated with the elliptical host
of NGC 1052. \\

\noindent {\it NGC 1097 (L2/S1.5)}. --- An analysis of the {\it ASCA}\ data 
is presented in Iyomoto et al. (1996). The {\rosat} HRI image shows a 
distinct nucleus and weak extended emission due to a circumnuclear 
star-forming ring (P\'erez-Olea \& Colina 1996).  A recent {\chandra} 
observation shows that the hard X-ray emission is dominated by the nucleus
(Y. Terashima, \& A.~S. Wilson 2001, private communication).\\

\noindent {\it NGC 1365 (S1.8)}. --- The results of the {\asca} observations, 
performed in 1994 August and 1995 January, are presented in Iyomoto et al. 
(1997). A serendipitous source is detected at 1\farcm5 SW of the nucleus 
in the second observation, but the separation is too close to allow the 
two sources to be separated with GIS, which has a broader PSF than SIS.

Iyomoto et al. (1997) generated five spectra: (1) SIS spectrum of the
nucleus in 1994, (2) SIS spectrum of the nucleus in 1995, (3) SIS
spectrum of the serendipitous source in 1995, (4) GIS spectrum of the
nucleus in 1994, and (5) GIS spectrum of the nucleus plus the
serendipitous source in 1995, fitted simultaneously. The
spectral models they used were: (a) a sum of a power law, Gaussian, and RS
plasma for (1), (2), and (4) (nuclear component), (b) a power law for
(3) (serendipitous source), and (c) a sum of the nuclear component and the
serendipitous source model for (5). The spectral parameters for SIS and
GIS were assumed to be same. No spectral variability was assumed
between the 1994 and 1995 observations.

We applied a similar technique to fit both the nucleus and the
serendipitous source. We fitted the three spectra (SIS spectrum of the
nucleus [(1)+(2)], SIS spectrum of the serendipitous source (3), and
GIS spectrum of the nucleus plus the serendipitous source [(4)+(5)])
simultaneously. The radii used to extract the spectra are the same
as those in Iyomoto et al. (1997). In the present analysis, the background
spectra were extracted from a source-free region in the same field of view.  
Various models were examined for the AGN component, as in the other galaxies. 
The best-fit model consists of a power law, a Gaussian, and a RS component. 
We confirmed the presence of a strong Fe line (EW = $1.9^{+1.0}_{-0.8}$ keV) 
at a center energy higher than 6.4 keV ($6.59^{+0.04}_{-0.05}$ keV).

The X-ray flux obtained in a later {\sax} observation (
Risaliti et al. 2000) is higher by a factor of 6 compared to {\asca}, and an
absorbed ({\NH} = $4\times10^{23}$ {\pcm}) direct continuum from the
nucleus appeared. The Fe line center energy in the {\sax} observation
is slightly lower than 6.4 keV.

A soft-band image obtained with the {\rosat} HRI may be extended compared
to the PSF (Stevens, Forbes, \& Norris 1999; but see also Komossa, \& Schultz 
1998). Plausible origins of the soft component are soft thermal gas due to 
starburst activity or scattered emission from the AGN.\\

\noindent {\it NGC 1386 (S2)}. --- Detailed results are discussed in Iyomoto 
et al. (1997). In the NGC 1386 field, non-uniform diffuse emission due to
the Fornax cluster is seen. In the present paper, we measured the temperature 
and brightness of the cluster component using a region which is located the 
same distance from the cluster center as is NGC 1386, and then added it to the 
spectral model instead of subtracting it from the data in order to improve 
the photon statistics.

A power law absorbed with {\NH} = $4.5\times10^{23}$ {\pcm} is seen in our 
{\asca} spectrum. Maiolino et al. (1998), on the other hand, modeled the 
{\sax} spectrum of this source using a cold reflection model. The difference 
in spectral models, however, does not necessarily imply true spectral 
variability, because the photon statistics of the {\sax} spectrum are very 
limited.\\

\noindent {\it NGC 2273 (S2)}. --- There is a serendipitous source 2$^\prime$ 
south of NGC 2273 in both the SIS and GIS images. The position of this source
coincides with the radio source 064542.0+605210. We extracted SIS
spectra of NGC 2273 and this radio source using circular regions with
a radius of 1\farcm5 and 1\farcm0, respectively. We made one GIS spectrum 
using an extraction radius of 4$^\prime$ which contains both
sources since the separation is too small to be resolved with the GIS.
These three spectra were fitted simultaneously. The models
applied are (a) a partially covered power-law model with two Gaussians
(Fe~K$\alpha$ and Fe~K$\beta$) modified by absorption along the line
of sight for the SIS spectrum of NGC 2273, (b) an absorbed power-law for
the SIS spectrum of the radio source, and (c) a combination of (a) and
(b) for the GIS spectrum. The normalizations for the SIS model and GIS
model were fitted separately, while the other spectral parameters were
tied. The ratio of the normalizations of the two sources for GIS was
assumed to be same as those of SIS. The results of the fits are shown in
Table 4. The best-fit spectral parameters for the radio source are
$\Gamma\,=\,1.85^{+0.30}_{-0.24}$, {\NH} = 0 ($<7.0\times10^{20}$) {\pcm},
and $f$(2--10 keV) = $1.2\times10^{-13}$ {\eps}{\pcm}. In Figure~1{\it b}, 
the steeper power law corresponds to the spectrum of the radio source.

The spectrum of NGC 2273 shows a heavily absorbed continuum ({\NH} =
$1.1\times10^{24}$ {\pcm}) and a strong fluorescent Fe~K line with EW
= 1.0 keV. The large EW of the Fe line indicates that the hard X-ray
emission is dominated by reflection from cold material. Our modeling
of the continuum shows that transmitted emission through a column
density of $10^{24}$ {\pcm} is also present.  We tried a pure
reflection component instead of a heavily absorbed power law, but this
gave a worse fit ($\Delta\chi^2$ = +16).  Maiolino et al. (1998)
detected a strong (EW = $3.8\pm1.1$ keV) fluorescent Fe~K line using
{\sax}, and they suggested that the source contains a significant cold
reflection component. Pappa et al. (2001) recently analyzed the same
{\asca} data set. Their preferred model is consistent with ours,
although our spectrum has better signal-to-noise ratio because we used
a larger extraction radius for the GIS data. \\

\noindent {\it NGC 2639 (S1.9)}. --- {\asca} observations of this Seyfert~1.9 
galaxy were published by Wilson et al. (1998). We used SIS0+SIS1 and GIS2+GIS3
spectra, while Wilson et al. fitted only GIS3 and SIS1 because of a
large offset angle from the optical axis for GIS2 and SIS0 data. Inclusion 
of data from all the detectors improved the photon statistics, particularly in 
the low-energy band which strongly affects the power-law slope. The best fit 
was achieved with a partially covered power-law model.  The derived photon 
index is somewhat steeper than normal for Seyferts, perhaps an indication 
of the presence of an additional soft component. Therefore, we tried a model 
consisting of a power law plus a RS thermal plasma. The abundance of the RS 
component was assumed to be 0.1 or 0.5 solar, and a Gaussian line was 
added. This model also successfully reproduced the observed spectrum. In this 
case, however, a heavily absorbed component is not necessary. On the other 
hand, the large EW ($>$ 1.1 keV) of the Fe emission line suggests that the 
nucleus is highly obscured. The observed hard emission thus can be interpreted 
as scattered emission. If this is the case, the intrinsic luminosity should be 
one or two orders of magnitude higher than the observed luminosity.\\

\noindent {\it NGC 2655 (S2)}. --- The {\it ASCA} spectrum
of this Seyfert~2 galaxy clearly shows the presence of a heavily absorbed
component and an additional soft component. A RS + power-law model fit gave 
a negative photon index.  Although such a very flat spectral slope
is expected from ``cold'' reflection, the lack of a strong Fe~K fluorescent 
line rules out this possibility. Therefore, we tried three
alternative models: a partially covered power-law model, a double
power-law model, and a RS + partially covered power-law model. The
best-fit photon index in the partially covering model ($\Gamma = 2.6$) is 
steeper than the canonical slope of Seyfert galaxies. This
slope is primarily determined by the soft-energy band, which has better
photon statistics, and suggests that there exists a softer component in
addition to a hard power law from the AGN. If we adopt a double
power-law model, the photon index of the soft component becomes
$\Gamma = 2.4$, where we fixed the photon index of the hard component
at $\Gamma = 2.0$. In this model, the absorption column density for
the hard component is {\NH} = $4.0\times10^{23}$ {\pcm}. Finally, we
tried to add a RS component to a partial covering model. We assumed a
temperature of $kT=0.65$ keV and an abundance of 0.1 solar since these
parameters were not well constrained. The absorption column densities for
the RS component and the less absorbed power law were assumed to be equal to 
the Galactic value. The model improved by $\Delta \chi^2$ = --16.3 for one 
additional parameter (normalization of the RS component) compared to a partial 
covering model. The best-fit photon index and absorption column are 
$\Gamma = 1.26^{+0.57}_{-0.69}$ and {\NH} = $4.5^{+2.0}_{-1.4}\times10^{23}$ 
{\pcm}.\\

\noindent {\it NGC 3031 (S1.5)}. --- {\asca} observed this galaxy many
times; see Iyomoto and Makishima (2001), who analyzed 16 data sets,
for the observation log and a timing analysis. Ishisaki et al. (1996)
presented detailed results of the observations between 1993 May 1 to 
1995 April 1, while Serlemitsos et al. (1996) describe three
observations done on 1993 April 16, April 25, and May 1.

We present a combined spectrum of the three observations performed on 1994 
October 21, 1995 April 1, and 1995 October 24 (observations \#9, 10, and 11 
in Iyomoto \& Makishima 2001).  More recent observations were made using  
unusual observation modes (lower spectral and spatial resolution but higher 
timing resolution for GIS). In the earlier observations, the nearby source 
SN~1993J was bright, and we therefore omitted these in the present analysis.

In our spectral fits, we discarded the energy range below 1 keV for the SIS 
data since the spectra of the SIS and GIS deviate in this energy range, most 
likely because of a calibration problem of the SIS in the soft-energy band. 
This problem is visible only in bright objects such as NGC 3031, and possibly 
NGC 4579 and NGC 5033.

A soft thermal component is not clearly seen in our spectra presumably
due to the brighter hard component in our spectrum, and we assumed the
temperature obtained by Ishisaki et al. (1996).  The width of the
Gaussian component for the Fe line was assumed to be narrow since the
line width was not well constrained. The broad ($\sigma\,\approx$ 0.2
keV) or possibly multiple-component Fe line previously reported by
Ishisaki et al.  (1996) and Serlemitsos et al. (1996) is not clearly
seen in our spectra, presumably due to the limited photon
statistics. The {\sax} observation of Pellegrini et al. (2000b) gave
upper limits of 0.3~keV for the width of an Fe line.  The line
centroid energy and EW we obtained are consistent with the {\asca}
results by Ishisaki et al. (1996) and Serlemitsos et al. (1996) and
the {\sax} results of Pellegrini et al. (2000b). Pellegrini et
al. (2000b) obtained an upper limit of 42~eV for the EW of an Fe line
at 6.4 keV. \\

%F2-10 Ishisaki et al.  Serlemitsos et al. 1.4e-11

\noindent {\it NGC 3079 (S2)}. --- {\asca} results are presented in Ptak et
al. (1999) and Dahlem, Weaver, \& Heckman (1998).  Although hard X-ray
emission is detected, we found no clear evidence for the presence of an 
AGN from the present data. The small {\LX}/{\LHa} ratio (Terashima et
al. 2000a) indicates that the AGN component, if the dominant ionizing source 
of the optical emission lines, should be heavily obscured in the energy
band below 10 keV.  The upper limit on the EW of an Fe K emission line is 
large ($\sim$2 keV), not inconsistent with a highly obscured nucleus. A recent 
{\sax} observation detected a strong Fe K emission line and highly absorbed
hard X-ray emission (Iyomoto et al. 2001).  Such a spectral shape gives 
clear evidence for the presence of a heavily obscured AGN.

The low-energy portion of the {\asca} spectrum shows strong emission
lines arising from $\alpha$-processed elements, which are most likely 
associated with hot gas arising from powerful starburst activity. A variable
abundance model gives a significantly better fit than the solar case 
($\Delta \chi^2$ = --16.3).

The absorption column for the hard component depends on the model adopted for 
the soft component. Our best-fit value is 
{\NH} = $1.7^{+1.8}_{-1.2}\times10^{22}$ {\pcm}.  Dahlem et al. (1998) used a 
two-temperature MEKAL plasma plus a partially covered power law and obtained 
{\NH} = $6\times10^{22}$ {\pcm}. Ptak et al. (1999), assuming a RS + power-law 
model, derived a very small column, with a typical upper limit of 
{\NH} $\approx$ few$\times 10^{22}$ {\pcm}.  Since the hard component seen 
by {\asca} is probably a combination of scattered emission from the AGN and
contributions from the starburst region (X-ray binaries, supernovae, and hot 
gas), the amount of absorption to be attributed to the AGN is highly 
ambiguous. \\

\noindent {\it NGC 3147 (S2)}. --- A detailed analysis of the {\asca} data is 
given in Ptak et al. (1996). We confirmed their detection of a strong Fe~K
emission line. Although this object is optically classified as a Seyfert~2 
galaxy, no strong absorption is seen in the X-ray spectrum.  The relatively 
strong Fe~K emission line (EW = $490^{+220}_{-230}$ eV) could be an indication 
that the observed X-rays are scattered emission and that the nucleus is 
obscured below 10 keV. However, the {\LX}/{\LHa} ratio (32) falls in the range 
of unobscured AGNs.  The ratio of X-ray to [\ion{O}{3}] $\lambda$5007 
luminosity is also typical of those observed in Seyfert~1 nuclei (Bassani et 
al. 1999). Some LLAGNs with low absorption in our sample also show similarly
large EWs for the Fe line ($\sim$300 eV). For example, the X-rays from 
NGC 4579 (EW = 490 eV) and NGC 5033 (EW = 306 eV) must be dominated by direct 
emission because the sources are variable. This suggests that NGC 3147, too, 
could be largely unobscured.  Future observations to search for variability or 
emission lines due to a scattering medium are crucial to distinguish between 
these two competing possibilities.\\

\noindent {\it NGC 3507 (L2)}. --- This object is very faint, and a power-law 
model describes the spectrum well. By adding a RS component, the fit improved 
by $\Delta\chi^2 = -3.2$ for one additional parameter (normalization of the 
RS component); we assumed a temperature of $kT$ = 0.65 keV and an abundance of 
0.1 solar because the photon statistics are inadequate to constrain these 
parameters.  The image in the 2--10 keV band is consistent with the PSF, 
although the constraint is not tight. A Gaussian fit to the observed image 
yielded $\sigma=0.5^{+0.8}_{-0.5}$ arcmin.\\

\noindent {\it NGC 3607 (L2)}. --- The soft and hard band images (0.5--2 keV, 2--7
keV, 4--10 keV) are clearly extended compared to the PSF. We found no
clear evidence for the presence of an AGN in this LINER 2.  The
probable origin of the extended hard X-ray emission is discrete
sources in the host galaxy. The hot gas in this elliptical galaxy has been 
analyzed by Matsushita et al. (2000).\\

\noindent {\it NGC 3998 (L1.9)}. --- {\asca} results are presented in Ptak et
al. (1999). An Fe~K line is marginally detected at 6.4 keV.  A recent {\sax} 
observation (Pellegrini et al. 2000a) has shown that the spectrum from 0.1 
to 100 keV is well represented by a power law.  The Fe~K line was not detected 
in the {\sax} spectrum to an EW upper limit of 40 eV. \\

\noindent {\it NGC 4192 (T2)}. --- An analysis of the {\asca} data is 
presented in Terashima et al. (2000c). Since the object is very faint, we 
estimated its spectral shape using the hardness ratio (2--7 keV)/(0.5--2 keV). 
Assuming the Galactic absorption and a power-law spectrum, we obtain 
a photon index of $\sim1.7$.\\

\noindent {\it NGC 4203 (L1.9)}. --- An analysis of the {\asca} data is 
presented in Iyomoto et al. (1998). There exists a serendipitous source located
2$^\prime$ SE of the nucleus of NGC 4203. In the present analysis, we extracted
the SIS and GIS spectra using a circular aperture with a radius of 1\farcm2 and
1\farcm5, respectively. {\rosat} PSPC and HRI images show a nucleus
as well as some extended emission (Halderson et al. 2001). The {\chandra} 
image published by Ho et al. (2001) is dominated by the nucleus.\\

\noindent {\it NGC 4258 (S1.9)}. --- We analyzed four {\asca} observations 
obtained in 1993 and 1996. Makishima et al. (1994) reported the discovery of an
obscured X-ray nucleus using the 1993 observation; Ptak et al. (1999)
also discussed the same data set. We reanalyzed this data set in greater
detail, as well as three others taken in 1996. Reynolds, Nowak, \& Maloney 
(2000) performed a deep ($\sim200$ ks) observation in 1999.

The {\asca} spectrum from 1993 is very complex. In the following fits,
we assume the Galactic absorption column density for the soft thermal
components. We examined the canonical model which consists of a soft
thermal plasma and a hard component. This simple model failed to
explain many of the emission lines seen in the region 0.5--3 keV as
well as the broad-band continuum shape. A single-temperature fit to
the soft X-ray spectrum resulted in $kT\,\approx\,0.65$ keV, but
residuals remained near the locations of the H-like O K, He-like and
H-like Mg K, and Fe L emission lines in region 0.7--1.4 keV.  This
suggests that the plasma has multiple components, with temperatures
cooler and hotter than 0.65 keV, or that a temperature gradient is
present. Therefore, we tried a model composed of a two-temperature RS
plasma plus a hard component. This model gave a better fit to the soft
thermal emission. However, systematic convex-shaped residuals remained
in the 3--10 keV region, which can be attributed to a deficit in the
model around 2--3 keV and to an {\NH} value too small and a photon
index (for the hard component) too steep at the $\chi^2$ minimum. A
satisfactory fit can be achieved with a model consisting of a
two-temperature RS plasma, a thermal bremsstrahlung component, and an
absorbed power law. The abundances of the two RS components were
assumed to be identical. The temperature of the thermal bremsstrahlung
component was fixed at 4 keV. An Fe line is seen at
$6.66^{+0.20}_{-0.07}$ keV (source rest frame) with EW =
$108^{+73}_{-64}$ eV.  The spectral parameters for the AGN continuum
are consistent with those derived by Makishima et al. (1994) based on
the GIS data only. Note that Ptak et al. (1999) used the canonical
model and obtained a photon index of $\sim$1.2 and {\NH} $\approx$
$7\times10^{22}$ {\pcm}. This model resulted in significant systematic
residuals (see Figs.~2{\it q}\ and 2{\it r}\ in Ptak et al. 1999),
suggesting that the true photon index should be steeper and {\NH}
larger.

The complicated soft thermal emission in NGC 4258 is probably due to the 
superposition of its extended galactic halo and interstellar medium, 
which is shock heated by the complex system of jets emanating from the 
nucleus (Pietsch et al. 1994; Cecil, Wilson, \& De Pree 1995).  A possible 
origin for the medium-hard emission we modeled with a $kT$ = 4 keV thermal 
bremsstrahlung component might be the integrated emission from low-mass X-ray 
binaries. Future X-ray spectroscopy with better spatial resolution will be 
able to resolve these multiple X-ray emission components (see, e.g., Wilson, 
Yang, \& Cecil 2001 for a {\chandra} image).

NGC 4258 was observed three times in 1996 (see Table 2). We fitted the
combined spectrum of these three data sets, after discarding the low-energy 
($<$0.7~keV) portion of the SIS data to minimize calibration
uncertainties. The {\asca} spectrum is well fitted with the canonical
model. The flux of the hard component is about twice that observed in 1993, and
the energy band above 2 keV is now dominated by the AGN emission. This
variability provides additional evidence for the presence of an
LLAGN. The medium-hard component introduced to fit the 1993 spectrum was not 
required for the 1996 data. The soft component can be represented by a
single-temperature RS model, in contrast to the 1993 result. This is probably 
attributable to (1) the degraded energy resolution of the SIS, (2) the limited
energy band used in the spectral fits, and (3) the brighter hard component
in 1996.  The absorption column for the AGN component decreased from 
{\NH} = $1.3 \times10^{23}$ {\pcm} in 1993 to $7\times10^{22}$ {\pcm} in 1996. 
An Fe emission line is detected at $6.31^{+0.09}_{-0.10}$ keV (source rest 
frame) with EW = $54^{+25}_{-27}$ eV.  The line centroid energy decreased from 
6.66 keV to 6.31 keV in the interval of three years, while the continuum 
luminosity increased during the same period.  A similar behavior of Fe line 
variability has been reported for NGC 4579 (Terashima et al. 2000c). The Fe 
line energy and EW measured by {\sax} are $6.57\pm0.20$ keV and $85\pm65$ eV,
respectively, but it is unclear whether these values are different compared 
to those measured with {\asca}.  \\

\noindent {\it NGC 4261 (L2)}. --- An {\asca} spectrum is presented in
Sambruna, Eracleous, \& Mushotzky (1999) and Matsumoto et
al. (2001). This FR~I radio galaxy has a pointlike hard X-ray source
with a relatively high X-ray luminosity.  The {\LX}/{\LHa} ratio of
this type~2 LINER suggests that it is predominantly ionized by an
LLAGN (see Paper II).  Sambruna et al. (1999) claimed that they
marginally detected an Fe~K line.  They assumed a rest energy of 6.4
keV in their spectral fits.  On the other hand, Matsumoto et
al. (2001) obtained only an upper limit of EW = 400 eV.  We fitted the
Fe line in our {\asca} spectrum with a free line centroid energy. The
line energy we obtained, $E\,=\,6.85^{+0.08}_{-0.15}$ keV, is higher
than observed in luminous AGNs (typically $E\,\approx\,6.4$ keV).

The soft-band image is extended to $r\,>$ 20$^{\prime}$ and is most likely 
sampling emission associated with the elliptical host and the galaxy group 
of which NGC 4261 is a member.  In the spectral fits, we made a background 
spectrum using an annular region around the nucleus in which diffuse emission 
is clearly present.  The soft X-ray flux presented here, therefore, should be
regarded as a lower limit to the true flux from the diffuse component.
The radial surface brightness profile of the soft-band image is not well
fitted by a model consisting of a Gaussian plus a constant background.
This is probably due to the presence of bright soft X-ray emission from the 
elliptical host galaxy and the surrounding group. The $\chi^2$ value for
the fit of the 2--10 keV image is also poor, probably for the same reason.
The 4--10 keV image, by contrast, is consistent with the PSF.  Inspection of
archival {\chandra} data shows that the hard-band image indeed is dominated
by the nucleus and that the soft-band image is extended.\\

%Davis, D.~S., Mushotzky, R.~F., Mulchaey, J.~S., Worrall, D.~M., Birkinshaw, M., \& Burstein, D. 1995, ApJ, 444, 582 \\% n4261

\noindent {\it NGC 4374 (L2)}. --- The X-ray images in the soft and hard bands 
are clearly extended.  The contribution of the AGN to the hard emission should
be small; this is confirmed by the {\chandra} of Ho et al. (2001). The most 
likely origin of the extended hard emission is X-ray binaries in the host 
galaxy, as discussed in Matsushita et al. (1994). The hot gas component in 
this elliptical galaxy has been analyzed by Matsumoto et al. (1997), Buote \& 
Fabian (1998), and Matsushita et al. (2000).\\

\noindent {\it NGC 4438 (L1.9)}. --- The LINER~1.9 nucleus of this galaxy is 
a very weak X-ray source. The image in the 2--10 keV band is consistent with 
the PSF, while the 0.5--2 keV image is clearly extended.  The {\rosat} HRI 
image is also extended (Halderson et al. 2001).  The {\LX}/{\LHa} ratio (0.8) 
is too low to account for the optical line emission, unless the nucleus is 
heavily obscured at energies above 2 keV.  \\

\noindent {\it NGC 4450 (L1.9)}. --- The {\rosat} PSPC image is
dominated by a pointlike nucleus (Halderson et al. 2001).  The
serendipitous source seen in the PSPC image ($\sim$3$^{\prime}$ NE of
the nucleus, Komossa et al. 1999) is not clearly present in the {\it
ASCA} data. \\

\noindent {\it NGC 4501 (S2)}. --- The image in the 0.5--2 keV band is 
extended.  A Gaussian fit to the radial profile yielded 
$\sigma=1.0\pm0.4$ arcmin. In the hard band (2--10 keV and 4--10 keV),
the radial profiles are consistent with the PSF ($\sigma=0.7\pm0.4$ and
$0.6^{+1.3}_{-0.6}$ arcmin, respectively). The {\rosat} HRI image is
extended (Halderson et al. 2001). The X-ray spectrum shows no indication of 
significant absorption ({\NH} $<\,2.3\times10^{22}$ {\pcm}), and there is no 
strong evidence for Fe~K emission (adding a narrow Gaussian at 6.4 keV gives 
$\Delta \chi^2 = - 2.9$), although the EW upper limit is large ($<2310$ eV). 
The {\LX}/{\LHa} ratio (17) lies in the range for unobscured AGNs (Terashima 
et al. 2000a). These facts suggest that this Seyfert~2 galaxy is not 
heavily obscured in the X-ray band. Alternatively, it is possible that the 
hard X-ray emission comes mainly from X-ray binaries, whatever contribution 
from an obscured AGN being strongly diluted, as in the case of NGC 5194 
(Terashima et al. 1998b; Terashima \& Wilson 2001; Fukazawa et al. 2001).\\

\noindent {\it NGC 4565 (S1.9)}. --- Mizuno et al. (1999) presented {\asca} 
results for NGC 4565. The nucleus and an off-nuclear source 0\farcm8 
away, previously detected with {\rosat} (PSPC: Vogler, Pietsch, \& Kahabka 
1996; HRI: Mizuno et al. 1999, Halderson et al. 2001), are seen in the {\asca} 
SIS images, but they are not resolved in the GIS images. The source coincident 
with the nucleus appears pointlike in a {\rosat} HRI image (Halderson et al. 
2001). The off-nuclear source is $\sim$2 times brighter than the nucleus. As 
the spatial resolution of {\asca} is insufficient to resolve the two sources, 
we extracted a single spectrum for both sources. Among the models we examined, 
the most successful fit was obtained with a partially covered power law plus a 
RS plasma.  A partial covering model without the RS component yielded a 
slightly worse result ($\Delta \chi^2$ = +4.6). A simple power-law model 
seems inappropriate in view of the systematic, wavy residuals it generated.
We also attempted a multicolor-disk blackbody model (Mitsuda et al. 1984),
which was examined by Mizuno et al. (1999), and obtained results
consistent with theirs. The best-fit parameters for this model are 
$kT_{\rm in} = 1.49\pm0.08$ keV and {\NH} = $1.2^{+2.5}_{-1.2}\times10^{20}$
{\pcm}; $\chi^2$ = 151.4 for 112 dof.

Inspection of the one-dimensional projected profiles in the hard and
soft-energy bands shows that the two sources have similar spectral
hardness. Thus, the nucleus and the off-nuclear source each
contributes about one-third and two-thirds, respectively, to the total
luminosity ($1.8\times10^{40}$ {\eps}). The intrinsic luminosity
depends on the assumed model. Tables 13 and 14 show the intrinsic flux
and luminosity for the case of a partially covered power law plus RS
model.

Mizuno et al. (1999) interpreted that both sources are luminous accreting 
black hole binaries (see also Makishima et al. 2000). They suggest that the 
derived absorption column density ({\NH} $<2\times10^{21}$ {\pcm}) is too low 
for the nucleus of an edge-on galaxy. We argue that an LLAGN is a likely 
origin for the nuclear source. First, the nucleus shows an optical spectrum
classified as a type~1.9 Seyfert (Ho et al.  1997a). The detection of broad
H$\alpha$ emission provides strong support for the presence of an AGN. Second, 
the {\LX}/{\LHa} ratio (20) is in good agreement with those of LLAGNs and
luminous AGNs (Terashima et al. 2000a). Third, the internal reddening
determined from the Balmer decrement, $E(B-V)_{\rm int}$ = 0.47 mag 
(Ho et al. 1997a),  corresponds to {\NH} = $2.7\times10^{21}$ {\pcm} for the 
conversion $E(B-V)$ = {\NH}/($5.8\times10^{21}$ {\pcm}) mag (Bohlin, Savage, 
\& Drake 1978). This value of {\NH} is consistent with the mild absorption
observed in the X-ray spectrum. Finally, the radio properties further
support the AGN interpretation: the nucleus contains a compact radio core 
which has a flat spectrum (Nagar et al. 2000; Falcke et al. 2000) and is 
variable (Falcke et al. 2001; Ho \& Ulvestad 2001). Further X-ray observations
with higher spatial resolution would be extremely useful. \\

\noindent {\it NGC 4569 (T2)}. --- Detailed analysis of an {\asca} observation 
is presented in Terashima et al. (2000c). The {\asca} image in the hard
band (2--7 keV) is clearly extended compared to the PSF. This implies 
that the luminosity of the nucleus (before correction for absorption) is much 
lower than the observed luminosity. The recent {\chandra} observation 
of Ho et al. (2001) shows that the nucleus is surrounded by other sources 
of comparable brightness; for a power-law spectrum with $\Gamma = 1.8$ and 
{\NH} = $2\times10^{20}$ {\pcm}, the X-ray luminosity of the nucleus is 
$2.6\times10^{39}$ {\eps} in the 2--10 keV band.  This luminosity is about a 
factor of 4 smaller than the {\asca} luminosity in the same energy band.  \\

\noindent {\it NGC 4579 (S1.9/L1.9)}. --- Detailed analyses of three {\asca}
observations are presented in Terashima et al. (1998a, 2000c).  A {\chandra} 
image in the hard band is dominated by the nucleus (Ho et al. 2001).  \\

\noindent {\it NGC 4594 (L2)}. --- {\asca} results are briefly discussed in
Nicholson et al. (1998). These authors reported that adding an RS component to 
the power-law model did not affect the fit appreciably.  By contrast, we 
find that adding an RS component significantly improves the fit 
($\Delta \chi^2=-33.8$ for three additional parameters). Our result agrees 
well with the independent analyses of the same data set by Serlemitsos et al. 
(1996), Ptak et al. (1999), and Roberts, Schurch, \& Warwick (2001). Fabianno 
\& Juda (1997) and Roberts et al. (2001) detected a nuclear point source which
dominates the soft X-ray luminosity in the {\rosat} HRI image. A {\chandra}
image in the hard band is dominated also by the nucleus (Ho et al. 2001).\\

\noindent {\it NGC 4636 (L1.9)}. --- NGC 4636 was observed with {\asca} in 
1993 and 1995--1996. We analyzed only the newer data set because the second
observation is much deeper than the first. In the spectral fits, all four 
spectra from SIS and GIS were fitted simultaneously since the photon 
statistics are very good.  This elliptical galaxy has a very bright extended 
X-ray halo with a temperature of $\sim$0.8 keV. The canonical model
consisting of soft thermal emission and a hard component was not
acceptable because the single-temperature RS model poorly fits the
low-energy portion of the spectrum.  This is probably due to several 
complications, including the possible presence of temperature and abundance 
gradients, nonsolar abundance ratios, and uncertainties in the model, in 
particular the treatment of the Fe~L emission lines. The thermal emission 
of this galaxy has been extensively discussed by Matsushita et al. (1997) and 
Matsushita, Ohashi, \& Makishima (2000), and we will not address it here.

We fitted a simple power-law model to the region 4--10 keV, the latter 
chosen to minimize contamination from thermal emission; the best fit gives
$\Gamma=1.67^{+0.64}_{-0.60}$ and $\chi^2$ = 103.7 for 92 dof.  The upper 
limit on the Fe~K line shown in Table~10 was estimated from the same model.  
We tried to fit the spectrum with a thermal bremsstrahlung model instead of a 
power law and obtained $kT = 13.5 (>5.7)$ keV ($\chi^2$ = 103.9, 92 dof).  
These results are in good agreement with the composite spectrum of elliptical 
galaxies given by Matsumoto et al. (1997): thermal bremsstrahlung with $kT$ = 
$12.0^{+29.3}_{-5.5}$ keV or a power law with $\Gamma\,=\,1.8\pm0.4$.

The hard-band image above 4 keV is clearly extended compared to the
PSF. Therefore, the contribution of an AGN to the hard component in
this galaxy is small, if any. The {\LX}/{\LHa} ratio calculated using 
the {\it integrated}\ X-ray luminosity (89) is in the range of unobscured
AGNs.  But the hard X-ray emission is dominated by sources other than 
an AGN, and the {\LX}/{\LHa} ratio for the {\it nucleus}\ alone should be much 
smaller. The X-ray output of the nucleus is probably insufficient to 
drive the optical line emission. The powering source of the optical emission 
lines and the origin of the broad H$\alpha$ line are thus still puzzling.

The most likely origin of the majority of the hard emission is X-ray binaries
distributed over the host galaxy (see Matsushita et al. 1994). This idea has 
been confirmed by recent {\it Chandra} observations.  The high-resolution 
image reveals an extensive population of discrete point sources, while the 
nucleus is not clearly detected (Loewenstein et al. 2001).\\

\noindent {\it NGC 4639 (S1.0)}. --- A detailed analysis of the
{\asca} data is presented in Ho et al. (1999). The reanalysis of the
same data set in this paper shows a hint of an Fe~K line. The {\rosat}
HRI image shows a pointlike nucleus (Koratkar et al. 1995). A
{\chandra} image in the hard band is dominated by the nucleus.\\

\noindent {\it NGC 4736 (L2)}. --- {\asca} and {\rosat} results are given by
Roberts, Warwick, \& Ohashi (1999) and Roberts et al. (2001). They reported a 
marginal detection of an ionized Fe~K emission line. We found a possible hint 
of Fe~K emission in the {\asca} spectrum, although the equivalent width in our 
fit is lower than previously obtained.  The line center energy is consistent
with a He-like ionization state for Fe, but neutral Fe cannot be ruled out. 
Very extended emission is seen in the {\rosat} HRI image (Cui, Feldkhun, \& 
Braun 1997; Halderson et al. 2001). There are several bright X-ray sources in 
an archival {\chandra} image of the nuclear region. The {\asca} flux is 
the superposition of these sources and is not dominated by the nucleus.\\

\noindent {\it NGC 4941 (S2)}. --- The observed luminosity of this Seyfert 2 
galaxy before correction of absorption is only $6.4\times10^{39}$ {\eps},
which is among the weakest Seyfert~2s observed thus far in the X-rays.  
Comparing with the results obtained with {\sax} (Maiolino et al. 1998)
suggests variability of the absorption column. \\

\noindent {\it NGC 5005 (L1.9)}. --- The {\asca} data show the presence of a
pointlike hard X-ray nucleus with an X-ray luminosity consistent with
that expected from the H$\alpha$ luminosity (Terashima et
al. 2000a). Furthermore, comparison between the {\asca} and {\sax} data 
(Table 15) indicates that the 2--10 keV flux has varied by a factor of
2.4 between the two observations.  These characteristics strongly suggest that 
the nuclear source is an LLAGN.  A soft X-ray image taken with the {\rosat} HRI
shows an extended component ($\sim13$\% of the total flux; Rush \&
Malkan 1996) and is presumably associated with the soft thermal plasma
emission seen in our {\asca} spectrum.\\

\noindent {\it NGC 5033 (S1.5)}. --- A detailed analysis of the {\asca} data is
presented in Terashima et al. (1999). The {\rosat} HRI image is pointlike 
(Koratkar et al. 1995), as is the {\chandra} image (Ho et al. 2001)\\

\noindent {\it NGC 5194 (S2)}. --- Results of the first {\asca} observation is
presented in Terashima et al. (1998b). The second {\asca} observation
performed in 1994 is discussed in Ptak et al. (1999) and Fukazawa et
al. (2001). In this paper, we concentrate on the first observation in
1993 because the SIS data in 1994 suffered from serious telemetry
saturation. The hard-band image is extended and indicates that the AGN does
not dominate the hard X-ray flux. The detection of a strong
fluorescent Fe~K line gives strong evidence for the presence of a heavily
obscured AGN. Long-term variability in the hard X-ray band also
supports the presence of an AGN (Table 15; see also Fukazawa et
al. 2001). A recent {\chandra} spectrum that spatially isolates the nucleus 
confirms the presence of a strong Fe~K fluorescent line (Terashima \& Wilson
2001).  The Compton-thick nature of the nucleus is further suggested by the 
{\it BeppoSAX} detection of an absorbed 
({\NH} = $5.6^{+4.0}_{-1.6}\times10^{24}$ {\pcm}) power-law continuum above 
10 keV (Fukazawa et al. 2001). \\

\noindent {\it NGC 7217 (L2)}. --- The image in the 2--10 keV band is
probably extended. A Gaussian fit to the radial surface brightness
profile yielded $\sigma=0.80\pm0.40$ arcmin. The flux from an AGN, even if 
present, thus should be lower than the observed flux.  With {\LX}/{\LHa} = 1.2, 
the observed X-ray power is insufficient to drive the optical line emission.
The possibility of a Compton-thick AGN may be ruled out by the relatively
low upper limit of on the EW of a fluorescent Fe~K line ($<460$ eV).
The soft-band (0.5--2 keV) image is consistent with being pointlike
($\sigma=0.20^{+0.19}_{-0.16}$ arcmin), but the {\rosat} HRI image shows 
extended emission (Roberts et al. 2001).\\

\noindent {\it NGC 7743 (S2)}. --- This object is very faint, and we calculated
the X-ray fluxes in the 0.5--2 keV and 2--10 keV band by fitting the
one-dimensional projection of the SIS images along the axis connecting the 
nucleus with a serendipitous source located at 
(23$^h$44$^m$49$^s$, 09$^{\circ}$56$^\prime$35$^{\prime\prime}$) (J2000)
on the same CCD chip. We used a width of 3\farcm2 to make the projected
profile. This is the same technique as applied to NGC 4192 by Terashima
et al. (2000c). We estimated the spectral shape of the nucleus by using 
the hardness ratio (2--10 keV)/(0.5--2 keV). Assuming the Galactic absorption
({\NH} = 5.3$\times10^{20}$ {\pcm}) and a power-law spectrum, we obtained a 
photon index of $1.89^{+0.16}_{-0.13}$ (errors are at 1$\sigma$). Although 
this value is consistent with the canonical spectrum of Seyfert galaxies, the 
spectral shape might be more complicated, as is the case for the 
other galaxies in our sample, and we cannot constrain the intrinsic
absorption column density of the AGN. Using the Galactic absorption and 
$\Gamma = 1.9$, we derive X-ray fluxes of $4.6\times10^{-14}$ {\eps}{\pcm} in 
the 0.5--2 keV band and $7.3\times10^{-14}$ {\eps}{\pcm} in the 2--10 keV 
band, which correspond to X-ray luminosities of $3.3\times10^{39}$ {\eps} 
and $5.2\times10^{39}$ {\eps}, respectively.

NGC 7743 is the only object in our low-luminosity Seyfert sample which does 
not show any signature of an AGN (see Paper II). We need future observations 
to see whether this object is a Compton-thick AGN or truly a Seyfert~2 
nucleus without an accretion-powered source.

\section{Summary}

We have systematically analyzed archival {\it ASCA}\ data for a large 
sample of LINERs and low-luminosity Seyfert galaxies in order to derive a 
homogeneous database of X-ray properties suitable for investigating the 
physical nature of low-level activity in the centers of nearby galaxies.  
This paper defines the sample, discusses the observations and reductions, and 
presents the basic quantitative measurements of the spectral, spatial, and 
variability properties.  The next paper of this series discusses the 
implications of these results for a variety of issues concerning the nature of
low-luminosity AGNs.

\acknowledgments

The authors are grateful to all the {\asca} team members. 
Y.~T. and N.~I. are supported by the Japan Society for the Promotion of
Science Postdoctoral Fellowships for Young Scientists.  
L.C.H acknowledges financial support through NASA grants GO-06837.01-95A,
AR-07527.02-96A, and AR-08361.02-97A from the Space Telescope Science
Institute (operated by AURA, Inc., under NASA contract NAS5-26555).
This research has made use of data obtained from the High Energy
Astrophysics Science Archive Research Center (HEASARC), provided by
NASA's Goddard Space Flight Center and the NASA/IPAC Extragalactic
Database (NED) which is operated by the Jet Propulsion Laboratory,
California Institute of Technology, under contract with NASA.

\small

%\begin{table*}
\begin{deluxetable}{llllll}
        \tablecaption{The Sample\label{tbl-1}}
\tablewidth{10cm}
\tabletypesize{\footnotesize}
\tablehead{
\colhead{Name}    &
\colhead{Other Name} & 
\colhead{Hubble Type}	& 
\colhead{Distance}	& 
\colhead{$cz$}          & 
\colhead{Spectral Class$^a$} \\
	& 		& 		& (Mpc)		& (km s$^{-1}$) & \\
}
%\hline
\startdata
NGC 315	 &	& E+:	    & 65.8	& 4942		& L1.9	\\
NGC 404  &	& S0-       &  2.4      & --43	 	& L2	\\
NGC 1052 &	& E4	    & 17.8      & 1470		& L1.9	\\
NGC 1097 &	& SBS3      & 14.5      & 1275		& S1.5/L2 \\
NGC 1365 &	& SBbb	    & 16.9	& 1636		& S1.8	\\
NGC 1386 &	& SB0+	    & 16.9	& 868		& S2	\\
NGC 2273 &	& SBa:	    & 28.4      & 1871		& S2    \\
NGC 2639 &	& Sa?	    & 42.6	& 3336		& S1.9	\\
NGC 2655 &	& SAB0/a    & 24.4	& 1404		& S2	\\
NGC 3031 & M81	& Sab	    &  1.4	& --34		& S1.5  \\
NGC 3079 &	& SBc spin  & 20.4      & 1125		& S2    \\
NGC 3147 &	& Sbc	    & 40.9	& 2820		& S2    \\
NGC 3507 &	& SBb	    & 19.8	& 979		& L2	\\
NGC 3607 &	& S0:	    & 19.9	& 935		& L2	\\
NGC 3998 &	& S0?	    & 21.6      & 1040		& L1.9  \\
NGC 4111 &	& S0+:spin  & 17.0      & 807		& L2	\\
NGC 4192 & M98	& SABab     & 16.8      & --142		& T2	\\
NGC 4203 &	& SAB0$-$:  &  9.7      & 1086		& L1.9  \\
NGC 4258 & M106	& SABbc	    &  6.8	& 448		& S1.9  \\
NGC 4261 &	& E2+       & 35.1	& 2238		& L2    \\
NGC 4374 & M84	& E1        & 16.8	& 1060		& L2    \\
NGC 4438 &	& S0/a:     & 16.8      & 71		& L1.9  \\
NGC 4450 &	& Sab       & 16.8      & 1954		& L1.9  \\
NGC 4457 &	& SAB0/a    & 17.4	& 882		& L2	\\
NGC 4501 & M88	& Sb	    & 16.8	& 2281		& S2	\\
NGC 4565 &	& Sb? spin  &  9.7      & 1282		& S1.9  \\
NGC 4569 & M90	& SABab	    & 16.8      & --235		& T2    \\
NGC 4579 & M58	& SABb	    & 16.8      & 1519		& S1.9/L1.9 \\
NGC 4594 & M104	& Sa spin   & 20.0      & 1024		& L2	\\
NGC 4636 &	& E0+	    & 17.0      & 938		& L1.9	\\
NGC 4639 &	& SABbc	    & 16.8	& 1010		& S1.0	\\
NGC 4736 & M94	& Sab	    &  4.3      & 308		& L2	\\
NGC 4941 &	& SABab	    &  6.4      & 1108		& S2	\\
NGC 5005 &	& SABbc	    & 21.3      & 946		& L1.9	\\
NGC 5033 &	& Sc	    & 18.7      & 875	 	& S1.5	\\
NGC 5194 & M51	& Sbc pec   &  7.7      & 463		& S2  	\\
NGC 7217 &	& Sab	    & 16.0      & 952		& L2  	\\
NGC 7743 &	& SB0+	    & 24.4	& 1710		& S2	\\
\enddata
%\hline
%\end{tabular} 
%\tablecomments{
\tablenotetext{a}{Spectral class of the nucleus taken from Ho et
al. (1997a) except for NGC 1097 (Phillips et al. (1984);
Storchi-Bergmann et al. (1993)), NGC 1365, NGC 1386 (V\'eron-Cetty \&
V\'eron (1986)), and NGC 4941 (Storchi-Bergmann \& Pastriza (1989)),
where L = LINER, S = Seyfert, T = ``transition object''
(LINER/\ion{H}{2}), 1 = type 1, 2 = type 2, and a fractional number
between 1 and 2 denotes various intermediate types.  }
%\end{table*}
\end{deluxetable}

\begin{deluxetable}{llllrlr}
	\tablecaption{Observation Log\label{tbl-2}}
\tablewidth{15cm}
\tabletypesize{\footnotesize}
\tablehead{
\colhead{Name}	& 
\colhead{Date$^a$}	& 
\colhead{SIS Mode}	& 
\multicolumn{2}{c}{SIS} & 
\multicolumn{2}{c}{GIS}\\
\colhead{}	&	
\colhead{}	&
\colhead{}	& 
\colhead{count rate}	& 
\colhead{exposure} & 
\colhead{count rate}	& 
\colhead{exposure} \\
\colhead{}	&
\colhead{}	&
\colhead{}	& 
\colhead{(counts s$^{-1}$)} & 
\colhead{(ks)}	&  
\colhead{(counts s$^{-1}$)} & 
\colhead{(ks)}\\
}
\startdata
NGC 315 &1996 Aug 05	&1CCD Faint 		& 0.031	& 33.5	& 0.018	& 39.7\\ % LD$^c$ OF
NGC 404	&1997 Jul 21 	&1CCD Faint 		& ---	& 12.3	& ---	& 13.4\\
	&1998 Feb 06 	&1CCD Faint 		& ---	& 27.2	& ---	& 29.7\\
NGC 1052&1996 Aug 11	&1CCD Faint 		& 0.046	& 36.5	& 0.041	& 37.7\\
NGC 1097&1994 Jan 12	&2CCD Faint		& 0.079	& 36.7	& 0.055	& 42.0\\
NGC 1365&1994 Aug 12	&1CCD Faint		& 0.019	& 7.9	& 0.024	& 9.5\\
	&1995 Jan 25	&4CCD Faint/Bright LD$^c$ 0.48 keV	& 0.017	& 35.7	& 0.034$^b$& 39.7\\
NGC 1386&1995 Jan 26	&4CCD Faint/Bright LD$^c$ 0.55 keV	& 0.0053& 35.8	& 0.0063& 40.1\\
NGC 2273&1996 Oct 20	&1CCD Faint LD$^c$ 0.48 keV	& 0.004	& 36.1	& 0.013	& 38.4\\
NGC 2639&1997 Apr 16	&1CCD Faint		& 0.008	& 28.3	& 0.005	& 32.1\\% LD$^c$ OF
NGC 2655&1998 Oct 29	&2CCD Faint/Bright LD$^c$ 0.55 keV	& 0.012	& 32.6	& 0.009	& 39.6\\ %0101
NGC 3031&1994 Oct 21	&1CCD Faint		& 0.35	& 44.0	& 0.19	& 48.3\\
	&1995 Apr 01	&1CCD Faint LD$^c$ 0.41 keV	& 0.50	& 18.1	& 0.28	& 21.4\\
	&1995 Oct 24	&1CCD Faint LD$^c$ 0.41 keV	& 0.44	& 35.5	& 0.25	& 36.4\\
NGC 3079&1993 May 09 	&1CCD Faint/Bright	& 0.022	& 37.2	& 0.008	& 41.1\\
NGC 3147&1993 Sep 29	&4CCD Faint/Bright	& 0.048	& 21.4	& 0.045	& 38.6\\ % LD$^c$ OF
NGC 3507&1998 Nov 30	&1CCD Faint 		& 0.006	& 40.2	& 0.003	& 43.5\\
NGC 3607&1996 May 26	&2CCD Faint/Bright	& 0.011	& 68.5	& 0.007	& 64.5\\ % LD$^c$ OF 1212/3030
NGC 3998&1994 May 10	&2CCD/1CCD Faint	& 0.29	& 24.3	& 0.18	& 39.7\\
NGC 4111&1997 Dec 07 	&1CCD Faint LD$^c$ 0.48 keV & 0.008	& 14.4	& 0.004	& 15.6\\
	&1997 Dec 15 	&1CCD Faint LD$^c$ 0.48 keV & 0.008	& 19.0	& 0.003	& 21.2\\
NGC 4192&1997 Dec 17	&2CCD Faint/Bright 	& ---	& 3.9	& ---	& 4.3\\
	&1997 Dec 23	&2CCD Faint/Bright 	& ---	& 13.3	& ---	& 14.3\\
NGC 4203&1993 Dec 17	&2CCD Faint		& 0.035	& 35.1	& 0.024	& 38.1\\
NGC 4258&1993 May 15	&4CCD Faint/Bright 	& 0.075	& 36.5	& 0.088	& 40.1\\
	&1996 May 23	&1CCD Faint		& 0.16	& 24.2	& 0.12	& 31.5\\
	&1996 Jun 05	&1CCD Faint		& 0.17	& 29.1	& 0.13	& 33.5\\
	&1996 Dec 18	&1CCD Faint		& 0.16	& 28.4	& 0.13	& 30.4\\
NGC 4261&1996 Jun 23	&1CCD Faint		& 0.039	& 57.6	& 0.026	& 61.3\\
NGC 4374&1993 Jul 04	&2CCD Faint/Bright	& 0.075	& 37.4	& 0.025	& 21.7\\ %LVL OF 0202/0202
NGC 4438&1995 Dec 24 	&2CCD/1CCD Faint	& 0.012	& 21.5	& 0.004	& 20.6\\ %LVL OF
	&1996 Jan 05	&1CCD Faint/Bright	& 0.014	& 21.6	& 0.004	& 22.6\\ %LVL OF
NGC 4450&1995 Jun 20	&1CCD Faint LD$^c$ 0.41 keV	& 0.026	& 37.4	& 0.015	& 36.7\\
NGC 4457&1998 Jun 14	&1CCD Faint 		& 0.01	& 42.1	& 0.005	& 44.0\\
NGC 4501&1997 Jun 22	&2CCD Faint LD$^c$ 0.55 keV	& 0.019	& 35.7	& 0.013	& 35.4\\
NGC 4565&1994 May 28	&2CCD Faint		& 0.056$^b$& 37.4	& 0.036$^b$& 38.8\\% LVL OF
NGC 4569&1997 Jun 24	&1CCD Faint 		& 0.03	& 21.9	& 0.006	& 21.0\\
	&1997 Jul 06	&1CCD Faint 		& 0.03	& 19.0	& 0.007	& 20.4\\
NGC 4579&1995 Jun 25	&2CCD Faint/Bright	& 0.15	& 32.0 	& 0.090	& 30.9\\
	&1998 Dec 18	&1CCD Faint		& 0.14	& 18.6	& 0.14	& 19.6\\
	&1998 Dec 28	&1CCD Faint		& 0.14	& 17.6	& 0.12	& 19.4\\
NGC 4594&1994 Jan 20	&2CCD Faint LD$^c$ 120 ADU 	& 0.075	& 18.7 	& 0.060	& 19.7\\
NGC 4636&1995 Dec 28	&1CCD Faint LD$^c$ 0.48 keV	& 0.22	& 253.6	& 0.083	& 187.5\\
NGC 4639&1997 Dec 17	&1CCD Faint		& 0.029	& 31.6	& 0.019	& 33.1\\ 
	&1997 Dec 23	&1CCD Faint		& 0.030	& 35.2	& 0.018	& 38.0\\ 
NGC 4736&1995 May 25	&1CCD Faint LD$^c$ 0.41 keV	& 0.073	& 28.4	& 0.042	& 31.2\\
NGC 4941&1996 Jul 19	&1CCD Faint 		& 0.010	& 18.2	& 0.009	& 17.9\\
	&1997 Jan 08 	&1CCD Faint		& 0.008	& 17.9	& 0.009	& 17.8\\
NGC 5005&1995 Dec 13	&1CCD Faint		& 0.024	& 35.9	& 0.012	& 38.3\\
NGC 5033&1995 Dec 14	&1CCD Faint		& 0.16	& 36.1	& 0.11	& 38.6\\
NGC 5194&1993 May 11	&4CCD Faint/Bright	& 0.044	& 34.6	& 0.020	& 38.5\\
NGC 7217&1995 Nov 19	&2CCD Faint/Bright LD$^c$ 0.48 keV	& 0.008	& 81.0	& 0.004	& 80.9\\
NGC 7743&1998 Dec 09	&1CCD Faint		& ---	& 36.2	& ---	& 37.0\\ 
\enddata
\tablenotetext{a}{Observation start date.}
\tablenotetext{b}{Count rate includes a nearby source.}
\tablenotetext{c}{Level discriminator is enabled.}
\end{deluxetable}

\begin{table*}
\begin{center}
	\caption{Results of Power-Law Fits}
	\label{table:PLfit}
\begin{tabular}{lllll}
\hline \hline
Name	& \NH		& $\Gamma$ 	& $\chi^2$/dof	& \\%notes\\
	& ($10^{22}$~{\pcm}) &		&		& \\
\hline
NGC 315	& 0.0		& 1.74				& 158.4/71	& \\
NGC 1052& 0.0		& 0.20				& 240.8/111	& \\
NGC 1097& 0.059		& 1.92				& 285.4/146	& \\
NGC 1365& 0.0		& 2.0				& 276.0/180	& \\
NGC 1386& 0.0		& 2.5				& 81.1/40	& \\
NGC 2273& 0.0		& 1.25				& 113.4/59	&\\%1\\
NGC 2639& 0($<$0.16)	& $2.1^{+0.6}_{-0.3}$		& 59.5/53	& \\
NGC 2655& 0.0		& 1.1				& 189.8/60	& \\
NGC 3031& 0.0($<$0.009)	& $1.814^{+0.016}_{-0.013}$	& 439.1/515	& \\%2
NGC 3079& 0.0		& 2.34				& 134.4/90	& \\
NGC 3147& $0.062^{+0.05}_{-0.024}$ & $1.82^{+0.10}_{-0.09}$ & 249.6/243	&\\%1\\
NGC 3507 & 0($<0.072$)	& $1.71^{+0.24}_{-0.22}$	& 27.1/25	& \\
NGC 3607 & 0.0		& 2.57				& 156.3/63	& \\
NGC 3998 & $0.082\pm0.012$ & $1.90^{+0.03}_{-0.04}$ & 307.1/279 	&\\%1\\
NGC 4111 & 0.0		& 3.1				& 48.7/24	& \\
NGC 4203& $0.022(<0.053)$ & $1.78^{+0.07}_{-0.08}$	& 50.6/71	& \\
NGC 4258 (1993)& 0.0		& 1.3			& 2835/172	& \\
NGC 4258 (1996)& 0.0		& 0.22			& 6435/303	& \\
NGC 4261& 0.0		& 2.3				& 456.4/183 	&\\
NGC 4374& 0.0		& 3.1				& 170.4/81	&\\
NGC 4438& 0.0		& 2.68				& 58.0/42 & \\%1.381
NGC 4450& 0($<0.043$)	& $1.89^{+0.14}_{-0.08}$	& 72.1/70	& \\
NGC 4457& 0.0		& 2.3				& 50.1/27	& \\
NGC 4501& 0.0		& 1.98				& 114.0/76	&\\%1.20
NGC 4552& 0.0		& 2.35				& 155.7/82	&\\%1.19
NGC 4565& $0.22\pm0.04$	& $1.81^{+0.07}_{-0.08}$	& 163.7/112	&\\
NGC 4569& 0.0		& 2.1				& 101.1/36	&\\
NGC 4579 (1995)& $0.043^{+0.035}_{-0.012}$ & $1.80^{+0.08}_{-0.04}$ & 227.3/203 &\\%2 \\%1.12
NGC 4579 (1998)& 0.031 ($<$0.071)	& $1.80^{+0.06}_{-0.05}$	& 201/226 \\%1.15	
NGC 4594 & 0.056	& 1.61				& 138.7/88	& \\
NGC 4636& 0.82		& 8.2				& 11495/541	& \\
NGC 4639& $0.069^{+0.041}_{-0.038}$ & $1.66\pm0.10$	& 146.4/123	&\\% 2\\
NGC 4736& 0.0		& 2.0				& 293.7/151	& \\
NGC 4941& 0.0		& 0.52				& 134.9/55	& \\
NGC 5005& 0.0		& 2.05				& 144.7/60	& \\
NGC 5033& $0.087\pm 0.017$ & $1.72\pm 0.04$		& 173.1/186	&\\% 1\\
NGC 5194& 0.0		& 2.7				& 547.2/156	& \\
NGC 7217& 0.0		& 1.77				& 91.5/58	& \\
\hline
\end{tabular}
\end{center}
\end{table*}

\clearpage

%\begin{table}[hbt]
\begin{table}
\begin{center}
	\caption{Results of Partially Covered Power-Law Fits}
	\label{table:PCfit}
\begin{tabular}{llllllll}
\hline \hline
Name	& $N^1_{\rm H}$		& $N^2_{\rm H}$		& Covering		& $\Gamma$ & $\chi^2$/dof &\\% notes\\
	& ($10^{22}$~{\pcm}) & ($10^{22}$~{\pcm})       & Fraction	        &                         &\\
\hline
NGC 1052& $0.0(<0.034)$	& $9.5\pm1.6$	& $0.77^{+0.05}_{-0.07}$ & $1.11^{+0.14}_{-0.41}$ & 128.6/107	&\\% 1\\
NGC 2273& 0.0($<0.070$)	& $108^{+10}_{-9}$ &$0.983^{+0.003}_{-0.004}$ & $1.54^{+0.12}_{-0.11}$ & 69.1/57 &\\% 1\\
NGC 2639& 0.08($<$0.32)	& $33^{+125}_{-30}$	& $0.89^{+0.08}_{-0.81}$& $2.8^{+1.0}_{-0.6}$ & 45.8/49	&\\% 1 \\
NGC 2655& 0.0($<0.046$)	& $44^{+12}_{-9}$ & $0.978^{+0.010}_{-0.019}$ & $2.6\pm0.4$	& 74.5/58\\
NGC 4565& $0.30\pm0.07$	& $2.9^{+0.8}_{-0.9}$	& $0.65^{+0.10}_{-0.15}$ & $2.60\pm0.32$ 	& 140.6/110 	& \\
%NGC 4941& $0.18(<0.56)$	& $100^{+50}_{-30}$ & $0.962^{+0.023}_{-0.050}$	& $1.50^{+0.58}_{-0.48}$	& 43.5/51	&\\% 1\\
NGC 4941& $0.16^{+0.14}_{-0.11}$ & $99^{+12}_{-11}$ & $0.966^{+0.006}_{-0.007}$ & $1.48^{+0.14}_{-0.15}$	& 45.5 / 51 \\ 
\hline
\end{tabular}
\end{center}
\end{table}

%\begin{table}[tbh]
\begin{table}
\begin{center}
	\caption{Results of Thermal Bremsstrahlung Fits}
	\label{table:Bremsfit}
\begin{tabular}{lllll}
\hline \hline
Name	& \NH		& $kT$ 	& $\chi^2$/dof	&\\% notes\\
	& ($10^{22}$~{\pcm}) & (keV)		&		& \\
\hline
NGC 315	& 0.0			& 6.2			& 179.9/71	& \\
NGC 1097& 0.0			& 5.3			& 418.5/147	& \\
NGC 1365& 0.0			& 3.7			& 330.2/180	& \\
%NGC 1667& $0<0.070$		& $1.1^{+0.6}_{-0.4}$	& 14.0/10	& \\
NGC 2655& 0.0			& 200			& 193.3/60	& \\
NGC 3031& $0.0(<0.0007)$		& $5.16^{+0.15}_{-0.14}$& 757.5/515	& \\
NGC 3079& 0.0			& 2.1			& 182.7/90	& \\
NGC 3147& $0(<0.019)$		& $6.5^{+0.5}_{-1.2}$	& 256.0/243	&\\% 1\\
NGC 3507& $0(<0.054)$		& $6.3^{+6.8}_{-2.6}$	& 28.7/25	& \\
NGC 3607& 0.0			& 2.0			& 240.4/64	& \\
NGC 3998& $0.003 (<0.010)$	& $5.5\pm0.3$	& 373.8/279	& \\%1 $\sigma=0$\\
NGC 4111& 0.0			& 0.65			& 74.3/24	& \\
NGC 4203& 0.0$(<0.005)$		& $6.1^{+0.9}_{-0.7}$	& 76.5/71	& \\
NGC 4261& 0.0			& 2.0			& 619.4/183	&\\%1.245
NGC 4374& 0.0			& 0.75			& 220.1 / 81	& \\
NGC 4450& $0.0 (<0.009)$		& $3.7^{+0.8}_{-0.6}$	& 93.0/70	& \\
NGC 4457& 0.0			& 3.1			& 70.7/27	& \\
NGC 4501& 0.0			& 4.2			& 132.6/76	& \\%1.146
NGC 4552& 0.0			& 2.1			& 204.8/82	& \\%1.11
NGC 4565& $0.14\pm0.03$		& $6.4^{+1.1}_{-0.8}$	& 148.2/112	& \\
NGC 4636& 0.33			& 0.25			& 13008/541\\
NGC 4639& $0.005 (<0.037)$	& $9.3^{-1.8}_{+2.1}$	& 149.0/123	&\\% 2\\
NGC 4569& 0.0			& 3.5			& 119.8/36	& \\
NGC 4579 (1995)& 0.0($<$0.004) 	& $5.9^{+0.2}_{-0.7}$	& 260.3/203	&\\% 2\\ 
NGC 4579 (1998)& 0.0($<$0.005)	& $5.9\pm0.4$		& 232.5/226	& \\
NGC 4594& 0.0 ($<$0.023)	& $10.2^{+2.0}_{1.8}$	& 133.7/88	& \\
NGC 4636& 0.26			& 0.32			& 13038/541	& \\
NGC 4639& $0.005 (<0.039)$	& $9.3^{+2.5}_{-1.9}$	& 149.4/123	& \\
NGC 4736& 0.0			& 3.1			& 450.4/151	& \\
NGC 5005& 0.0			& 2.8			& 188.2/60	& \\
NGC 5033& $0.018\pm0.013$	& $7.9^{+0.8}_{-0.7}$	& 197.8/186	& \\
NGC 5194& 0.0			& 0.8			& 748.0/156	& \\
NGC 7217& 0.0			& 5.2			& 92.6/58\\
\hline
\end{tabular}
\end{center}
\end{table}

\begin{deluxetable}{llllllll}
\tablewidth{14cm}
\tabletypesize{\footnotesize}
	\tablecaption{Results of Power-Law + Raymond-Smith Model Fits}
	\label{table:RSfit}
\tablehead{
\colhead{Name}	& \NH & $kT$			& Abundances	& \NH		& $\Gamma$ 	& $\chi^2$/dof\\
	& ($10^{22}$~{\pcm}) & (keV)	& (solar units)	& ($10^{22}$~{\pcm}) &		& \\
}
\startdata
NGC 315	& 0.059(f)	& $0.77^{+0.05}_{-0.07}$& 0.5(f) ($>0.12$) & $0.49^{+0.41}_{-0.33}$ & $1.73^{+0.28}_{-0.25}$ & 89.5/69	& \\
NGC 1052& 0.030(f)	& $1.0^{+2.2}_{-0.3}$	& $0.04(<0.10)$	& $20.0^{+7.0}_{-8.3}$	& $1.67^{+0.57}_{-0.40}$ & 104.4/104	& \\
	&		&			&		& $2.5^{+2.7}_{-1.4}$ &	(CF$^a$ $0.77^{+0.08}_{-0.10}$) & \\
NGC 1097& 0.019(f)	& $0.75^{+0.06}_{-0.12}$& $0.077(>0.040)$ & $0.12^{+0.21}_{-0.08}$ & $1.66^{+0.13}_{-0.11}$ & 175.3/143 &\\
NGC 1365& 0.015(f)	& $0.85^{+0.05}_{-0.07}$	& 0.14($>0.071)$	& 0.54($<2.0$)	& $1.12^{+0.43}_{-0.37}$	& 186.7/176\\
NGC 1386& 0.014(f)	& $1.09^{+0.30}_{-0.22}$	& $0.136^{+0.396}_{-0.128}$	& $45^{+33}_{-22}$		& 1.7(f)	& 62.1/38\\
NGC 2639& 0.027(f)	& $0.81^{+0.27}_{-0.35}$ & 0.1(f) & 0.0($<$3.1)		& $1.64^{+0.91}_{-1.18}$	& 45.3/49	& \\
	& 0.027(f)	& $0.80^{+0.27}_{-0.40}$ & 0.5(f) & 0.0($<$0.31)		& $1.92^{+0.70}_{-0.37}$	& 45.6/49	& \\
NGC 2655& 0.021(f)	& 0.65(f)		& 0.1(f)  & $40^{+22}_{-13}$ 	& $1.2^{+0.6}_{-0.7}$	& 58.0/58\\
	&		&			&	  & 0.0 (f)		& (CF$^a$ $0.91^{+0.05}_{-0.13}$) & \\  
NGC 3031& 0.042(f)	& 0.86(f)		& 0.1(f) & 0.0($<$0.053)		& $1.796^{+0.027}_{-0.028}$	& 437.4/514\\ 
	& 0.042(f)	& 0.86(f)		& 0.5(f) & 0.0($<$0.036)		& $1.803^{+0.027}_{-0.020}$	& 437.5/514\\ 
NGC 3079& 0.008(f)	& $0.33^{+0.11}_{-0.06}$ & 0.1(f) & 0.0($<0.09$) & $1.87\pm0.18$ & 105.1/88	& \\
	& 0.008(f)	& $0.32^{+0.10}_{-0.06}$ & 0.5(f) & 0.0($<0.06$) & $1.93\pm0.17$ & 106.3/88	& \\
NGC 3507& 0.016(f)	& 0.65(f)		&  0.1(f) & $1.1^{+1.0}_{-0.9}$		& $2.3^{+1.0}_{-0.7}$	& 23.9/24	& \\
	& 0.016(f)	& 0.65(f)		&  0.5(f) & $0.96^{+0.85}_{-0.70}$	& $2.3^{+0.9}_{-0.6}$	& 25.3/24	& \\
NGC 3607& 0.015(f)	& $0.73^{+0.06}_{-0.07}$ & 0.1(f) & $1.6^{+1.8}_{-1.2}$	& $1.62^{+0.72}_{-0.42}$	& 43.1/61	& \\
	& 0.015(f)	& $0.70^{+0.09}_{-0.06}$ & 0.5(f) & 0.02($<0.6$)	& $1.37^{+0.30}_{-0.28}$	& 42.8/61	& \\
NGC 4111& 0.014(f)	& $0.65^{+0.12}_{-0.14}$ & 0.1(f) & 0.0($<3.0$)		& $0.92^{+1.03}_{-0.62}$	& 14.4/22	& \\
	& 0.014(f)	& $0.66\pm0.10$		& 0.5(f)  & 0.0($<0.47$)	& $1.36^{+0.46}_{-0.44}$	& 16.1/22	& \\
NGC 4258 (1993)$^b$& 0.012(f)	& $0.27^{+0.02}_{-0.04}, 0.77^{+0.05}_{-0.06}$ & $0.26^{+0.03}_{-0.05}$ & $12.7^{+0.9}_{-1.1}$ & $1.65^{+0.10}_{-0.16}$	& 197.3/164 \\
NGC 4258 (1996)& 0.012(f)& $0.69\pm0.02$		& $0.042^{+0.007}_{-0.006}$	& $6.6\pm0.3$	& $1.55^{+0.06}_{-0.07}$	& 355.6/298\\
NGC 4261& 0.016(f)	& $0.82^{+0.02}_{-0.03}$ & $0.17^{+0.03}_{-0.02}$ & $0.17(<0.39)$ & $1.30^{+0.07}_{-0.06}$ & 201.7/178	&\\% 2\\
NGC 4374& 0.026(f)	& $0.78^{+0.03}_{-0.05}$ & $0.12^{+0.31}_{-0.06}$ & 0($<2.2$)	& $1.29^{+0.81}_{-0.77}$	& 69.1/78 &\\
NGC 4438& 0.02(f)	& $0.79^{+0.07}_{-0.15}$ & 0.1(f) & 1.4($<5.3$)	& $2.0^{+1.5}_{-1.0}$		& 35.0/40	& \\
	& 0.02(f)	& $0.76^{+0.08}_{-0.13}$ & 0.5(f) & 0.0($<0.18$)	& $2.0^{+0.4}_{-0.3}$		& 34.9/40	& \\
NGC 4450& 0.024(f)	& $0.64^{+0.21}_{-0.31}$ & 0.1(f) & 0.0($<$0.082) & $1.75^{+0.18}_{-0.17}$ & 63.9/68	& \\
	& 0.024(f)	& $0.63^{+0.22}_{-0.32}$ & 0.5(f) & 0.0($<$0.062) & $1.81^{+0.16}_{-0.13}$ & 64.2/68	& \\
NGC 4457& 0.018(f)	& $0.68^{+0.12}_{-0.16}$ & 0.1(f) & $1.1(<3.8)$	& $1.7^{+1.2}_{-0.8}$	& 18.0/25\\
	& 0.018(f)	& $0.66^{+0.10}_{-0.15}$ & 0.5(f) & $0.15(<1.2)$& $1.5^{+0.7}_{-0.4}$	& 18.2/25\\
NGC 4501& 0.011(f)	& $0.79^{+0.10}_{-0.16}$ & 0.1(f) & $0.66(<2.3)$& $1.49^{+0.74}_{-0.51}$ & 93.6/74	& \\ %1.23
	& 0.011(f)	& $0.77^{+0.10}_{-0.14}$ & 0.5(f) & 0.0($<$0.90)& $1.48^{+0.53}_{-0.30}$ 	& 93.8/74 	& \\ %1.23
NGC 4565$^c$& 0.038(f)	& $1.23^{+0.95}_{-0.30}$ & 0.1(f) & $2.6^{+1.1}_{-1.2}$ & $2.51^{+0.35}_{-0.33}$	& 136.5/108 & \\
	&		&			&		& $0.37^{+0.35}_{-0.15}$  & (CF$^a$ $0.70^{+0.09}_{-0.15}$)	& & \\
	& 0.038(f)	& $1.36^{+1.56}_{-0.37}$ & 0.5(f) & $2.3^{+1.0}_{-0.9}$ & $2.48^{+0.31}_{-0.29}$	& 136.0/108 & \\
	&		& 			&		& $0.24\pm0.09$		& (CF$^a$ $0.68^{+0.20}_{-0.13}$)	& \\
NGC 4569& 0.029(f)	& $0.66\pm0.09$  & 0.1(f)	& $1.5^{+1.2}_{-0.8}$	& $2.18^{+0.71}_{-0.66}$ & 54.9/32 & \\
	& 0.029(f)	& $0.67\pm0.09$  & 0.5(f)	& $1.1^{+0.9}_{-0.8}$	& $2.17^{+0.62}_{-0.64}$ & 55.1/32 & \\
NGC 4579 (1995)& 0.031(f)	& $0.90^{+0.11}_{-0.05}$ & 0.5(f)	& $0.04\pm0.03$	& $1.72\pm0.05$		& 192.4/201	&\\% 2 \\
NGC 4579 (1998)& 0.031(f)	& 0.90(f)		& 0.5(f)	& 0.04 ($<$0.13) & $1.81\pm0.06$	& 200.5/225	&\\% 1 \\
NGC 4594& 0.035(f)	& $0.64^{+0.18}_{-0.15}$ & $0.05^{+0.07}_{-0.02}$ & $0.73^{+0.36}_{-0.33}$ & $1.89\pm0.17$ & 104.9/85 & \\
NGC 4636& 0.018(f)	& 0.77			& 0.31			& 0.15		& 1.32			& 1398.8/538\\
NGC 4736& 0.011(f)	& $0.61^{+0.05}_{-0.10}$ & $0.08^{+0.50}_{-0.06}$ & 0.0($<1.3$) & $1.56^{+0.12}_{-0.15}$ & 165.3/146	& \\
NGC 5005& 0.011(f)	& $0.76^{+0.07}_{-0.08}$ & $0.06^{+0.09}_{-0.02}$ & 0.10($<0.86$) & $0.97\pm0.37$ 	& 71.3/57	& \\
NGC 5194& 0.013(f)	& $0.64^{+0.03}_{-0.04}$ & $0.040^{+0.012}_{-0.009}$ & $2.1^{+1.2}_{-1.0}$ & $1.60^{+0.49}_{-0.47}$	& 186.4/151 &\\% 1 \\
NGC 7217& 0.11(f)	& $0.76^{+0.10}_{-0.13}$ & 0.1(f) & $1.6^{+1.6}_{-1.0}$		& $2.40^{+0.79}_{-0.51}$&71.2/57\\
	& 0.11(f)	& $0.74^{+0.07}_{-0.13}$ & 0.5(f) & $0.92^{+0.74}_{-0.62}$	& $2.22^{+0.54}_{-0.44}$&71.3/57\\
\enddata
\tablenotetext{a}{
The power-law component is assumed to be covered partially.}
\tablenotetext{b}{
Two-temperature Raymond-Smith model is assumed for the soft
component.}
\tablenotetext{c}{
Fitting results of the nucleus + the off-center source.
}
\end{deluxetable}

%\begin{table}[htb]
\begin{table}
\begin{center}
	\caption{Results of Thermal Bermsstrahlung + Raymond-Smith Model Fits\tablenotemark{a}}
	\label{table:RSBrefit}
\begin{tabular}{lllllllll}
\hline \hline
Name	& \NH (galactic) & $kT$			& Abundances	& \NH		& $kT$	 & $\chi^2$/dof\\
	& ($10^{22}$ {\pcm})	& (keV)		& (solar units)	& ($10^{22}$ {\pcm})	& (keV)	& \\
\hline
NGC 315 & 0.059(f)	& $0.77^{+0.05}_-0.07{}$	& 0.5(f) ($>$0.11)	& $0.29^{+0.30}_{-0.23}$	& $10.4^{+22.6}_{-4.2}$	& 90.9/69 & \\
NGC 1097& 0.019(f)	& $0.78^{+0.05}_{-0.07}$	& $0.065^{+0.045}_{-0.024}$	& $0.10^{+0.16}_{-0.07}$ & $12.2^{+5.3}_{-3.1}$	& 181.3/143 & \\
NGC 1365& 0.015(f)	& $0.85\pm0.03$		& $0.14\pm0.02$		& $0.72^{+0.33}_{-0.27}$		& $>14$	& 186.9/176\\
NGC 3031& 0.042(f)	& 0.86(f)		& 0.1(f)	& $0.0 (<0.009)$	& $7.53^{+0.46}_{-0.43}$& 510.2/514\\
	& 0.042(f)	& 0.86(f)		& 0.5(f)	& $0.0 (<0.009)$	& $6.40^{+0.28}_{-0.27}$& 564.7/514\\
NGC 3079& 0.008(f)	& $0.33^{+0.09}_{-0.05}$& 0.1(f)	& $0.0 (<0.06)$	& $5.0^{+2.7}_{-1.4}$	& 109.5/88	& \\
	& 0.008(f)	& $0.32^{+0.07}_{-0.05}$& 0.5(f)	& $0.0 (<0.003)$	& $4.4^{+1.9}_{-1.2}$	& 112.8/88	& \\
NGC 3507& 0.016(f)	& 0.65(f)		& 0.1(f)	& $0.71 (<1.4)$	& $4.1^{+45}_{-2.2}$	& 25.0/24	& \\
	& 0.016(f)	& 0.65(f)		& 0.5(f)	& $0.0 (<1.1)$	& $9.5^{+39}_{-7.4}$	& 25.9/24	& \\
NGC 3607& 0.015(f)	& $0.73^{+0.06}_{-0.08}$& 0.1(f)	& $1.3^{+1.3}_{-1.2}$	& 19($>6.0$)		& 43.1/61	& \\
	& 0.015(f)	& $0.73^{+0.06}_{-0.08}$& 0.5(f)	& 0.0($<$0.60)		& 30($>9.3$)		& 43.0/61	& \\
NGC 4111& 0.014(f)	& $0.65^{+0.13}_{-0.14}$& 0.1(f)	& 0.49($<$2.8)		& $>5.0$		& 14.9/22	& \\
	& 0.014(f)	& $0.66\pm0.09$ 	& 0.5(f)	& 0.0($<$0.43)		& $>6.5$		& 16.4/22	& \\
NGC 4261& 0.016(f)	& $0.82\pm0.02$		& $0.17^{+0.03}_{-0.02}$ & $0.13(<0.35)$ & $>44$		& 203.1/178 	&\\% 2\\
NGC 4374& 0.026(f)	& $0.78^{+0.04}_{-0.05}$ & $0.12\pm0.03$& 0.0($<$1.9)	& $>7.3$		& 69.3 / 78	& \\
NGC 4438& 0.020(f)	& $0.79^{+0.07}_{-0.15}$& 0.1(f)	& 0.98 ($<$4.2)	& $6.2 (>2.0)$	& 35.1/40\\
	& 0.020(f)	& $0.77^{+0.07}_{-0.12}$& 0.5(f)	& 0.0 ($<$0.19)	& $4.5^{+5.1}_{-1.8}$	& 36.7/40\\
NGC 4450& 0.024(f)	& $0.67^{+0.20}_{-0.27}$& 0.1(f)	& 0.0 ($<$0.047)	& $7.2^{+12}_{-2.3}$	& 68.6/68&\\%1.00
	& 0.024(f)	& $0.64^{+0.18}_{-0.34}$& 0.5(f)	& 0.0 ($<$0.026)	& $5.6^{+2.4}_{-1.4}$	& 72.2/68 &\\ %1.05
NGC 4457& 0.018(f)	& $0.68^{+0.12}_{-0.16}$& 0.1(f)	& 0.84($<$3.1)	& 11 ($>2.8$)	& 17.9/25\\
	& 0.018(f)	& $0.66^{+0.10}_{-0.14}$& 0.5(f)	& 0.07($<$0.83)	& 19 ($>4.3$)	& 18.1/25\\
NGC 4501& 0.011(f)	& $0.75^{+0.14}_{-0.12}$& 0.1(f)	& 0.37 ($<$1.7)	& $34 (>5.2)$		& 94.0/74 &\\ %1.23
	& 0.011(f)	& $0.78^{+0.13}_{-0.13}$& 0.5(f)	& 0.0 ($<$0.53)	& $19 (>6.7)$		& 94.2/74 &\\ %1.23
%NGC 4552& 0.026(f)	& $0.71^{+0.06}_{-0.07}$& 0.1(f)	& $0.94(<1.4)$	& $9.9^{+65}_{-5.7}$	& 85.5/80	& \\%1.29
%	& 0.026(f)	& $0.70\pm0.06$		& 0.5(f)	& $0.063(<0.52)$& $8.6^{+13}_{-4.1}$	& 84.7/80	& \\%1.29
%NGC 4565& \\
%NGC4569& 0.02(f)	& 0.67			& 0.12		& 0.78		& 8.1($>3.4$)& 54.1/31\\
NGC 4569& 0.029(f)	& $0.66^{+0.09}_{-0.12}$& 0.1(f)	& $1.1\pm0.8$	& $5.7^{+23}_{-2.4}$	& 56.2/32\\
	& 0.029(f)	& $0.66^{+0.08}_{-0.20}$& 0.5(f)	& $0.62^{+0.67}_{-0.59}$	& $6.0^{+41}_{-2.9}$	& 56.5/32\\
NGC 4579 (1995)& 0.031(f)	& $0.89^{+0.12}_{-0.04}$	& 0.5(f)	& $0 (<0.010)$	& $7.9^{+1.2}_{-0.9}$	& 200.3/201 &\\
NGC 4579 (1998)& 0.031(f)	& 0.89 (f)	& 0.5(f)	& 0.0 ($<$0.082)	& $6.8^{+0.6}_{-1.0}$		& 213.6/225 & \\
NGC 4594& 0.035(f)	& $0.62^{+0.09}_{-0.21}$& $0.053 (>0.026)$	& $0.42^{+0.30}_{-0.37}$ & $7.5^{+3.1}_{-1.8}$ &104.2/85&\\ 
NGC 4636& 0.018(f)	& 0.77		& 0.31			& 0.0	& 10(f)	& 1418.1/539\\
NGC 4736& 0.011(f)	& $0.61^{+0.06}_{-0.09}$ & $0.048^{+0.041}_{-0.027}$ & 0.0($<$1.0) & $16^{+12}_{-6}$	& 164.1/146	& \\
NGC 5005& 0.011(f)	& $0.74\pm0.08$	& $0.072^{+0.11}_{-0.036}$	& $0.12(<1.0)$	& $>31$	& 72.6/57\\
NGC 5194& 0.013(f)	& $0.64^{+0.03}_{-0.04}$ & $0.041\pm0.009$	& $1.8^{+0.8}_{-0.9}$	& 28($>$8.1)	& 187.4/151\\
NGC 7217& 0.11(f)	& $0.74^{+0.11}_{-0.14}$ & 0.1(f)	& $1.0^{+1.2}_{-0.7}$		& $4.0^{+3.5}_{-1.6}$	& 70.1/57\\
	& 0.11(f)	& $0.70^{+0.12}_{-0.11}$ & 0.5(f)	& $0.46^{+0.61}_{-0.35}$	& $4.5^{+3.0}_{-1.6}$	& 69.6/57\\
\hline
\end{tabular}
\tablenotetext{a}{Entries followed by ``(f)'' indicate that the parameter is fixed.}
\end{center}
\end{table}

%\begin{table}[htb]
%\begin{center}
\begin{deluxetable}{llllllllll}
\tablewidth{15cm}
\tabletypesize{\footnotesize}
	\tablecaption{Results of Power-Law + Variable Abundance Raymond-Smith Model Fits\tablenotemark{a}}
	\label{table:vRSfit}
\tablehead{
%\begin{tabular}{llllllllll}
%\hline \hline
\colhead{Name} & \NH 	                & $kT$	& Abundance	& Abundance	& \NH		& $\Gamma$	& $\chi^2$/dof	&\\% notes\\
	& (10$^{22}$cm$^{-2}$)	& (keV)	& (O,Ne,Mg,Si)	& (Fe)	& (10$^{22}$cm$^{-2}$)	& 	& & \\
}
%\hline
\startdata
NGC 315	& 0.059(f) & $0.77^{+0.05}_{-0.07}$	& 0.5(f) & 0.32 ($>0.17$) & $0.61^{+0.59}_{-0.42}$ & $1.70^{+0.29}_{-0.27}$ & 88.9/68\\
NGC 1052& 0.030(f) & $0.84^{+2.9}_{-0.14}$ & 0.1(f)	& 0.051 $<0.12$	& $1.9^{+3.6}_{-0.9}$	& $1.55^{+0.49}_{-0.41}$ & 104.8/104\\
	&	&			&		& 		& $18^{+7}_{-5}$ & (CF\tablenotemark{b} $0.78^{+0.06}_{-0.11}$) & \\
NGC 1097& 0.019(f) & $0.65\pm0.08$	& 0.5(f)	& $0.11^{+0.07}_{-0.04}$	& $0.13^{+0.10}_{-0.07}$ & $1.67^{+0.09}_{-0.10}$ & 167.6/143 & \\
NGC 1365& 0.015(f)	& $0.84^{+0.04}_{-0.06}$	& 0.5(f)	& $0.31^{+1.14}_{-0.15}$	& 0.0($<$1.0)		& $1.08^{+0.28}_{-0.34}$	&  186.7/176\\
NGC 3079& 0.008(f) & $0.62^{+0.09}_{-0.12}$ & 0.5(f) & $0.048^{+0.026}_{-0.019}$ & $1.7^{+1.8}_{-1.2}$	& $1.96^{+0.74}_{-0.71}$ & 88.8/87 & \\
NGC 4261& 0.016(f) & $0.81^{+0.03}_{-0.02}$ & 0.5(f)	& $0.35\pm0.04$	& 0.044($<0.18$)	& $1.37^{+0.07}_{-0.06}$	& 205.7/178	& \\
NGC 4374& 0.026(f) & $0.77^{+0.04}_{-0.06}$ & 0.5(f)	& $0.20^{+0.18}_{-0.06}$	& 0.0($<2.6$)	& $1.34^{+0.61}_{-0.77}$& 67.4/78\\
NGC 4450& 0.024(f) & $0.65^{+0.18}_{-0.30}$ & 0.5(f) & $0.09(>0.03)$ 	& 0.0($<$0.32) 	& $1.64^{+0.30}_{-0.39}$ & 62.8/68 & \\
NGC 4501& 0.011(f) & $0.77^{+0.11}_{-0.17}$ & 0.5(f) & $0.16(>0.035)$	& $0.59 (<3.5)$	& $1.43^{+0.87}_{-0.55}$ & 93.3/73 & \\%1.22
%NGC 4552& 0.026(f) & $0.70\pm0.06$	& 0.5(f)	& $0.70(>0.18)$	& $1.82^{+0.45}_{-0.40}$ & $83.9/79$	& \\%1.29\\
NGC 4565& 0.038(f) & $0.80^{+1.08}_{-0.38}$ & 0.5(f) & $0.034(<0.092)$	& $3.4^{+1.5}_{-1.2}$	& $2.71^{+0.43}_{-0.53}$ & 133.2/107 &\\% 3\\
	&	   &				&	&		& $0.52^{+0.38}_{-0.30}$ & (CF$^a$ $0.70^{+0.11}_{-0.18}$)	&\\
NGC 4579 (1995) & 0.031(f) & $0.89^{+0.12}_{-0.08}$ & 0.5(f)	& $0.4(>0.12)$	& $0.04\pm0.03$	& $1.72\pm0.05$ & 192.6/201 &\\% 2\\
NGC 4594 & 0.035(f) & $0.62^{+0.08}_{-0.11}$ & 0.5(f) & $0.11^{+0.08}_{-0.03}$ & $0.73\pm0.29$ 	& $1.89\pm0.16$ 	& 100.2/85		& \\
NGC 4636& 0.018(f)& 0.76		& 1(f)	& 0.60	& 0.046	& 1.60	& 1315.9/538 \\
NGC 4736& 0.011(f)& $0.61^{+0.05}_{-0.08}$	& 0.5(f) & $0.26\pm0.04$	& 0.0 ($<$0.031) & $1.62^{+0.10}_{-0.08}$ & 166.6/146 	&\\% 1\\
NGC 5005& 0.011(f)& $0.72\pm0.09$	& 0.5(f) & $0.14^{+0.22}_{-0.06}$	& 0.0($<$0.27)	& $1.06^{+0.24}_{-0.18}$	& 73.8/57	& \\
NGC 5194& 0.013(f) & $0.60^{+0.05}_{-0.04}$ 	& $0.15^{+0.06}_{-0.06}$ & $0.046^{+0.016}_{-0.012}$	& $2.2^{+1.2}_{-1.3}$ & $1.65^{+0.42}_{-0.47}$	& 177.4/150	& \\
\enddata
\tablenotetext{a}{Entries followed by ``(f)'' indicate that the parameter is fixed.}
\tablenotetext{b}{The power-law component is assumed to be partially covered.}
\end{deluxetable}

%\begin{table}[htb]
\begin{table}
\begin{center}
	\caption{Summary of Fe K Emission-Line Parameters}
	\label{table:Fepar}
\begin{tabular}{llllll}
\hline \hline
Name	& Center Energy\tablenotemark{a} 	& Line Width	& Equivalent Width& $\Delta\chi^{2}$\tablenotemark{b}& Notes\tablenotemark{c}\\
	& (keV) 		& (keV)		& (eV)			& \\
\hline
NGC 1052& $6.35\pm0.08$		& 0(f)		& $180^{+80}_{-90}$	&12.1	& 5\\
NGC 1365& $6.59^{+0.04}_{-0.05}$& 0.002($<$0.12)& $1900^{+1000}_{-800}$ &36.4	& 2\\
NGC 1386& $6.50^{+0.34}_{-0.16}$& 0(f)		& $910^{+2890}_{-770}$	& 5.7	& 2\\
NGC 2273& $6.33^{+0.05}_{-0.03}$ & 0(f)		& $1040^{+440}_{-460}$	&20.1	& 4\\
NGC 2639& 6.4(f)		& 0(f)		& $1490^{+11110}_{-1270}$& 3.3	& 4\\
	& 6.4(f)		& 0(f)		& $3130^{+2270}_{-2000}$& 6.5 	& 2\\
NGC 3031& $6.59^{+0.22}_{-0.13}$& 0(f)		& $106^{+59}_{-56}$	&11.0	& 2\\
NGC 3147& $6.49\pm0.09$ 	& 0(f)		& $490^{+220}_{-230}$	&12.9	& 1\\
NGC 3998& $6.41^{+0.12}_{-0.19}$ & 0(f)		& $85^{+81}_{-71}$	& 3.7	& 1\\
NGC 4258 (1993)& $6.66^{+0.20}_{-0.07}$	& 0(f)	& $108^{+73}_{-64}$ 	& 7.4	& 2\tablenotemark{d}\\
NGC 4258 (1996)& $6.31^{+0.09}_{-0.10}$ & 0(f)	& $54^{+25}_{-27}$ 	&10.6	& 2\\
NGC 4261& $6.85^{+0.08}_{-0.15}$ & 0(f)		& $550^{+300}_{-310}$	& 8.5	& 2\\
NGC 4579 (1995)& $6.73^{+0.13}_{-0.12}$ & $0.17^{+0.11}_{-0.12}$ & $490^{+180}_{-190}$	& 20.0	& 2\\
NGC 4579 (1998)& $6.39\pm0.09$	& 0.0 ($<$0.16)	& $250^{+105}_{-95}$	& 17.8	& 2\\
NGC 4639& $6.67^{+0.16}_{-0.23}$& 0(f)		& $520^{+320}_{-300}$	& 7.7	& 1\\
NGC 4736& $6.51^{+0.42}_{-0.19}$& 0(f)		& $334^{+233}_{-239}$	& 5.1	& 2\\
NGC 4941& $6.35^{+0.04}_{-0.10}$ & 0(f)		& $568^{+222}_{-227}$	&14.9	& 4\\
NGC 5033& $6.43^{+0.13}_{-0.08}$ & 0.08 ($<$0.23)& $306^{+116}_{-119}$	&22.5	& 1\\
NGC 5194& $6.34^{+0.04}_{-0.13}$ & 0(f)		& $910^{+350}_{-360}$	& 17.8	& 2\\
\hline
\end{tabular}
\end{center}
%\tablecomments{{\it a}: Line center energy is corrected for redshift, ]
\tablenotetext{a}{Line center energy is corrected for redshift.}
\tablenotetext{b}{Improvement of $\chi^2$ by adding a Gaussian component.}
\tablenotetext{c}{Models applied: 1 = PL model, 2 = PL + RS model, 3 = PL + variable RS model, 
                  4 = partial covering PL model, 5 = partial covering PL + RS model.}
\tablenotetext{d}{See text for details.}

\end{table}

%\begin{table}[htb]
\begin{table}
\begin{center}
	\caption{Upper Limits on the Equivalent Width of Fe K Emission}
	\label{table:fe_ew}
\begin{tabular}{lccccccccccc}
\hline \hline
Name		& Fe K 6.4 keV & Fe K 6.7 keV 	& $\Delta\chi^{2}$ (6.4 keV)\tablenotemark{a} & $\Delta\chi^{2}$ (6.7 keV)\tablenotemark{a}& Notes\tablenotemark{b} & \\
		& EW (eV)	& EW (eV) \\
\hline
NGC 315		& $<380$	& 325($<1030$)		& 0.0	& 0.7	& 2&\\
NGC 404		& ...		& ...			& ...	& ...	& \tablenotemark{c}\\
NGC 1097	& $<160$ 	& $<210$ 		& 0.0 	& 0.1	& 3\\
NGC 2655	& $<270$	& $125 (<420)$		& 0.0	& 0.6	& 5\\
NGC 3079 	& $930(<1960)$ 	& 1320 ($<2670$)	& 2.4	& 2.5	& 2\\
NGC 3507	& $193 (<2830)$	& $70 (<3580)$		& 0.0	& 0.0	& 2\\
NGC 3607	& $590 (<1260)$	& $334 (<1220)$		& 2.1	& 0.4	& 2\\
NGC 4111	& $<1860$	& $<2800$		& 0.0	& 0.0	& 2& \\
NGC 4192	& ...		& ...			& ...	& ...	& \tablenotemark{c}\\
NGC 4203	& $<310$	& $<270$		& 0.0	& 0.0	& 1& \\
NGC 4374	& $<720$	& 610 ($<1780$)		& 0.0	& 0.9	& 2\\
NGC 4438	& $<1300$	& $<3900$		& 0.0	& 1.0	& 2\\
NGC 4450	& $550(<1200)$ & 610($<1400)$ 		& 2.5	& 1.7	& 2& \\
NGC 4457	& ...		& ...			& ...	& ...	& \tablenotemark{c}\\
NGC 4501	& $1200^{+1110}_{-1070}$ & $910^{+900}_{-870}$ 	& 2.9	& 2.6	& 2&\\
NGC 4565	& $240(<620)$	& $<350$		& 1.7	& 0.1	& 5& \\
NGC 4569	& $<1800$	& $<4800$		& 0.0	& 0.0	& 2& \\
NGC 4594	& $<150$  	& $<260$ 		& 0.0	& 0.0	& 3& \\
NGC 4636	& $250(<540)$	& $<120$		& 2.7	& 0.0	& 3& \\
NGC 5005	& $200(<810)$	& $<510$		& 0.9	& 0.0	& 2&\\
NGC 7217	& $<460$	& $<1100$		& 0.0	& 0.0	& 2&\\
NGC 7743	& ...		& ...			& ...	& ...	& \tablenotemark{c}\\
\hline \hline
\end{tabular}
\end{center}
%\tablecomments{}
\tablenotetext{a}{ Improvement of $\chi^2$ by adding a Gaussian component.}
\tablenotetext{b}{Models applied: 1 = PL model, 2 = PL + RS model, 3 = PL + variable RS 
model, 4 = partial covering PL model, 5 = partial covering PL + RS model.}
\tablenotetext{c}{Low signal-to-noise ratio.} 

\end{table}

\begin{deluxetable}{llllll}
\tabletypesize{\footnotesize}
\tablewidth{11cm} 
\tablecaption{Summary of the Spectral Parameters for the Soft Component
	\label{table:summary_hard}
}
\tablehead{
\colhead{Name}	& \colhead{Spectral}	& \colhead{$kT$}& \colhead{Abundances}& \colhead{Fe Abundance}	& \colhead{Model\tablenotemark{a}}\\
	&Class 	& (keV)		& (solar units)	& (solar units)	& \\
}
\startdata
NGC 315	& L1.9	& $0.77^{+0.05}_{-0.07}$ & 0.5(f)	& ...	& PL+RS\\
NGC 1052& L1.9	& $1.0^{+2.2}_{-0.3}$	& 0.04($<$0.10)	& ...	& PC+RS+GA\\
NGC 3998& L1.9	& ...			& ...		& ...	& PL+GA\\
NGC 4203& L1.9	& ...			& ...		& ...	& PL\\
NGC 4438& L1.9	& $0.79^{+0.07}_{-0.15}$& 0.1(f)	& ...	& PL+RS\\
	&	& $0.76^{+0.08}_{-0.13}$& 0.5(f)	& ...	& PL+RS\\
NGC 4450& L1.9	& $0.64^{+0.21}_{-0.31}$& 0.1(f)	& ...	& PL+RS\\
	&	& $0.63^{+0.22}_{-0.32}$& 0.5(f)	& ...	& PL+RS\\
NGC 4579 (1995)& S1.9/L1.9 & $0.90^{+0.11}_{-0.05}$	& 0.5(f)& ...	& PL+RS+GA\\
NGC 4579 (1998)& S1.9/L1.9 &  0.90(f)	& 0.5(f)	& ...	& PL+RS+GA\\
NGC 4636& L1.9	& 0.76			& 1(f)		& 0.60	& PL+RS\\
NGC 5005& L1.9	& $0.76^{+0.07}_{-0.08}$& $0.06^{+0.09}_{-0.02}$& ...& PL+RS\\
\tableline
NGC 404	& L2	& ...			& ...		& ...	& \tablenotemark{b}\\
NGC 3507& L2	& 0.65(f)		& 0.1(f)	& ...	& PL+RS\\
	&	& 0.65(f)		& 0.5(f)	& ...	& PL+RS\\
NGC 3607& L2	& $0.73^{+0.06}_{-0.07}$& 0.1(f)	& ...	& PL+RS\\
	&	& $0.70^{+0.09}_{-0.06}$& 0.5(f)	& ...	& PL+RS\\
NGC 4111& L2	& $0.65^{+0.12}_{-0.14}$& 0.1(f)	& ...	& PL+RS\\
	&	& $0.66\pm0.10$		& 0.5(f)	& ...	& PL+RS\\
NGC 4192& T2	& ...			& ...		& ...	& \tablenotemark{b}\\
NGC 4261& L2	& $0.82^{+0.02}_{-0.03}$& $0.17^{+0.03}_{-0.02}$& ...& PL+RS+GA\\
NGC 4374& L2	& $0.78^{+0.03}_{-0.05}$& $0.12^{+0.31}_{-0.06}$ & ... & PL+RS\\
NGC 4457& L2	& $0.68^{+0.12}_{-0.16}$& 0.1(f)	& ...	& PL+RS\\
	&	& $0.66^{+0.12}_{-0.16}$& 0.5(f)	& ...	& PL+RS\\
NGC 4569& T2	& $0.66\pm0.09$		& 0.1(f)	& ...	& PL+RS\\
	&	& $0.67\pm0.09$		& 0.1(f)	& ...	& PL+RS\\
NGC 4594& L2	& $0.62^{+0.08}_{-0.11}$& 0.5(f)	& $0.11^{+0.08}_{-0.03}$ & PL+vRS\\
NGC 4736& L2	& $0.61^{+0.05}_{-0.10}$& $0.08^{+0.50}_{-0.06}$ & ...& PL+RS+GA\\
NGC 7217& L2	& $0.76^{+0.10}_{-0.13}$& 0.1(f)	& ...	& PL+RS\\
	&	& $0.74^{+0.07}_{-0.13}$& 0.5(f)	& ...	& PL+RS\\
\tableline
NGC 1097& S1.5/L2	& $0.65\pm0.08$		& 0.5(f)	& $0.11^{+0.07}_{-0.04}$ & PL+vRS\\
NGC 1365& S1.8	& $0.85^{+0.05}_{-0.07}$& 0.14 ($>$0.071)& ...	& RS+PL+GA\\
NGC 2639& S1.9	& $0.81^{+0.27}_{-0.35}$& 0.1(f)	& ...	& RS+PL+GA\\
	&	& $0.80^{+0.27}_{-0.40}$& 0.5(f)	& ...	& RS+PL+GA\\
NGC 3031& S1.5	& 0.86(f)		& 0.1(f)	& ...	& RS+PL+GA\\
	&	& 0.86(f)		& 0.5(f)	& ...	& RS+PL+GA\\
NGC 4258 (1993)& S1.9	& $0.27^{+0.02}_{-0.04}$	& $0.26^{+0.03}_{-0.05}$ & ...	& PL+RS+RS+BRE+GA\\*
		& 	& $0.77^{+0.05}_{-0.06}$ \\
NGC 4258 (1996)& S1.9	& $0.69\pm0.02$	& $0.042^{+0.007}_{-0.006}$	& ...	& PL+RS+GA\\
NGC 4565& S1.5	& $1.23^{+0.95}_{-0.30}$& 0.1(f)	& ...	& PC+RS\\
	&	& $1.36^{+1.56}_{-0.37}$& 0.5(f)	& ...	& PC+RS\\
NGC 4639& S1.0	& ...			& ...		& ...	& PL+GA\\
NGC 5033& S1.5	& ...			& ...		& ...	& PL+GA\\
\tableline
NGC 1386& S2	& $1.09^{+0.30}_{-0.22}$& $0.136^{+0.396}_{-0.128}$	& ...	& PL+RS+GA\\
NGC 2273& S2	& ...			& ...		& ...	& PC+GA\\
NGC 2655& S2	& 0.65(f)		& 0.1(f)	& ...	& PC+RS\\
NGC 3079& S2	& $0.62^{+0.09}_{-0.12}$& 0.5(f)	& $0.048^{+0.026}_{-0.019}$	& PL+vRS\\
NGC 3147& S2	& ...			& ...		& ...	& PL+GA\\
NGC 4501& S2	& $0.79^{+0.10}_{-0.16}$& 0.1(f)	& ...	& PL+RS\\
	&	& $0.77^{+0.10}_{-0.14}$& 0.5(f)	& ...	& PL+RS\\
NGC 4941& S2	& ...			& ...		& ...	& PC+GA\\
NGC 5194& S2	& $0.60^{+0.05}_{-0.04}$& $0.15\pm0.06$	& $0.046^{+0.016}_{-0.012}$	& PL+vRS+GA\\
NGC 7743& S2	& ...			& ...		& ...	& ...\tablenotemark{b}\\
\enddata
\tablenotetext{a}{
Model: 
PL = power law, 
PC = partially covered power law, 
BRE = thermal bremsstrahlung,
RS = Raymond-Smith,
vRS = variable abundance Raymond-Smith,
GA = Gaussian. 
The values in parentheses are the assumed abundance of the RS component.
}
\tablenotetext{b}{
Low signal-to-noise ratio.
}
\end{deluxetable}

\begin{deluxetable}{lllll}
\tabletypesize{\footnotesize}
\tablewidth{11cm}
\tablecaption{Summary of the Spectral Parameters for the Hard Component.
	\label{table:summary_hard}
}
\tablehead{
\colhead{Name}	& \colhead{Spectral}	& \colhead{{\NH}}& \colhead{Photon Index} & \colhead{Model\tablenotemark{a}}\\
         	&          Class                     & ($10^{22}$ \pcm )	&		& \\
}
\startdata
NGC 315	& L1.9	& $0.49^{+0.41}_{-0.33}$ & $1.73^{+0.28}_{-0.25}$ & PL+RS\\
NGC 1052& L1.9	& $2.5^{+2.7}_{-1.4}$	& $1.67^{+0.57}_{-0.40}$ & PC+RS+GA\\
	&	& $20.0^{+7.0}_{-8.3}$	& (CF $0.77^{+0.08}_{-0.10}$)\\
NGC 3998& L1.9	& $0.082\pm0.012$	& $1.90^{+0.03}_{-0.04}$ & PL+GA\\
NGC 4203& L1.9	& 0.022($<$0.053)	& $1.78^{+0.07}_{-0.08}$ & PL\\
NGC 4450& L1.9	& 0($<$0.082)		& $1.75^{+0.18}_{-0.17}$ & PL+RS (0.1)\\
	&	& 0($<$0.062)		& $1.81^{+0.16}_{-0.13}$ & PL+RS (0.5)\\
NGC 4579 (1995)	& S1.9/L1.9 & $0.04\pm0.03$ 	& $1.72\pm0.05$ & PL+RS+GA\\
NGC 4579 (1998)	& S1.9/L1.9 & 0.04($<0.13$)	& $1.81\pm0.06$	& PL+RS+GA\\
NGC 4636& L1.9	& 0.018(f)		& $1.67^{+0.64}_{-0.60}$& PL\tablenotemark{b}\\
NGC 5005& L1.9	& 0.10($<$0.86)		& $0.97\pm0.37$		& PL+RS\\
\tableline
NGC 404	& L2	& ...			& ...			& ...\tablenotemark{c}\\
NGC 3507& L2	& $1.1^{+1.0}_{-0.9}$	& $2.3^{+1.0}_{-0.7}$	& PL+RS (0.1)\\
	& 	& $0.96^{+1.0}_{-0.9}$	& $2.3^{+0.9}_{-0.6}$	& PL+RS (0.5)\\
NGC 3607& L2	& $1.6^{+1.8}_{-1.2}$	& $1.62^{+0.72}_{-0.42}$& PL+RS (0.1)\\
	&	& 0.02($<$0.6)		& $1.37^{+0.30}_{-0.28}$& PL+RS (0.5)\\
NGC 4111& L2	& 0($<$3.0)		& $0.92^{+1.03}_{-0.62}$& PL+RS (0.1)\\
	&	& 0($<$0.47)		& $1.36^{+0.46}_{-0.44}$& PL+RS (0.5)\\
NGC 4192& T2	& 0.027(f)		& $1.70^{+0.19}_{-0.16}$& PL\tablenotemark{c,d}\\
NGC 4261& L2	& 0.17($<$0.39)		& $1.30^{+0.07}_{-0.06}$ & PL+RS+GA\\
NGC 4374& L2	& 0($<2.2$)	 	& $1.29^{+0.81}_{-0.77}$& PL+RS\\
NGC 4569& T2	& $1.5^{+1.2}_{-0.8}$	& $2.18^{+0.71}_{-0.66}$& PL+RS (0.1)\\
	&	& $1.1^{+0.9}_{-0.8}$	& $2.17^{+0.62}_{-0.64}$& PL+RS (0.5)\\
NGC 4594& L2	& $0.73\pm0.29$		& $1.89\pm0.16$		& PL+vRS\\

NGC 4736& L2	& 0($<1.3$)		& $1.56^{+0.12}_{-0.15}$& PL+RS+GA\\
NGC 7217& L2	& $1.6^{+1.6}_{-1.0}$	& $2.40^{+0.79}_{-0.51}$& PL+RS (0.1)\\
	&	& $0.92^{+0.74}_{-0.62}$& $2.22^{+0.54}_{-0.44}$& PL+RS (0.5)\\
\tableline
NGC 1097& S1.5/L2	& $0.13^{+0.10}_{-0.07}$ & $1.67^{+0.09}_{-0.10}$& PL+vRS\\
NGC 1365& S1.8	& 0.54($<2.0$)		& $1.12^{+0.43}_{-0.37}$& PL+RS+GA\\
NGC 2639& S1.9	& 0($<$3.1)		& $1.64^{+0.91}_{-1.18}$& PL+RS+GA (0.1)\\
	&	& 0($<$0.31)		& $1.92^{+0.70}_{-0.37}$& PL+RS+GA (0.5)\\
NGC 3031&S1.5	& 0($<0.053$)		& $1.796^{+0.027}_{-0.028}$	&PL+RS+GA (0.1)\\
	&	& 0($<0.036$)		& $1.803^{+0.027}_{-0.020}$	&PL+RS+GA (0.5)\\
NGC 4258 (1993) & S1.9	& $12.7^{+0.9}_{-1.1}$ & $1.65^{+0.10}_{-0.16}$	& PL+RS+RS+BRE+GA \\
NGC 4258 (1996) & S1.9	& $6.6\pm0.3$	& $1.55^{+0.06}_{-0.07}$	& PL+RS+GA\\
NGC 4565& S1.5	& $0.37^{+0.35}_{-0.15}$& $2.51^{+0.35}_{-0.33}$	& PC+RS (0.1)\\*
	&	& $2.6^{+1.1}_{-1.2}$ & (CF $0.70^{+0.09}_{-0.15}$) & \\
	&	& $0.24\pm0.09$	& $2.48^{+0.31}_{-0.29}$	& PC+RS (0.5)\\*
	&	& $2.3^{+1.0}_{-0.9}$ & (CF $0.68^{+0.20}_{-0.13}$)& \\
NGC 4639& S1.0	& $0.069^{+0.041}_{-0.038}$ & $1.66\pm0.10$	& PL+GA\\
NGC 5033& S1.5	& $0.087\pm 0.017$ 	& $1.72\pm 0.04$ & PL+GA\\
\tableline
NGC 1386& S2	& $45^{+33}_{-22}$	& 1.7(f)	& PL+RS+GA\\
NGC 2273& S2	& 0($<0.070$)		& $1.54^{+0.12}_{-0.11}$ & PC+GA\\
	&	& $108^{+10}_{-9}$	& (CF $0.983^{+0.003}_{-0.004}$ & \\
NGC 2655& S2	& 0.021(f)		& $1.2^{+0.6}_{-0.7}$	& PC+RS (0.1)\\
	&	& $40^{+22}_{-13}$	& (CF $0.91^{+0.05}_{-0.13}$) & \\
NGC 3079& S2	& $1.7^{+1.8}_{-1.2}$	& $1.96^{+0.74}_{-0.71}$ & PL+vRS\\
NGC 3147& S2	& $0.062^{+0.05}_{-0.024}$ & $1.82^{+0.10}_{-0.09}$ & PL+GA\\
NGC 4501& S2	& $0.66(<2.3)$		& $1.49^{+0.74}_{-0.51}$ & PL+RS (0.1)\\
	&	& $0(<0.90)$		& $1.48^{+0.53}_{-0.30}$ & PL+RS (0.5)\\
NGC 4941& S2	& $0.16^{+0.14}_{-0.11}$& $1.48^{+0.14}_{-0.15}$ & PC+GA\\
	& 	& $99^{+12}_{-11}$ 	& (CF $0.966^{+0.006}_{-0.007}$) & \\ 
NGC 5194& S2	& $2.2^{+1.2}_{-1.3}$ 	& $1.65^{+0.42}_{-0.47}$ & PL+vRS+GA\\
NGC 7743& S2	& 0.053(f)		& $1.89^{+0.16}_{-0.13}$ & PL\tablenotemark{c,d}\\
\enddata
\tablenotetext{a}{
Model: 
PL = power law, 
PC = partially covered power law, 
BRE = thermal bremsstrahlung,
RS = Raymond-Smith,
vRS = variable abundance Raymond-Smith,
GA = Gaussian. 
The values in parentheses are the assumed abundance of the RS component.
}
\tablenotetext{b}{
Power-law fit in the 4--10 keV band.
}
\tablenotetext{c}{
Low signal-to-noise ratio.
}
\tablenotetext{d}{
The photon index is determined from a hardness ratio assuming a
power-law spectrum and Galactic absorption. Errors are at the
1$\sigma$ level.
}
\end{deluxetable}

\begin{deluxetable}{lllllll}
\tabletypesize{\footnotesize}
\tablewidth{13cm}
        \tablecaption{Summary of X-ray Fluxes\tablenotemark{a}}
        \label{table:flux}
\tablehead{
\colhead{Name}	& \multicolumn{2}{c}{Total}	& \multicolumn{2}{c}{Power Law}	& \multicolumn{2}{c}{Raymond-Smith}\\
		& (observed)		& (observed) 	& (observed)	& (intrinsic)	& (observed)	& (intrinsic)\\
        & 0.5--2~keV	& 2--10~keV	& 2--10 keV	& 2--10 keV	& 0.5--4 keV	& 0.5--4 keV}
%\hline
\startdata
NGC 315		& 0.450		& 0.908		& 0.892		& 0.935		& 0.271	& 0.325\\
NGC 404		& $<0.011$	& $<0.066$	& ...		& ...		& ...	& ...\\
NGC 1052	& 0.349		& 4.78		& 4.72		& 9.94		& 0.316	& 0.344\\
NGC 1097	& 1.27		& 2.08		& 2.04		& 2.07		& 0.551	& 0.587\\
NGC 1365	& 0.506		& 1.16		& 1.12		& 1.16		& 0.458	& 0.479\\
NGC 1386	& 0.258		& 0.388		& 1.27		& 1.27		& 0.159	& 0.165\\
NGC 2273	& 0.0684	& 1.25		& 1.25		& 8.71		& ... 	& ...\\
NGC 2639	& 0.167		& 0.323		& 0.323		& 2.14		& ...	& ...(PC)\tablenotemark{b}\\
		& 0.157		& 0.190		& 0.182		& 0.182		& 0.089	& 0.097 (0.1)\tablenotemark{c}\\
		& 0.150		& 0.246		& 0.243		& 0.244		& 0.048	& 0.053 (0.5)\\
NGC 2655	& 0.210		& 1.04		& 1.03		& 3.03		& 1.46	& 1.57 (0.1)\\
NGC 3031	& 10.1		& 18.3		& 18.3		& 18.4		& 0.271	& 0.306 (0.1)\\
NGC 3079	& 0.463		& 0.457		& 0.431		& 0.504		& 0.441	& 0.454\\
NGC 3147	& 0.792		& 1.64		& 1.64		& 1.67		& ...	& ...\\
NGC 3507	& 0.090		& 0.161		& 0.158		& 0.178		& 0.055	& 0.058 (0.1)\\
		& 0.085		& 0.157		& 0.156		& 0.173		& 0.042	& 0.045 (0.5)\\
NGC 3607	& 0.276		& 0.298		& 0.281		& 0.318		& 0.269	& 0.286	(0.1)\\ 
		& 0.263		& 0.309		& 0.303		& 0.304		& 0.175	& 0.186 (0.5)\\
NGC 3998	& 4.56		& 8.06		& 8.06		& 8.13		& ...	& ...\\
NGC 4111	& 0.206		& 0.260		& 0.252		& 0.252		& 0.173	& 0.181 (0.1)\\
		& 0.194		& 0.225		& 0.222		& 0.222		& 0.128	& 0.134	(0.5)\\
NGC 4192	& 0.057		& 0.112		& ...		& ...		& ...	& ...\\
NGC 4203	& 1.19		& 2.05		& 2.05		& 2.06		& ...	& ...\\
NGC 4258 (1993) & 2.06		& 7.13		& 6.55		& 12.3		& 2.38	& 2.48\\
NGC 4258 (1996)	& 1.83		& 14.4		& 14.3		& 29.9		& 1.90	& 1.98\\
NGC 4261	& 0.837		& 1.04		& 0.988		& 1.00		& 0.711	& 0.745\\
NGC 4374	& 1.46		& 0.760		& 0.650		& 0.651		& 1.46	& 1.58\\	
NGC 4438	& 0.402		& 0.278		& 0.243 & 0.276	& 0.400	& 0.426 (0.1)\\
		& 0.406		& 0.272		& 0.262 & 0.263	& 0.189	& 0.201 (0.5)\\
NGC 4450	& 0.444		& 0.658		& 0.654		& 0.655		& 0.0857& 0.0932(0.1)\\
		& 0.443		& 0.646		& 0.644		& 0.646		& 0.0565& 0.0613(0.5)\\
NGC 4457	& 0.175		& 0.251		& 0.242		& 0.265		& 0.153	& 0.163	(0.1)\\
		& 0.166		& 0.260		& 0.257		& 0.261		& 0.100	& 0.106 (0.5)\\
NGC 4501	& 0.343		& 0.575		& 0.552		& 0.576		& 0.280	& 0.290 (0.1)\\
		& 0.340		& 0.574		& 0.567		& 0.569		& 0.142	& 0.147 (0.5)\\
NGC 4565	& 0.715		& 1.64		& 1.60		& 2.20		& 0.286	& 0.320\\
NGC 4569	& 0.264		& 0.301		& 0.288		& 0.338		& 0.235	& 0.256 (0.1)\\
		& 0.250		& 0.302		& 0.296		& 0.332		& 0.198	& 0.218 (0.5)\\
NGC 4579 (1995)	& 2.18 		& 4.27		& 4.24		& 4.27		& 0.319 & 0.347 (0.5)\\
NGC 4579 (1998) & 3.01		& 5.86		& 5.84		& 5.86		& 0.174	& 0.189 (0.5)\\
NGC 4594	& 1.22		& 2.78		& 2.74		& 2.94		& 0.689	& 0.776\\% (vRS fit)\\
NGC 4636	& 5.43		& 0.768		& 0.479		& 0.482		& 5.51	& 5.82\\
NGC 4639	& 0.436		& 1.07		& 1.07		& 1.07		& ...	& ...\\
NGC 4736	& 1.32		& 1.93		& 1.91		& 1.91		& 0.54	&0.57\\
NGC 4941	& 0.089		& 1.31		& 1.31		& 9.88		& ...	& ...\\
NGC 5005	& 0.426		& 0.730		& 0.701		& 0.706		& 0.353	&0.366\\
NGC 5194	& 1.26		& 0.919		& 0.854		& 0.999		& 1.28	& 1.34\\
NGC 5033	& 2.40		& 5.49		& 5.49		& 5.53		& ...	& ...\\
NGC 7217	& 0.127		& 0.212		& 0.202		& 0.244		& 0.104	&0.144 (0.1)\\
		& 0.121		& 0.215		& 0.211		& 0.234		& 0.074	& 0.102 (0.5)\\
NGC 7743	& 0.046		& 0.073		& ...		& ...		& ...	& ...\\
\enddata
%\hline
%\end{tabular}
%\end{center}
\tablenotetext{a}{In units of 10$^{-12}$ ergs cm$^{-2}$ s$^{-1}$.}

\tablenotetext{b}{Partial covering model without a Gaussian component.}

\tablenotetext{c}{The values in parentheses denote the assumed abundance of the R-S component.}
\end{deluxetable}

\begin{deluxetable}{lllllll}
\tablewidth{13cm}
\tabletypesize{\footnotesize}
        \tablecaption{Summary of X-ray Luminosities\tablenotemark{a}}
        \label{table:luminosity}
\tablehead{
\colhead{Name}	& \multicolumn{2}{c}{Total}	 & \multicolumn{2}{c}{Power Law}	& \multicolumn{2}{c}{Raymond-Smith}\\
	& (observed) 	& (observed) 	& (observed)	& (intrinsic)	& (observed)	& (intrinsic)\\
	& 0.5--2~keV	& 2--10~keV	& 2--10~keV	& 2--10~keV	& 0.5--4~keV	& 0.5--4~keV}
\startdata
NGC 315		&23.4	& 47.2		& 46.3		& 48.6		& 14.1	& 16.9\\
NGC 404		&$<0.00076$ & $<0.0046$	& ...		& ...		& ...	& ...\\
NGC 1052	& 1.33	& 18.2		& 17.9		& 37.8		& 1.20 	& 1.31\\
NGC 1097	& 3.20	& 5.24 		& 5.16		& 5.22		& 1.39 	& 1.48\\
NGC 1365	& 1.73	& 3.98		& 3.84		& 3.98		& 1.57	& 1.64\\
NGC 1386	& 0.884	& 1.33		& 4.35		& 4.35		& 0.544	& 0.565\\
NGC 2273	& 0.662	& 12.1 		& 12.1 		& 84.3		& ... 	& ...\\
NGC 2639	& 3.64	& 7.03		& 7.03		& 46.6		& ...	& ... (PC)\tablenotemark{b}\\
		& 3.42	& 4.14		& 3.96		& 3.96		& 1.94	& 2.11 (0.1)\tablenotemark{c}\\
		& 3.27	& 5.36		& 5.29		& 5.31		& 1.05	& 1.14 (0.5)\\
NGC 2655	& 1.50	& 7.43		& 7.36		& 21.6		& 1.04	& 1.12 (0.1)\\
NGC 3031	& 0.238	& 0.430		& 0.430		& 0.433		& 0.0637& 0.0720 (0.1)\\
NGC 3079	& 2.31	& 2.28 		& 2.15		& 2.52		& 2.20	& 2.27\\
NGC 3147	& 1.59	& 32.9		& 32.9		& 33.5		& ...	& ...\\
NGC 3507	& 0.423	& 0.757		& 0.743		& 8.37		& 0.259	& 2.73 (0.1)\\
		& 0.400	& 0.739		& 0.733		& 8.14		& 0.198	& 2.12 (0.5)\\
NGC 3607	& 1.31	& 1.42		& 1.34 		& 1.51 		& 1.28	& 1.36 (0.1)\\
		& 1.24	& 1.47 		& 1.44		& 1.44		& 0.832	& 0.884 (0.5)\\
NGC 3998	& 25.5	& 45.1		& 45.1		& 45.5		& ...	& ...\\
NGC 4111	& 0.714	& 0.902		& 0.874		& 0.874		& 0.600	& 0.628 (0.1)\\
		& 0.673	& 0.780		& 0.769		& 0.769		& 0.444	& 0.464 (0.5)\\
NGC 4192	& 0.19	& 0.38		& ...		& ...		& ...	& ...\\
NGC 4203	& 1.34	& 2.32		& 2.32		& 2.32		& ...	& ...\\
NCG 4258(1993)	& 1.14	& 3.96		& 3.63		& 6.82		& 1.32	& 1.38\\
NCG 4258(1996)	& 1.02	& 8.03		& 7.95		& 16.6		& 1.05	& 1.10\\
NGC 4261	& 12.3	& 15.5		& 14.6		& 14.8		& 10.5	& 11.0\\
NGC 4374	& 4.93	& 2.57		& 2.20		& 2.20		& 4.68	& 5.08\\
NGC 4438	& 1.36	& 0.940		& 0.823		& 0.934		& 1.35	& 1.44 (0.1)\\
		& 1.37	& 0.919		& 0.889		& 0.890		& 0.641	& 0.682 (0.5)\\
NGC 4450	& 1.50	& 2.23 		& 2.21		& 2.22		& 0.290	& 0.315 (0.1)\\
		& 1.50	& 2.19 		& 2.18		& 2.19		& 0.191	& 0.208 (0.5)\\
NGC 4457	& 0.636	& 0.912		& 0.879		& 0.963		& 0.556	& 0.592 (0.1)\\
		& 0.603	& 0.944		& 0.934		& 0.948		& 0.363	& 0.385 (0.5)\\
NGC 4501	& 1.16	& 1.94		& 1.87		& 1.95		& 0.948	& 0.982 (0.1)\\
		& 1.15	& 1.94		& 1.92		& 1.93		& 0.481	& 0.498 (0.5)\\
NGC 4565	& 0.807	& 1.84 		& 1.81		& 2.48		& 0.323	& 0.361\\
NGC 4569	& 0.894	& 1.02		& 0.975		& 1.14		& 0.796	& 0.867\\
		& 0.847	& 1.02		& 1.00		& 1.12		& 0.671	& 0.738\\
NGC 4579 (1995)	& 7.40	& 14.4		& 14.4		& 14.4		& 1.08 	& 1.17 (0.5)\\
NGC 4579 (1998) & 10.2	& 19.8		& 19.8		& 19.8		& 0.589	& 0.640 (0.5)\\
NGC 4594	& 5.86	& 13.3		& 13.2		& 14.1		& 3.31	& 3.72 \\%(vRS)\\
NGC 4636	& 18.8	& 2.66		& 1.66		& 1.67		& 19.1	& 20.2\\% (vRS)
NGC 4639	& 1.48	& 3.62		& 3.62		& 3.62		& ...	& ...\\
NGC 4736	& 0.293	& 0.428		& 0.424		& 0.424		& 0.120	&0.126 \\
NGC 4941	& 0.044	& 0.644		& 0.644		& 4.86		& ...	& ...\\
NGC 5005	& 2.32	& 3.97		& 3.82		& 3.84		& 1.92	& 1.99\\
NGC 5033	& 10.1	& 23.0		& 23.0		& 23.2		& ...	& ...\\
NGC 5194	& 0.896	& 0.654		& 0.608		& 0.711		& 0.911	&0.953\\
NGC 7217	& 0.390	& 0.651		& 0.620		& 0.749		& 0.319	&0.442 (0.1)\\
		& 0.372	& 0.660		& 0.648		& 0.719		& 0.227	&0.313 (0.5)\\
NGC 7743	& 0.33	& 0.52		& ...		& ...		& ...	& ...\\
\enddata
\tablenotetext{a}{In units of 10$^{40} $ergs s$^{-1}$.}

\tablenotetext{b}{Partial covering model without a Gaussian component.}

\tablenotetext{c}{The values in parentheses denote the assumed abundance of the R-S component.}

\end{deluxetable}

\begin{table*}
\begin{center}
	\caption{Long-term Variability}
\begin{tabular}{clclr}
\tableline \tableline
Name	& Dates	& Flux\tablenotemark{a} 	& Instruments	& References\\
\tableline
NGC 1052& 1996 Aug 11	& 4.78			& \asca	& \\
	& 2000 Jan & 4.0			& \sax	& 1\\
NGC 1365& 1994 Aug 12, 1995 Jan 25\tablenotemark{b}	& 1.16	& \asca & \\
	& 1997 Aug	& 6.6			& \sax	& 2\\
NGC 1386& 1995 Jan 26 	& 0.39			& \asca & \\
	& 1996 Dec 10	& 0.24			& \sax	& 3\\
NGC 2273& 1996 Oct 20	& 1.25			& \asca	& \\
	& 1997 Feb 22	& 0.99\tablenotemark{c}		& \sax	& 3\\
NGC 3031& 1993 Apr 30--1998 Oct 20\tablenotemark{d} 	& 12-40	& \asca & 4\\
	& 1998 Jun 04	& 38			& \sax	& 5\\
NGC 3079& 1993 May 09	& 0.46			& \asca	& \\
	& 2000 May 26	& 0.34			& \sax	& 6\\
NGC 3998& 1988 Apr 30	& 15 			& \Ginga& 7\\
	& 1994 May 10	& 8.06			& \asca\\
	& 1999 Jun 29	& 12.0			& \sax	& 8\\
NGC 4258& 1993 May 15	& 7.13			& \asca\\
	& 1996 May 23--1996 Dec 18\tablenotemark{e} & 14.4	& \asca\\
	& 1998 Dec 19--22	& 8.0		& \sax	& 9\\
	& 1999 May 15--20	& 5.8		& \asca	& 10\\
NGC 4565& 1994 May 28	& 1.64\tablenotemark{f}		& \asca \\
	& ...	& 1.8\tablenotemark{c}			& \sax	& 11\\
NGC 4941& 1996 Jul 19, 1997 Jan 08\tablenotemark{b}	& 1.31	& \asca\\
	& 1997 Jan 22	& 0.66			& \sax	& 3\\
NGC 5005& 1995 Dec 13	& 0.73			& \asca\\
	& ...		& 0.3			& \sax	& 11\\
NGC 5033& 1984 Feb 03	& 4.7			& \exosat& 12\\
	& 1995 Dec 14	& 5.5			& \asca\\
NGC 5194& 1988 May 03	& 6.1			& \Ginga & 13\\
	& 1993 May 11	& 0.919			& \asca\\
	& 2000 Jan 18--20& 0.34			& \sax	& 14\\
\tableline
\end{tabular}
\end{center}
\tablerefs{
(1) Guainazzi et al. 2001; 
(2) Risaliti et al. 2000;
(3) Maiolino et al. 1998;
(4) Iyomoto \& Makishima 2001;
(5) Pellegrini et al. 2000a; 
(6) Iyomoto et al. 2001;
(7) Awaki et al. 1991;
(8) Pellegrini et al. 2000b; 
(9) Fiore et al. 2001;
(10) Reynolds et al. 2000;
(11) Risaliti et al. 1999;
(12) Turner \& Pounds 1989;
(13) Makishima et al. 1990;
(14) Fukazawa et al. 2001.
}
\tablenotetext{a}{Observed flux in the 2--10 keV band (not corrected for absorption),
in units of $10^{-12}$ ergs \ps \pcm.}
\tablenotetext{b}{The two observations were combined.}
\tablenotetext{c}{No description of a nearby source in the reference.}
\tablenotetext{d}{Sixteen observations during this period.}
\tablenotetext{e}{Three observations during this period.}
\tablenotetext{f}{Includes the flux from a nearby source.
}
\end{table*}
 
\end{document}